%% file: journal-v7.tex
\documentclass[11pt,epsf,amssymb,ulem,qsymbols]{article}
\pdfoutput=1

\usepackage{jcappub}

\usepackage{tabularx}
\usepackage{array}
\usepackage{graphics}
\usepackage{graphicx}
\usepackage{psfrag}
\usepackage{epsfig}
\usepackage{amsmath}
\usepackage{amssymb}
\usepackage{ulem}
\usepackage{setspace}
\usepackage{rotating}
\usepackage{colortbl}
\usepackage{tabularx}
\usepackage{longtable}
\usepackage{multirow}
\usepackage[utf8]{inputenc}
\usepackage{cases}
\usepackage{tikz}
\usepackage{epstopdf}
\usetikzlibrary{snakes}
\makeatletter

\numberwithin{equation}{section}

\usepackage{verbatim}

\input mylatex

\setlongtables

\newcommand{\ec}{\end{center}}
\newcommand{\bc}{\begin{center}}

\newcommand{\pdir}{p\kern -5.2pt\raise 0.2ex\hbox {/}}

\newcommand{\vdir}{v\kern -5.75pt\raise 0.15ex\hbox {/}}
\newcommand{\kdir}{k\kern -5.75pt\raise 0.15ex\hbox {/}}
\newcommand{\epsdir}{\epsilon\kern -5.0pt\raise 0.15ex\hbox {/}}
\newcommand{\bvdir}{\bar{v}\kern -5.75pt\raise 0.15ex\hbox {/}}
\newcommand{\Ddir}{D\kern -7.75pt\raise 0.20ex\hbox {/}}
\newcommand{\Adir}{A\kern -7.75pt\raise 0.20ex\hbox {/}}
\newcommand{\ldir}{l\kern -5.0pt\raise 0.2ex\hbox{/}}
\newcommand{\varepsdir}{\varepsilon\kern -5.5pt\raise 0.15ex\hbox{/}}

\makeatother

\def\mphia{m_{\phi_1}}
\def\mphib{m_{\phi_2}}
\def\la{\lambda_1}
\def\lb{\lambda_2}
\def\lc{\lambda_3}

\def\mO{\mathcal{O}}
\def\mZ{\mathcal{Z}}

\def\be{\begin{equation}}
\def\ee{\end{equation}}

\def\lsp{\quad}

\def\ZZ{\mZ_2 \times \mZ_2}
\def\ZZp{\mZ_2 \times \mZ_2^{'}}

\def\gt{\zeta}

\definecolor{straw}{rgb}{1,1,0.50}
\definecolor{lstraw}{rgb}{1,1,0.70}
\definecolor{red}{rgb}{1.00,0.00,0.00}
\definecolor{teal}{rgb}{0.70,0.70,0.90}
\definecolor{darkblue}{rgb}{0,0,0.50}
\definecolor{lblue}{rgb}{0.8,0.8,1}

\title{Multipartite Interacting Scalar Dark Matter in the light of updated LUX data}
\author{Subhaditya Bhattacharya,}
\author{Purusottam Ghosh,}
\author{Poulose Poulose}
\affiliation{Department of Physics, Indian Institute of Technology Guwahati\\Guwahati, Assam 781039, India}
\emailAdd{subhab@iitg.ernet.in}
\emailAdd{p.ghosh@iitg.ernet.in}
\emailAdd{poulose@iitg.ernet.in}

\abstract{We explore constraints on multipartite dark matter (DM) framework composed of singlet scalar DM interacting with the Standard Model (SM) through Higgs portal coupling. We compute relic density and direct search constraints including the updated LUX bound for two component scenario with non-zero interactions between two DM components in $\ZZp$ framework in comparison with the one having $\mO(2)$ symmetry. We point out availability of a significantly large region of parameter space of such a multipartite model with DM-DM interactions.}


\begin{document}
\maketitle
\flushbottom

\section{\label{sec-0}Introduction}

Dark matter (DM) is postulated to form a significant part of the universe. Evidences have come from different astrophysical and cosmological observations such as from the motion of stars in galaxies \cite{Oort:1932}, the motion of galaxies in clusters \cite{Zwicky1,Zwicky2,Rubin}, structure formation \cite{Klypin:1992sf}
and the inhomogeneities in the CMBR \cite{WMAP,PLANCK}. However, DM hasn't been discovered yet; although theoretical attempts have been made to accommodate DM in different extensions of the Standard Model (SM). 

We consider one of the simplest possibilities of DM models in this note and a comparative analysis is made to understand the distinction between single and multicomponent nature of DM through self interactions. 

A SM gauge singlet real scalar which is odd under a $\mZ_2$ symmetry, while the SM is even under $\mZ_2$, provides with a simplest dark matter candidate which has portal interactions through the SM Higgs. This model has been studied extensively for its simplicity and predictability \cite{s1,s2,s3,s4,s5,s6,s7,s8,s9,singlet-FIMP,singlet-Z3,singlet-other1,singlet-other2}. This model is also known to reduce the fine-tuning problem SM Higgs is plagued with \cite{singlet-ft}. One can accommodate as many of them as possible to extend to a multicomponent DM framework. The model can inherently carry a $\mO(N)$ symmetry for $N$ components on top of the individual $\mZ_2$ symmetry, which however makes all the DMs having equal masses and same coupling strength to SM \cite{multi-singlet1}. This particular feature restricts the model not to have non-zero interaction between different DM species, although such terms are present in the scalar potential. Hence, it results in no alternation to the thermal freeze out of each DM component compared to the single component scenario. The departure in allowed parameter space from the singlet scalar DM framework to such a multipartite framework is significant subject to the number of DM components, however equally restrictive. On the contrary, if one ignores the $\mO(N)$ symmetry the masses of different DM components and their interactions with the SM can be set to be different from each other. Further, the absence of $\mO(N)$ symmetry leads to  non-zero interactions between the DM components. We study such interacting multipartite scalar DM scenario, specifically for two component case. Generalisation to three or more components is straight forward, but not discussed in this report. We demonstrate the case when there are two distinct $\mZ_2$ symmetry on each of the components and call that a model under $\mZ_2 \times \mZ_2^{'}$. It is important to note that when we have two DM scalar singlets odd under same $\mZ_2$, i.e.  a framework with $\ZZ$, there is inevitably one DM component, lighter of the two; the heavier one will decay to the lighter one at tree level or through loop generated process, depending on the mass splitting between the components. 

Although the models have been discussed earlier \cite{multi-singlet2,multi-singlet3,multi-singlet4,multi-singlet5}, we perform a systematic comparative analysis of the DM-allowed parameter space of the two-component framework. We explore the viability of $\ZZp$ model for relic abundance constraints from WMAP \cite{WMAP}, PLANCK \cite{PLANCK} data, and direct detection constraint from XENON \cite{Xenon1,Xenon2}, LUX \cite{LUX} as well as from updated LUX bound \cite{LUX2}. We specifically indicate that direct search constraints can be shown to be somewhat relaxed in presence of DM-DM interactions in multicomponent framework, allowing the DM components to be delayed till XENON1T limit. 

The paper is organised as follows. We discuss the singlet scalar single and multicomponent DM frameworks in Sec.~\ref{model}. Thermal freeze out and corresponding Boltzmann Equations are discussed in Sec.~\ref{BE}. Parameter space scan on relic density of $\ZZp$ framework is discussed in Sec.~\ref{ZZp} and approximate analytical solution of such a case is discussed in Sec.~\ref{ZZpA}. Direct search constraints are discussed next in Sec.~\ref{DD}. Finally we conclude.

\section{Models for the Singlet Scalar DM}
\label{model}
\subsection{Single Component framework}
\label{single}

Let us first review the DM model with a real singlet scalar $ \phi $ \cite{s1}. The interaction with the SM can be written in general of the form $\frac{1}{\Lambda^{n-4}} \ocal_{DM} \ocal_{SM} $. Here $ \ocal_{SM} $ is gauge invariant operator composed of the SM fields, whereas $ \ocal_{DM} $ are those involving the DM fields. $\Lambda$ is the new physics scale in effective theory formalism, whose power is determined by the dimension of the operator product, $n$. Since we are interested in renormalisable interaction terms, we consider only operators with $n \le 4$. Stability of the DM is achieved through imposing a $Z_2$ symmetry, under which $ \phi $ is specified to be odd with other fields are taken to be even. This necessitates $\phi$ to appear in even powers in the operators. Thus, the scalar potential, including the DM $ \phi $ interacts with the SM through the Higgs portal term reads as \cite{sdm1,sdm2}
\begin{eqnarray}
V(H, \phi) = 
- \mu_{H}^2 H^{\dagger} H 
+ \lambda_{H} (H^{\dagger} H)^2 
+ \frac{1}{2}  \mu_{\phi}^2 \phi^2
+ \frac{1}{4!} \lambda_{\phi} \phi^4
 + \frac{1}{2} \lambda_{1} H^{\dagger} H \phi^2 \,,
\label{pot}
\end{eqnarray}
where $\lambda_1$ parametrizes the `Higgs-portal' interaction discussed above, with $H$ denoting the SM Higgs isodoublet.
The Lagrangian density for the scalar sector is then given by:
\begin{eqnarray}
\lcal_{\rm scal} = \frac{1}{2} \partial_{\mu} \phi \partial^{\mu} \phi + 
D_{\mu}H^{\dagger} D^{\mu} H 
- V(H, \phi)\,,
\end{eqnarray}
where $D_\mu$ is the covariant derivative related to the SM gauge symmetry. Electroweak symmetry breaking (EWSB) occurs via non-zero vacuum expectation value of the Higgs doublet $\left<H\right> = (0, v/\sqrt{2})^T$, with $v = 246~ \gev$.
On the other hand, unbroken $Z_2$ symmetry requires $\mu_\phi^2 > 0$, ensuring  $\left<\phi\right> = 0$,  which yields DM mass term $m_{\phi}^2 = \mu_{\phi}^2 +\frac{ \lambda_{1} v^2}{2} $. 

In order to stabilize the vacuum we require that the scalar potential in eq.~(\ref{pot}) 
is bounded from below. At the tree level it implies the following conditions \cite{vs1,vs2} 
\beq
\lambda_{\phi} > 0  \,,\lambda_{H} > 0 \,, \lambda_1^2 <  {\frac{2}{3}\lambda_{\phi}\lambda_{H}} ~. \quad
\label{stab_con}
\eeq

Tree-level unitarity constraints imply \cite{vs3}
\beq
\lambda_{\phi} < 8 \pi , \lsp
|\lambda_1| < 4 \pi\,~.
\label{unit_con}
\eeq



Finally, the condition that the global $\mZ_2$ symmetry remains unbroken requires $\mu_{\phi}^2 > 0$, which leads to $m_{\phi}^2 > \lambda_{1}v^2/2$.
DM relic density and direct search constraints are relevant for this model in a two dimensional parameter space denoted by the DM mass and the Higgs-portal coupling as 
\bea
\{m_{\phi},\lambda_1\} ~ .
\eea
The phenomenology and the allowed parameter space for such a model has been discussed by many authors \cite{s2,s3,s4,s5,s6,s7}. In this report we shall limit our discussion in citing some of the important outcomes of this model in the comparative analysis with two component interacting DM frameworks that we analyse shortly in details. 

\subsection{Multipartite framework with $\mathcal{O}(N)$ symmetry}
\label{multi1}

Simplest extension of a single component scalar DM framework to a multipartite case is obtained by assuming $N (\ge 2)$ such components of DM having same mass and coupling to the SM, replacing $\phi$ in above section by 
\bea
\vec{\phi} \equiv \{\phi_1,\phi_2,\phi_3...\phi_N\},
\eea
so that the scalar potential reads \cite{multi-singlet1}:
\bea
V(H, \varphi) = 
- \mu_{H}^2 H^{\dagger} H 
+ \lambda_{H} (H^{\dagger} H)^2 
+ \frac{1}{2}  \mu_{\vec{\phi}}^2 \vec{\phi}^2
+ \frac{1}{4!} \lambda_{\vec{\phi}} \left(\vec{\phi}^2 \right)^2
 + \frac{1}{2} \lambda_{1} H^{\dagger} H \vec{\phi}^2 \,.
\label{pot-O(n)}
\eea

Note here that the presence of an unbroken $\mO(N)$ symmetry in the Lagrangian implies the same couplings $\mu_{\vec{\phi}},~\lambda_{\vec{\phi}}$ and $ \lambda_1$ for all 
the DM components. This makes sure that each of the DM acquires same mass after EWSB as $m_{\vec{\phi}}^2=\mu_{\vec{\phi}}^2+\frac{\lambda_1 v^2}{2}$ and of course have same coupling 
to SM sector, $\lambda_1$. The term $ \left(\vec{\phi}^2 \right)^2$ in Eq. \ref{pot-O(n)} although indicate the presence of interaction vertices between different DM components, the fact that all the DM components acquire same mass, yields effectively zero DM-DM interaction cross-section and thus drastically reduces the phenomenological possibilities of this model. This multipartite DM framework has been discussed in great details in \cite{multi-singlet1} and we will refer to some of its important features in the following comparative analysis. It should be noted here that on top of the $\mO(N)$, a separate $\mZ_2$ symmetry has to be imparted to each of the DM components for their stability so that 
\bea
\phi_i \xrightarrow{\mZ_2} -\phi_i ~.
\eea
Vacuum stability of the potential dictates similar constraints as that of the single component case on the dimensionless couplings introduced here. 
This model has then three parameters as 
\bea
\{m_{\phi},~\lambda_1,~N\} ~,
\eea
where the cross-sections are essentially governed by the same DM mass and coupling.  The third parameter,  $N$ is the number of DMs considered in the model, 
which crucially changes the outcome, as we will see later. However it must be pointed out that, $\mO(N)$ symmetry is not a necessary requirement for the components to
be DM candidates. In the following, we shall explore some of the possibilities when the $\mO(N)$ is not respected. 

\subsection{Two Component framework with $\ZZp$ symmetry}
\label{2comp1}
When we break the global $\mO(N)$ symmetry as described above and impose a separate $\mZ_2$ on each of the components the scenario gets phenomenologically more interesting
with interaction between different DM candidates. The simplest multipartite framework is to consider a two component DM set up by imposing two distinguishable 
$\mZ_2 \times \mZ_2^\prime$ symmetry under which the two singlet scalars ($\phi_1,\phi_2$) stabilize. That means, under  $\mZ_2 \times \mZ_2^\prime$, different fields 
transform as: SM $ [+,+]$, $\phi_1 [-,+]$, $\phi_2 [+,-]$. The new Lagrangian involves a few more parameters compared to the previous case respecting $\mO(N)$ symmetry. 
The part of the Lagrangian involving only singlet scalar fields read \cite{multi-singlet2}

\begin{eqnarray}
\mathcal{L}_{DM}&=&\frac{1}{2}(\partial_\mu \phi_1)^2+\frac{1}{2}(\partial_\mu \phi_2)^2-\frac{1}{2}\mu_1^2 \phi_1^2-\frac{1}{2}\mu_2^2 \phi_2^2-\frac{1}{4}\lambda_3\phi_1^2\phi_2^2-\frac{1}{4!}\lambda_4\phi_1^4-\frac{1}{4!}\lambda_5\phi_2^4  ~.
\label{eq:z2z2p}
\end{eqnarray}

The part of the Lagrangian describing the interaction of DM fields with the SM Higgs fields is given by 

\begin{equation}
-\mathcal{L}_{SM-DM}=\frac{1}{2}\lambda_1\phi_1^2 H^{\dagger}H+\frac{1}{2}\lambda_2\phi_2^2 H^{\dagger}H.
\end{equation}

The scalar potential hence read:

\bea
V(\phi_1,\phi_2,H)&=&\frac{1}{2}\mu_1^2 \phi_1^2+\frac{1}{2}\mu_2^2 \phi_2^2+\frac{1}{4}\lambda_3\phi_1^2\phi_2^2+\frac{1}{4!}\lambda_4\phi_1^4+\frac{1}{4!}\lambda_5\phi_2^4 \nonumber \\ &&- \mu_{H}^2 H^{\dagger} H 
+ \lambda_{H} (H^{\dagger} H)^2 +\frac{1}{2}\lambda_1\phi_1^2 H^{\dagger}H+\frac{1}{2}\lambda_2\phi_2^2 H^{\dagger}H ~.
\label{pot-z2z2p}
\eea

Stability of this potential requires \cite{vs1,vs2}
\be
\lambda_{4} > 0  \,,\lambda_{5} > 0  \,,\lambda_{H} > 0 \, ,
\lambda_1^2 <  \frac{2}{3}\lambda_{4}\lambda_{H} \,, \lambda_2^2 < \frac{2}{3}\lambda_{5}\lambda_{H} \,, \lambda_3^2 < \frac{1}{9}\lambda_{4}\lambda_{5} \, ~.
\label{stab_conA}
\ee

We may note that, unlike the case where $\mO(N)$ symmetry is realised (Eq.~\ref{pot-O(n)}), here the couplings of $\phi_1$ and $\phi_2$ could be independent of each other.
After the electroweak phase transition, with the vacuum expectaion values of fields given by $\left<H\right> = (0, v/\sqrt{2})^T$, where $v = 246~\gev$, $\left<\phi_1\right> = \left<\phi_2\right>=0$, the part of the Lagrangian involving the DM fields takes the form

\begin{eqnarray}
\mathcal{L}_{DM}+\mathcal{L}_{SM-DM}&=&\frac{1}{2}(\partial_\mu \phi_1)^2+\frac{1}{2}(\partial_\mu \phi_2)^2-\frac{1}{2} m_{\phi_1}^2 \phi_1^2-\frac{1}{2}m_{\phi_2}^2\phi_2^2 \nonumber \\
&-& \frac{1}{4}\lambda_3\phi_1^2\phi_2^2 -\frac{1}{4!}\lambda_4\phi_1^4-\frac{1}{4!}\lambda_5\phi_2^4 \nonumber \\
&-& \frac{1}{4}\lambda_1\phi_1^2 h^2+\frac{1}{4}\lambda_2\phi_2^2 h^2 \nonumber\\
&-&\frac{1}{2}\lambda_1 v h\phi_1^2+\frac{1}{2}\lambda_2 v h\phi_2^2 ~.
\end{eqnarray}

Here the mass of the DM candidates are $m_{\phi_1}^2=(\mu_1^2 +\frac{\lambda_1v^2}{2})$ and  $m_{\phi_2}^2=(\mu_2^2 +\frac{\lambda_2v^2}{2})$. 
In the absence of $\mO(2)$ symmetry, the parameters $\lambda_i$ and $\mu_i^2$ are not necessarily the same for different fields, leading to possible mass splittings 
between different DM components. Presence of the mutual interaction  coupling, $\lambda_3$, which is absent in the scenario with $\mO(2)$ symmetry, along with the mass 
splitting could have dramatic effects, resulting in clearly distinguishable features. The main focus of this work is to highlight these features, which we shall do in the 
following sections in detail.

The relevant DM phenomenology of such a two-component framework is mainly dictated by the five parameters:
\be
\{m_{\phi_1},m_{\phi_2},\lambda_1,\lambda_2,\lambda_3\}.
\label{eq:paramset}
\ee
The parameter $\lambda_3$ solely determines the direct interaction between the two DM components. As  we shall demonstrate, presence of non-zero value of $\lambda_3$ marks a
significant departure in the allowed DM parameter space in terms of relic density and direct search constraints.  It is straightforward to extend such a model to multipartite
framework by considering $n$ such DM components $\{\phi_1,\phi_2,...\phi_n\}$ stabilised by $\mZ_2^{(1)}\times\mZ_2^{(2)}...\times \mZ_2^{(n)}$ symmetry, which involves $(n+n+{}^n{C_2})$ parameters over and above the SM parameters. In line with Eq.~\ref{eq:paramset} these parameters could be considered to be the masses and couplings denoted as
 
 \be
\{m_{\phi_1},..m_{\phi_n},\lambda_1,..\lambda_n,\lambda^{'}_{12},..\lambda^{'}_{(n-1)n}\}~,
\ee
where $\lambda_i$ denote the interactions of $\phi_i$ to the SM Higgs, whereas $\lambda^{'}_{ij}$ denotes the direct interaction between the DM components $\phi_i$ and $\phi_j$.
  Such a multi-component system with a large number of independent parameters is hard to analyse in a meaningful way, without further assumptions. However, all the essential 
features of such a scenario could be captured by a two-component case, and we shall limit ourselves to discuss the case of the two-component DM case in this work.

\section{Thermal freeze out and Boltzmann Equations}
\label{BE}

Our goal in this section is to determine the thermal freeze out of the DM components. We start with formulating the Boltzmann equations (BEQ) that govern cosmological evolution of the scalar singlets ($\phi$) DM in a single component framework. 

The BEQ for the single component framework reads \cite{Kolb:1990vq}: 
\bea
\dot{n}_{\phi}+3H n_{\phi} &=& - 
\int \frac{\gt_\phi d^3 p}{(2 \pi)^3 2E_p} \, \frac{\gt_\phi d^3 p'}{(2 \pi)^3 2E_p'}
\frac{\gt_{SM} d^3 q}{(2 \pi)^3 2E_q} \, \frac{\gt_{SM} d^3 q'}{(2 \pi)^3 2E_q'}
\delta^4 (p + p' - q - q')  \times \nonumber\\ 
&&  |\mathcal{M}_{\phi \phi \to SM SM}|^2  \left( \tilde{f}_\phi  \tilde{f}_\phi - \tilde{f}_\phi^{EQ} \tilde{f}_\phi^{EQ} \right) ~,
\label{be1}
\eea
where $n_{\phi}$ denote the number density of $\phi$, $n_{\phi}^{EQ}$  the corresponding equilibrium density, $\dot n_{\phi}$ is the time derivative of the number density,
$\mathcal{M}_{i\to f}$ is the amplitude for the process $i\to f$, including the spin average and symmetry factors, $H$ denotes the Hubble parameter and $\gt_\phi$ indicates the internal degrees of freedom associated to the particular DM species and for singlet scalar, we have $\gt_\phi=1$.
A phase space density $\tilde{f}_{\phi}$ and an equilibrium density $\tilde{f}_{\phi}^{EQ}$ are related to corresponding number densities as follows: 
\beq
n_\phi =  \int \frac{\gt_\phi d^3 p}{(2 \pi)^3 2E}  \tilde{f}_\phi,  \hspace{.4 cm}
n_\phi^{EQ} =  \int \frac{\gt_\phi d^3 p}{(2 \pi)^3 2E}  \tilde{f}_\phi^{EQ},    \hspace{.4 cm}   \tilde{f}_\phi^{EQ}  = \frac{1}{e^{E/T}-1} ~.
\eeq
To simplify BEQs we will use the thermally averaged cross section $\langle \sigma_{ a b \to c d}~ v \rangle$, 
defined as \cite{s2,Kolb:1990vq,Relic1}:
\bea
\langle \sigma_{ a b \to c d} v \rangle &\equiv & \frac{1}{\left(n_a^{EQ}n_b^{EQ}\right)}  
\int \frac{\gt_a d^3 p}{(2 \pi)^3 2E_p} \, \frac{\gt_b d^3 p'}{(2 \pi)^3 2E_p'}
\frac{\gt_{c} d^3 q}{(2 \pi)^3 2E_q} \, \frac{\gt_{d} d^3 q'}{(2 \pi)^3 2E_q'} \times\nonumber\\
&& \delta^4 (p + p' - q - q') |\mathcal{M}_{ab \to cd}|^2 e^{-(E_p+E_p')/T} \nonumber\\
 &= &  \int_{(m_a+m_b)^2}^{\infty}ds~ \frac{s\sqrt{s-(m_a+m_b)^2}~K_1(\frac{\sqrt{s}}{T})~(\sigma v)_{a b \to c d}}{16 T ~m_a^2m_b^2~K_2(\frac{m_a}{T})K_2(\frac{m_b}{T})} ~,
\label{eq:sigma-v1}
\eea

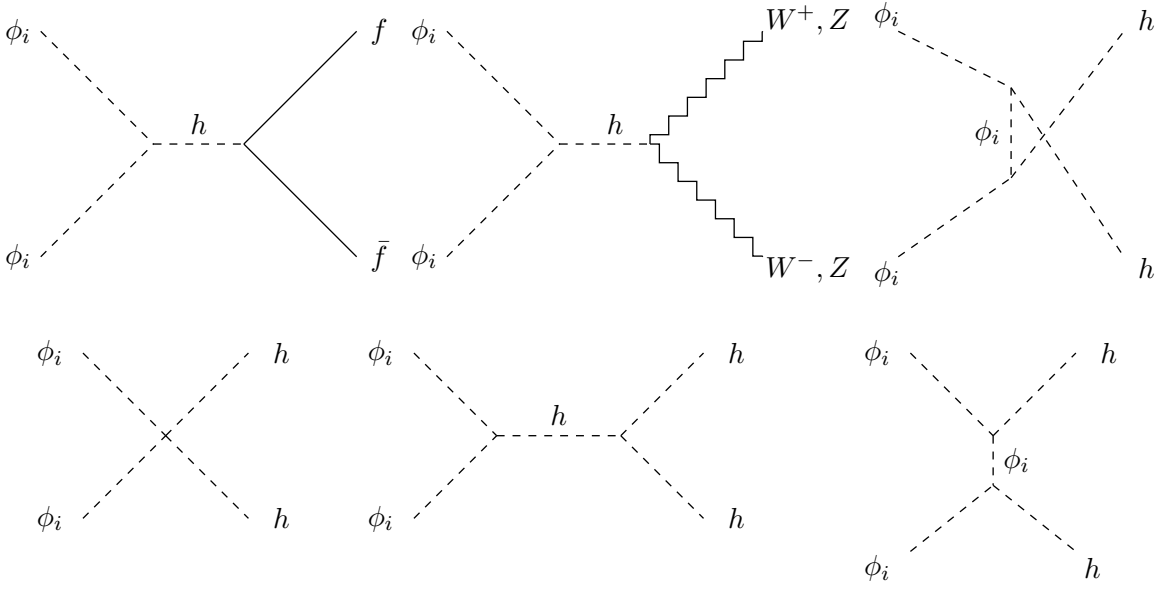
\begin{figure}
\begin{center}
    \begin{tikzpicture}[line width=0.5 pt, scale=1.5]
        \draw[dashed] (-5,1)--(-4,0);
	\draw[dashed] (-5,-1)--(-4,0);
	\draw[dashed] (-4,0)--(-3.2,0);
	\draw[solid] (-3.2,0)--(-2.2,1);
	\draw[solid] (-3.2,0)--(-2.2,-1);
	\node  at (-5.2,-1) {$\phi_i$};
	\node at (-5.2,1) {$\phi_i$};
	\node [above] at (-3.6,0) {$h$};
	\node at (-2.0,1.0) {$f$};
	\node at (-2.0,-1.0) {$\bar{f}$};
        \draw[dashed] (-1.4,1)--(-0.4,0);
	\draw[dashed] (-1.4,-1)--(-0.4,0);
	\draw[dashed] (-0.4,0)--(0.4,0);
	\draw[snake] (0.4,0)--(1.4,1);
	\draw[snake] (0.4,0)--(1.4,-1);
	\node  at (-1.6,-1) {$\phi_i$};
	\node at (-1.6,1) {$\phi_i$};
	\node [above] at (0.09,0) {$h$};
	\node at (1.8,1.1) {$W^{+},Z$};
	\node at (1.8,-1.1) {$W^{-},Z$};
        \draw[dashed] (2.6,1.0)--(3.6,0.5);
        \draw[dashed] (2.6,-1.0)--(3.6,-0.3);
        \draw[dashed] (3.6,0.5)--(3.6,-0.3);
	\draw[dashed] (3.6,-0.3)--(4.6,1.0);
	\draw[dashed] (3.6,0.5)--(4.6,-1.0);
	\node at (2.5,1.15) {$\phi_i$};
	\node at (2.5,-1.15) {$\phi_i$};
	\node [left] at (3.6,0.1) {$\phi_i$};
	\node at (4.8,1.1) {$h$};
	\node at (4.8,-1.1) {$h$};
     \end{tikzpicture}
 \end{center}
 \begin{center}
    \begin{tikzpicture}[line width=0.5 pt, scale=1.1]
        \draw[dashed] (-5,1)--(-4,0);
	\draw[dashed] (-5,-1)--(-4,0);
	\draw[dashed] (-4,0)--(-3,1);
	\draw[dashed] (-4,0)--(-3,-1);
	\node at (-5.4,1) {$\phi_i$};
	\node at (-5.4,-1) {$\phi_i$};
	\node at (-2.6,1) {$h$};
	\node at (-2.6,-1) {$h$};
        \draw[dashed] (-1,1)--(0,0);
	\draw[dashed] (-1,-1)--(0,0);
	\draw[dashed] (0,0)--(1.5,0);
	\draw[dashed] (1.5,0)--(2.5,1);
	\draw[dashed] (1.5,0)--(2.5,-1);
	\node  at (-1.4,-1) {$\phi_i$};
	\node at (-1.4,1) {$\phi_i$};
	\node [above] at (0.75,0) {$h$};
	\node at (2.9,1) {$h$};
	\node at (2.9,-1) {$h$};
	\draw[dashed] (5,1)--(6,0);
	\draw[dashed] (6,0)--(6,-0.6);
	\draw[dashed] (5,-1.4)--(6,-0.6);
	\draw[dashed] (6,0)--(7,1);
	\draw[dashed] (6,-0.6)--(7,-1.4);
	\node  at (4.6,1) {$\phi_i$};
	\node at (7.4,1) {$h$};
	\node [right] at (6,-0.3) {$\phi_i$};
	\node at (7.2,-1.6) {$h$};
	\node at (4.6,-1.6) {$\phi_i$};
     \end{tikzpicture}
 \end{center}
\caption{ Diagrams contributing to $\phi_i \phi_i$ annihilation to SM particles ($i=\{1,2\}$ in two component set up). }
\label{ann_diag}
 \end{figure}

 \begin{figure}
\begin{center}
    \begin{tikzpicture}[line width=0.5 pt, scale=1.1]
        \draw[dashed] (-5,1)--(-4,0);
	\draw[dashed] (-5,-1)--(-4,0);
	\draw[dashed] (-4,0)--(-3,1);
	\draw[dashed] (-4,0)--(-3,-1);
	\node at (-5.4,1) {$\phi_i$};
	\node at (-5.4,-1) {$\phi_i$};
	\node at (-2.6,1) {$\phi_j$};
	\node at (-2.6,-1) {$\phi_j$};
        \draw[dashed] (-1,1)--(0,0);
	\draw[dashed] (-1,-1)--(0,0);
	\draw[dashed] (0,0)--(1.5,0);
	\draw[dashed] (1.5,0)--(2.5,1);
	\draw[dashed] (1.5,0)--(2.5,-1);
	\node  at (-1.4,-1) {$\phi_i$};
	\node at (-1.4,1) {$\phi_i$};
	\node [above] at (0.75,0) {$h$};
	\node at (2.9,1) {$\phi_j$};
	\node at (2.9,-1) {$\phi_j$};
	\end{tikzpicture}
 \end{center}
 %
\caption{Diagrams contributing to DM conversion in $\ZZp$ model $(i,j=1,2;~ i \neq j)$.}
\label{fig:int}
 \end{figure}
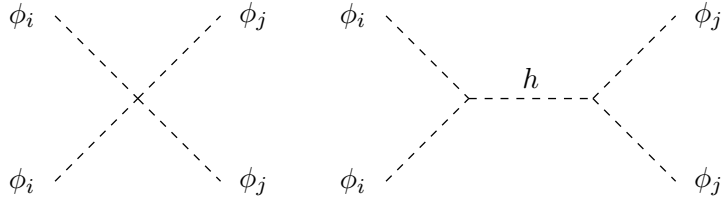

where in the last line we recast the thermal average cross-section in terms of the Mandelstam variable $s=(E_a+E_b)^2$ in the c.o.m frame, $K_{1,2}$ represent first and second Bessel functions in terms of appropriate variables and $T$ is the temperature of the universe. The threshold $s$ required is $s_0=(m_a+m_b)^2$ for the reactions to occur. Assuming kinetic equilibrium of DM with SM fields and neglecting possible effects of quantum statistics the BEQ for the single-component case simplify \cite{Kolb:1990vq}:
\bea
&& \dot{n}_{\phi}+3H n_{\phi} =
 -\langle \sigma_{ \phi \phi \to SM} v \rangle \left(n_{\phi}^2- n_{\phi}^{EQ} {}^2\right) ~.
\label{beq2}
\eea

The diagrams contributing to $\phi_i \phi_i$ annihilation to SM particles are shown in Fig.~\ref{ann_diag}. Dominant annihilation cross-section is obtained to gauge boson final states while the one to light fermions are small due to small Yukawa couplings. 
The corresponding cross sections are available in the literature (e.g. \cite{s4,s8}). In a single component framework, there is no scope for DM-DM interaction. While we move on to two-component framework, DM-DM interaction plays a crucial role and possible diagrams are shown in the Fig.~\ref{fig:int}.  

We now argue that in case of scalar DM interactions, instead of using the whole expression of thermal average cross-section $\langle \sigma v \rangle$ in BEQ, as in Eq.~\ref{eq:sigma-v1}, we can use $(\sigma v)$ at threshold $s=s_0=(m_a+m_b)^2$ as \cite{peskin}: 
\beq
(d\sigma~ v)_{ab \to cd} = \frac{|\mathcal{M}|_{ab \to cd}^2 ~dQ}{4 m_a m_b} ~,
\label{eq:sima-v2}
\eeq
where $dQ$ is the phase space differential and $|\mathcal{M}|^2$ is the matrix element square averaged and $m_{a,b}$ are the masses of annihilating particles. This essentially corresponds to the so called $s-$ wave contribution to annihilation cross-section and in presence of this term, the $v^2$ dependent terms ($p$ wave, for example) can be neglected. We will show that $(\sigma v)_{ab \to cd}$ is almost the same as to the thermal averaged $\langle\sigma v\rangle_{ab \to cd}$ even at low $x=\frac{m}{T}$. The advantage of using $(\sigma v)_{ab \to cd}$ is the absence of temperature dependence in it, which helps solving the BEQs relatively easier; particularly in case of coupled BEQ which we need for the DM analysis of the model. Hence, we rewrite the DM annihilation cross-sections to SM in terms of ($\sigma v$) adopting Eq.~\ref{eq:sima-v2} as follows \cite{s4,s8}:

\begin{figure}[htb!]
$$
\includegraphics[height=5.0cm]{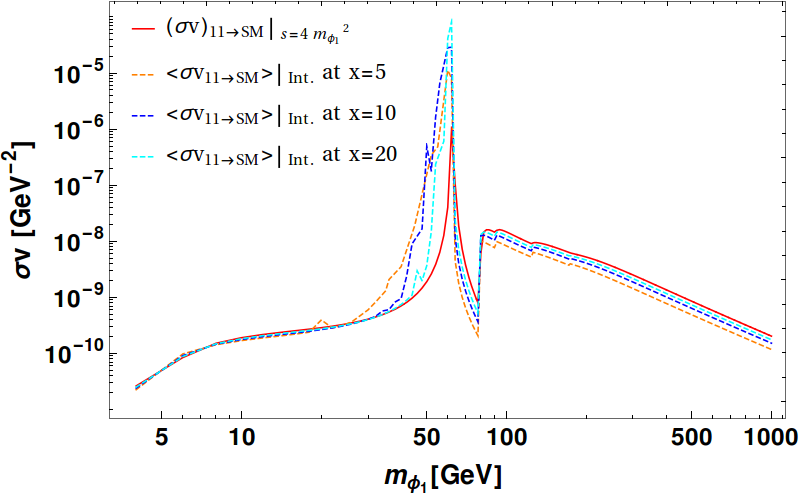}
\includegraphics[height=5.0cm]{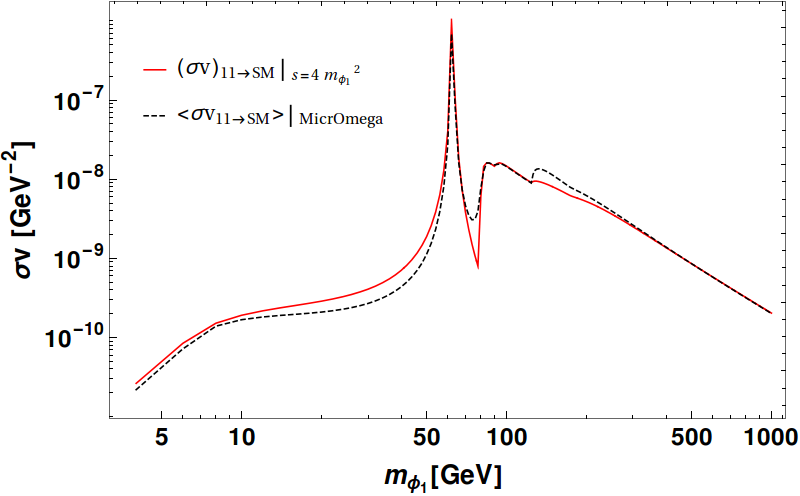}
 $$
 \caption{Left: Comparison of $(\sigma v)_{\phi_1 \phi_1 \to SM}$ (red thick line) as in Eq.~\ref{eq:sigmav-th} with thermal average cross-section $\langle \sigma v \rangle_{\phi_1\phi_1\to SM}$ as in Eq.~\ref{eq:sigma-v1}.  $\langle \sigma v\rangle$ is evaluated at $x=\frac{m_{\phi_1}}{T}=5~(\rm Orange ~dashed),~10~(\rm Blue~ dashed),~20~(\rm Cyan~ dashed)$. Right: Comparison of $(\sigma v)_{\phi_1 \phi_1 \to SM}$ (red thick line) to $\langle \sigma v \rangle_{\phi_1\phi_1\to SM}$ extracted from code {\tt MicrOMEGAs} (black dots). The plot is shown as a function of DM mass $m_{\phi_1}$ [GeV] for single component scalar singlet DM $\phi_1$. We choose the DM-SM coupling $\la=0.1$ for illustration. }
 \label{fig:sigmav-compare}
 \end{figure}
\begin{eqnarray}
(\sigma v)_{\phi_1 \phi_1 \rightarrow f \overline f}&=&\frac{1}{4\pi s \sqrt s} \frac{N_c\lambda_1^2 m_f^2}{(s-m_h^2)^2+m_h^2 \Gamma_h^2}(s-4m_f^2)^\frac{3}{2} ~,\nonumber\\
{(\sigma v)}_{\phi_1 \phi_1 \rightarrow W^+ W^-}&=&\frac{\lambda_1^2}{8\pi} \frac{s}{(s-m_h^2)^2+m_h^2 \Gamma_h^2}(1+\frac{12m_W^4}{s^2}-\frac{4m_W^2}{s})(1-\frac{4m_W^2}{s})^\frac{1}{2} ~,\nonumber\\
{(\sigma v)}_{\phi_1 \phi_1\rightarrow Z Z}&=&\frac{\lambda_1^2}{16\pi} \frac{s}{(s-m_h^2)^2+m_h^2 \Gamma_h^2}(1+\frac{12m_Z^4}{s^2}-\frac{4m_Z^2}{s})(1-\frac{4m_Z^2}{s})^\frac{1}{2} ~,\nonumber\\
{(\sigma v)}_{\phi_1 \phi_1 \rightarrow h h}&=&\frac{\lambda_1^2}{16\pi s}\left[1+\frac{3m_h^2}{(s-m_h^2)}-\frac{4\lambda_1 v^2}{(s-2m_h^2)}\right]^2 (1-\frac{4m_h^2}{s})^\frac{1}{2} ~,\nonumber\\
{(\sigma v)}_{\phi_1 \phi_1 \to SM}&=& (\sigma v)_{\phi_1 \phi_1 \rightarrow f \overline f}+{(\sigma v)}_{\phi_1 \phi_1 \rightarrow W^+ W^-}+ {(\sigma v)}_{\phi_1 \phi_1\rightarrow Z Z}\nonumber \\ &&+{(\sigma v)}_{\phi_1 \phi_1 \rightarrow h h} ~;
\label{eq:annihilationSM}
\end{eqnarray}
where $\sqrt s$ is the centre-of-mass energy and $\Gamma_h$~\cite{s5} denotes Higgs decay width at resonance. $N_c=3$ is the colour factor for quarks, for leptons it is unity. Similarly annihilation cross-sections for the second component: $(\sigma v)_ {\phi_2 \phi_2\to SM}$ can be written replacing $\lambda_1$ by $\lambda_2$ in Eq.~\ref{eq:annihilationSM}. 
In Fig.~\ref{fig:sigmav-compare}, we show that the $(\sigma v)_{\phi_1 \phi_1 \to SM}$ (as in Eq.~\ref{eq:annihilationSM}) in red thick line, closely mimics the thermal average annihilation cross-section $\langle \sigma v \rangle_{\phi_1\phi_1\to SM}$ (as in Eq.~\ref{eq:sigma-v1}). In the left panel of Fig.~\ref{fig:sigmav-compare}, we choose three different values of $x=\frac{m_{\phi_1}}{T}=5~(\rm Orange ~dashed),~10~(\rm Blue~ dashed),~20~(\rm Cyan~ dashed)$ to evaluate $\langle \sigma v \rangle_{\phi_1\phi_1\to SM}$. It clearly shows, that excepting some fluctuations near the resonance region ($\mphia=\frac{m_h}{2}$) for small values of $x$, thermal average cross-section agrees well to $(\sigma v)_{\phi_1 \phi_1 \to SM}$. In the right panel of Fig.~\ref{fig:sigmav-compare}, we compare the same $(\sigma v)_{\phi_1 \phi_1 \to SM}$ with the thermal average cross-section that code {\tt MicrOMEGAs} \cite{MO} generates (black dots) to further establish the claim. The annihilation cross-section is plotted as a function of DM mass $m_{\phi_1}$ [GeV] with a specific choice of $\la=0.1$. For other values of $\la$, the behaviour remains the same. In the expressions of $(\sigma v)_{\phi_1 \phi_1 \to SM}$ as in Eq.~\ref{eq:annihilationSM}, we have used threshold value of $s=s_0=4\mphia^2$. We will use it throughout the analysis and even if not explicit, we will mean 
\beq
(\sigma v)_{ab \to cd}\equiv (\sigma v)_{ab \to cd}|_{s=(m_a+m_b)^2} ~.
\label{eq:sigmav-th}
\eeq
For $\mZ_2\times \mZ_2^{'}$ model, the interaction between DM components ($\phi_1 \phi_1 \to \phi_2 \phi_2$) are governed by the diagrams in Fig.~\ref{fig:int}. Assuming $m_{\phi_1}>m_{\phi_2}$,
\begin{equation}
{(\sigma v)}_{\phi_1 \phi_1 \rightarrow \phi_2 \phi_2}=\frac{\sqrt{s-4m_{\phi_2}^2}}{8\pi s \sqrt s}\left[\frac{ v^4 \lambda_1^2\lambda_2^2}{(s-m_h^2)^2+m_h^2 \Gamma_h^2}+
\frac{ 2 (s-m_h^2) v^2 \lambda_1\lambda_2 \lambda_3}{ {(s-m_h^2)^2+m_h^2 \Gamma_h^2}}+\lambda_3^2\right] ~.
\label{eq:annihilationDM1}
\end{equation}
In absence of the contact interaction (first diagram in the top panel of Fig.~\ref{fig:int}), i.e. $\lambda_3 \to 0$, the expression simplifies to:
\be
{(\sigma v)}_{\phi_1 \phi_1 \rightarrow \phi_2 \phi_2}=\frac{ v^4 \lambda_1^2\lambda_2^2}{8\pi s \sqrt s}\frac{\sqrt{s-4m_{\phi_2}^2}}{(s-m_h^2)^2+m_h^2 \Gamma_h^2} ~.
\label{eq:annihilationDM1l30}
\ee
We would like to mention here that in $\mO(2)$ model, the DM-DM interactions as in Fig.~\ref{fig:int}, are all present, but the contribution is identically zero with masses of the two DM components identical.
\begin{figure}[htb!]
$$
\includegraphics[height=5.0cm]{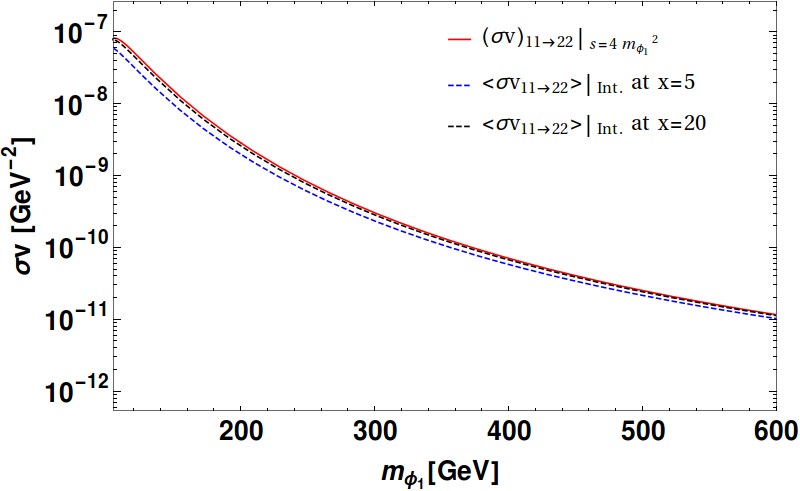}
\includegraphics[height=5.0cm]{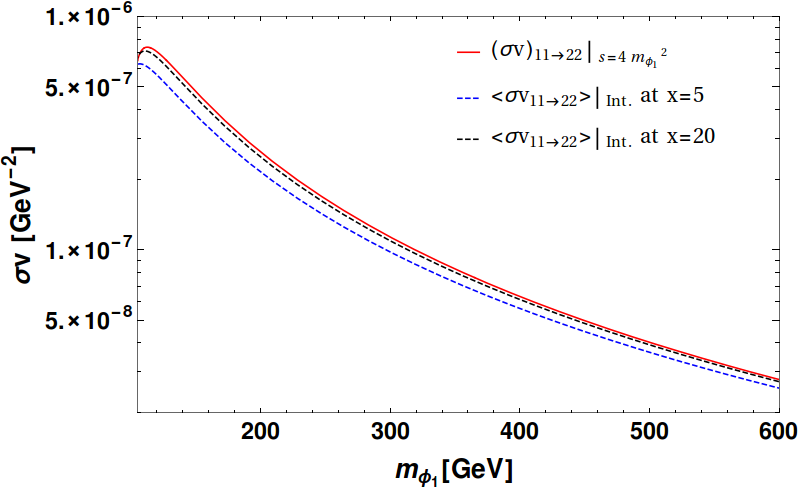}
 $$
 \caption{Comparison of $(\sigma v)_{\phi_1 \phi_1 \to \phi_2 \phi_2}$ (as in Eq.~\ref{eq:annihilationDM1}) in red thick line with thermal average cross-section $\langle \sigma v \rangle_{\phi_1 \phi_1 \to \phi_2 \phi_2}$ for DM-DM interactions in $\ZZp$ model is shown as a function of DM mass $m_{\phi_1}$ [GeV]. We choose parameters $\{\la,\lb,\lc,\mphib\}=\{0.5,0.5,0.01,100\}$ (left), $\{0.5,0.5,1.0,100\}$ (right) for illustration. $\langle \sigma v\rangle$ is evaluated at two different values: $x=\frac{m}{T}=5~(\rm Blue ~ dashed),~20~(\rm Black~ dashed)$. }
 \label{fig:sigmav-compare1}
 \end{figure}
For $m_{\phi_2}>m_{\phi_1}$, we similarly obtain $ {(\sigma v)}_{\phi_2 \phi_2 \rightarrow \phi_1 \phi_1}$. A detailed comparison of DM annihilation to SM with DM-DM interaction will be performed in the next section. In Fig.~\ref{fig:sigmav-compare1}, we demonstrate the comparison of $(\sigma v)_{\phi_1 \phi_1 \to \phi_2 \phi_2}$ (red line) to the thermally averaged DM-DM interactions $\langle \sigma v \rangle_{\phi_1 \phi_1 \to \phi_2 \phi_2}$ for two different choices of $x=\frac{m}{T}=5~(\rm Blue ~ dashed),~20~(\rm Black~ dashed)$. We illustrate two different parameter sets $\{\la,\lb,\lc,\mphib\}=\{0.5,0.5,0.01,100\}$ (left), $\{0.5,0.5,1.0,100\}$ (right); and both of them show a very good agreement of s-wave cross-section to the thermal average DM-DM interaction cross-section. 
\begin{figure}[htb!]
$$
\includegraphics[height=5.0cm]{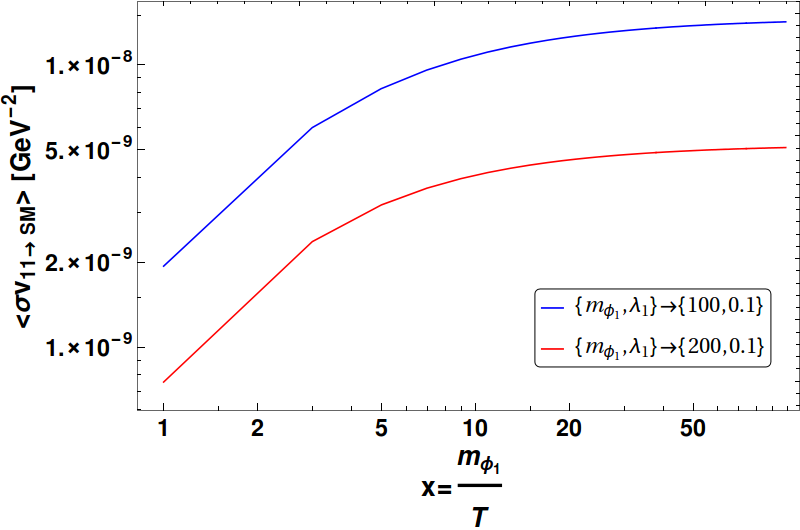}
\includegraphics[height=5.0cm]{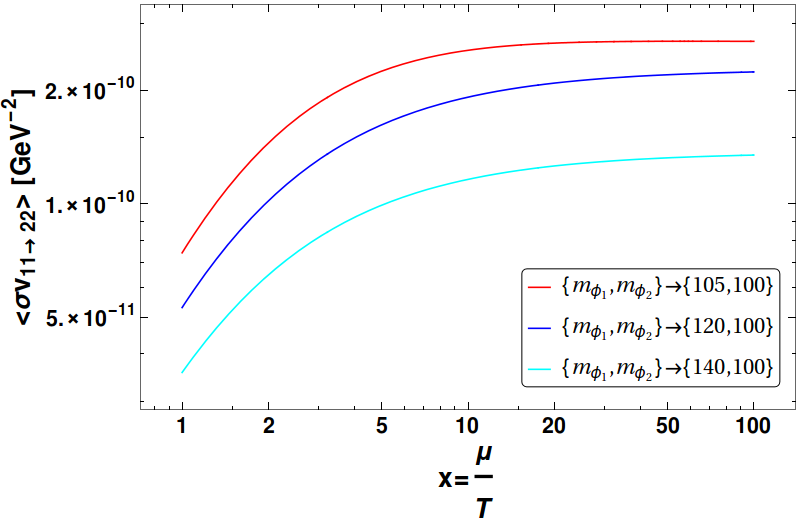}
 $$
 \caption{Variation of thermal average annihilation cross-section with $T$. Left: $\langle \sigma v \rangle_{\phi_1\phi_1\to SM}$ as in Eq.~\ref{eq:sigma-v1} for single component case is depicted with two different choices of DM mass $\mphia=\{100 ~(\rm Blue), ~ 200 ~(\rm Red)\}$ GeV for a fixed $\la=0.1$. Right: Variation of DM-DM interaction $\langle \sigma v \rangle_{\phi_1\phi_1\to \phi_2\phi_2}$ is shown for two component $\ZZp$ model for three different DM mass combinations: $\{\mphia,\mphib\}= \{105,100\}~(\rm Red),~ \{120,100\}~(\rm Blue), ~ \{140,100\}~(\rm Cyan)$; $\{\lambda_1,\lambda_2,\lambda_3\}=\{0.1,0.1,0.01\}$ are chosen for illustration.}
 \label{sigmav-T}
 \end{figure}

It is already understood that the thermal average cross-section inherently poses a temperature dependence, which is explicitly demonstrated in Fig.~\ref{sigmav-T}. Here  we show the variation in  $\langle \sigma v \rangle_{\phi_1\phi_1\to SM}$ with respect to temperature $T$ for single component case in the left panel and $\langle \sigma v \rangle_{\phi_1\phi_1\to \phi_2\phi_2}$ for the two component case in the right panel. We can see that the temperature dependence is more for small $x$ or large $T$ regions, although the variation is limited within an order of magnitude. For the two component case, we deliberately choose the masses of the DMs close to each other as the temperature dependence is more sensitive for such a scenario. However, we clearly see that for $x>5$, the annihilation cross-sections are almost independent of $T$ and we can safely use $(\sigma v)$ for this model.

Let us now turn to the coupled BEQs that govern the two component DM freeze out. We will recast the BEQs in terms of yield $Y=\frac{n_{\phi_i}}{s}$ so as to indicate the number density in comoving volume, where $n_{\phi_i}$ is the number density of $\phi_i$ and $s$ is the entropy density of the universe given by \cite{Kolb:1990vq}
\bea
s = \frac{2 \pi^2}{45} g_s (T) T^3 \,; \quad
g_s(T) = \sum_k r_k g_k \left (\frac{T_k}{T} \right )^3\theta (T-m_k) \,;
\eea
here $k$ runs over all particles, $T_k$ is the temperature of particle $k$
and $g_k$ its number of internal degrees of freedom, and
$ r_k=1\,(7/8)$ when $k$ is a bosons (fermion).

The BEQs can be written as a function of temperature ($T$) or in terms of $x=\frac{m}{T}$ where $m$ is the mass of the DM particle. However, using $x$ as a common variable for two-component case is problematic as mass $m$ here represents two different variables ($m_{\phi_1},m_{\phi_2}$) for two DM species. Hence, one way to obtain a common variable is to introduce a reduced mass $ \mu $ for the two component DM system and define $x$ with respect to reduced mass as $x=\frac{\mu}{T}$ where $\frac{1}{\mu}=\frac{1}{m_{\phi_1}}+\frac{1}{m_{\phi_2}}$ i.e $\frac{1}{x}=\frac{T}{\mu}=\frac{1}{x_1}+\frac{1}{x_2}$. In terms of reduced $x$ the BEQs read \cite{Yx-couple,Yx-couple1}:
\bea
  \frac{dY_1}{dx}&=& -0.264  M_{Pl}\sqrt g_*\frac{\mu}{x^2} \left[\langle \sigma v _{11\rightarrow SM}\rangle(Y_1^2-{Y_1^{EQ}}^2)+\langle
  \sigma v _{11\rightarrow 22}\rangle(Y_1^2-\frac{{Y_1^{EQ}}^2}{{Y_2^{EQ}}^2}Y_2^2)\Theta(m_{\phi_1}-m_{\phi_2})\right. \nonumber \\ 
  & &\left.-\langle \sigma v _{22\rightarrow 11}\rangle(Y_2^2-\frac{{Y_2^{EQ}}^2}{{Y_1^{EQ}}^2}Y_1^2)\Theta(m_{\phi_2}-m_{\phi_1})\right] ~,\nonumber\\
  \frac{dY_2}{dx}&=& -0.264  M_{Pl}\sqrt g_*\frac{\mu}{x^2} \left[\langle \sigma v _{22\rightarrow SM}\rangle(Y_2^2-{Y_2^{EQ}}^2)+\langle
  \sigma v _{22\rightarrow 11}\rangle(Y_2^2-\frac{{Y_2^{EQ}}^2}{{Y_1^{EQ}}^2}Y_1^2)\Theta(m_{\phi_2}-m_{\phi_1})\right. \nonumber \\ 
  & &\left.-\langle \sigma v _{11\rightarrow 22}\rangle(Y_1^2-\frac{{Y_1^{EQ}}^2}{{Y_2^{EQ}}^2}Y_2^2)\Theta(m_{\phi_1}-m_{\phi_2})\right] ~;
  \label{eq:BEQx}
 \eea
 
where the equilibrium distributions now recast in terms of $\mu$ has the form 
\be
Y_i^{EQ}(x)=0.145 \frac{g}{g_*}{x}^{\frac{3}{2}}(\frac{m_{\phi_i}}{\mu})^{\frac{3}{2}} e^{-x(\frac{m_{\phi_i}}{\mu})} ~.
\label{eq:Yeqx}
\ee
 In above equations, $M_{Pl}=\frac{1}{\sqrt G}=1.22 \times 10^{19} ~{\rm GeV}$ and $g_{*}=106.7 $ \cite{Kolb:1990vq}. One should note additional contributions due to DM-DM conversions in the coupled Eqs.~ \ref{eq:BEQx}. Depending on the mass hierarchy, one of $11 \to 22$ or $22 \to 11$ will contribute. We have used $\Theta$ functions appropriately to illustrate that fact. In the following analysis, we will demonstrate that in certain regions of the multipartite DM parameter space, DM-DM interaction is large enough to claim an alternation in DM freeze out and hence in relic density. 

 Even further simplification occurs to the coupled BEQ by pulling out the large numerical factors from $Y_i ~(i=1,2)$ in terms of modified $y_i$ as 
\be
y_i=0.264 M_{Pl} \sqrt {g_*} \mu  Y_i  ~;~~ y_i^{EQ}=0.264 M_{Pl} \sqrt {g_*} \mu  Y_i^{EQ} ~.
 \ee
 In terms of $y_i$ the BEQs for $\ZZp$ model:
 \bea
 \frac{d y_1}{dx}&=&-\frac{1}{x^2}\left[\langle{\sigma v}_{11\rightarrow SM}\rangle(y_1^2-{y_1^{EQ}}^2)+\langle{\sigma v}_{11\rightarrow 22}\rangle(y_1^2-
 \frac{{y_1^{EQ}}^2}{{y_2^{EQ}}^2} y_2^2)\Theta(m_{\phi_1}-m_{\phi_2})\right. \nonumber \\
 & &\left. -\langle{\sigma v}_{22\rightarrow 11}\rangle(y_2^2-\frac{{y_2^{EQ}}^2}{{y_1^{EQ}}^2} y_1^2)\Theta(m_{\phi_2}-m_{\phi_1})\right] ~,\nonumber \\
 \frac{d y_2}{dx}&=&-\frac{1}{x^2}\left[\langle{\sigma v}_{22\rightarrow SM}\rangle(y_2^2-{y_2^{EQ}}^2)-\langle{\sigma v}_{11\rightarrow 22}\rangle(y_1^2-
 \frac{{y_1^{EQ}}^2}{{y_2^{EQ}}^2} y_2^2)\Theta(m_{\phi_1}-m_{\phi_2})\right. \nonumber \\
 & &\left. +\langle{\sigma v}_{22\rightarrow 11}\rangle(y_2^2-\frac{{y_2^{EQ}}^2}{{y_1^{EQ}}^2} y_1^2)\Theta(m_{\phi_2}-m_{\phi_1})\right] ~.
 \label{eq:BEQyx} 
\eea
 We solve the coupled equations in Eq.~\ref{eq:BEQyx} numerically (as we will argue that solving the equations analytically is much harder) to find out the freeze out of the DM components and hence compute relic density of the DM species by \cite{Relic,Relic2,Relic3}
\begin{eqnarray}
\nonumber
 \Omega_1 h^2&=&\frac{854.45\times10^{-13}}{\sqrt g_*} \frac{m_{\phi_1}}{\mu}~ y_1\left[\frac{\mu}{m_{\phi_1}}x_{\infty}\right], \\
 \nonumber
\Omega_2 h^2&=&\frac{854.45\times10^{-13}}{\sqrt g_*} \frac{m_{\phi_2}}{\mu} ~y_2\left[\frac{\mu}{m_{\phi_2}}x_{\infty}\right], \\
\Omega_T h^2&=&\Omega_1 h^2+\Omega_2 h^2 ~.
\label{yomega}
\end{eqnarray}
In Eq.~\ref{yomega}, $y_1\left[\frac{\mu}{m_{\phi_1}}x_{\infty}\right]$ indicates the value of $y_1$ evaluated at $\frac{\mu}{m_{\phi_1}}x_{\infty}$, where $x_{\infty}$ indicates a very large value of $x$ after decoupling. In numerical analysis, $x \sim 100$ is good enough to indicate freeze out of scalar DM in such a model while we choose $x=500$ to be on the safe side. We will demonstrate freeze out of the DM components in the next section in different regions of parameter space to validate above claim. Also note here that the solution of the coupled BEQ yields $y_i$ as a function of $x=\frac{\mu}{T}$, while relic density of individual components should be expressed in terms of $x_i=\frac{m_i}{T}$. Hence, in Eq.~\ref{yomega}, we substitute $x= \frac{\mu}{T}=\frac{\mu}{m_{\phi_{i}}}x_i ~(i=1,2)$. 

Multipartite DM framework with $\mO(N)$ symmetry predicts degenerate DM scenario and hence do not yield non-zero DM-DM interactions as has already been stated. Hence, the BEQ mimics as single component BEQ as in Eq.~\ref{beq2}. The annihilation cross-sections also remain exactly the same although relic density in such a case is simply scaled by the number of DM components ($N$) all of which interact to SM in a similar way \cite{multi-singlet1}:
\bea
\Omega_T=N \Omega_i ~;
\label{eq:relicON}
\eea
 $\Omega_i$ is the relic density of an individual components $\phi_i$ given approximately by the inverse of annihilation cross-section of $\phi_i$ to SM as
 \bea
 \Omega_i = \frac{854.45\times10^{-13}}{\sqrt g_*} y_i[x_{\infty}]\simeq \frac{0.1 ~{\rm pb}}{\langle\sigma v\rangle_{i}} ~.
 \eea
In literature coupled BEQ for multi-partite scalar singlet DM have been mentioned in many cases \cite{multi-singlet1,multi-singlet2,multi-singlet3}, however an elaborate scan of the available parameter space has not been performed with DM-DM interactions kept on. In the following, we take up a systematic analysis of the two component DM scenario in $\ZZp$ model to find out accessible parameter space by relic density and direct search constraints and indicate possible distinctive features of having DM-DM interactions.
\section{Relic Density Analysis in $\ZZp$ model}
\label{ZZp}
%
\begin{figure}[htb!]
$$
 \includegraphics[height=5.8cm]{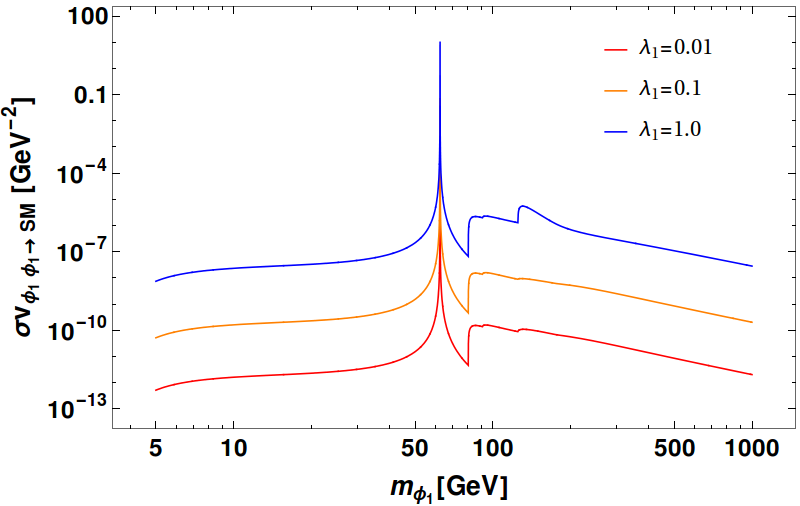}
 $$
$$
\includegraphics[height=5.0cm]{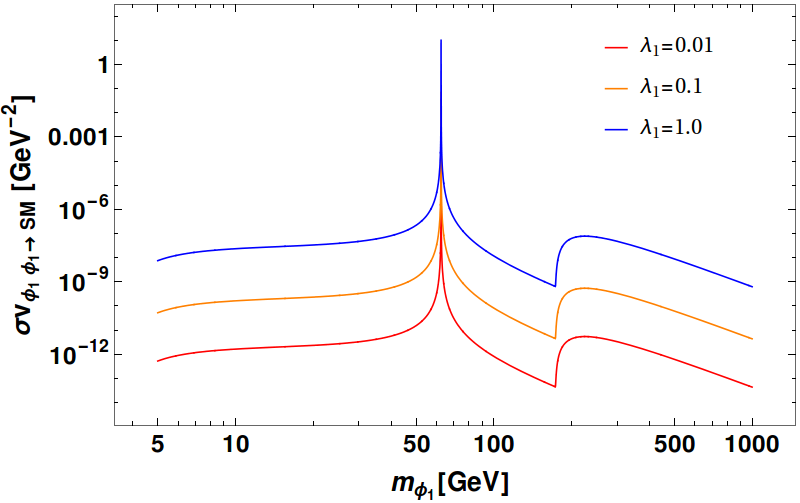}
\includegraphics[height=5.0cm]{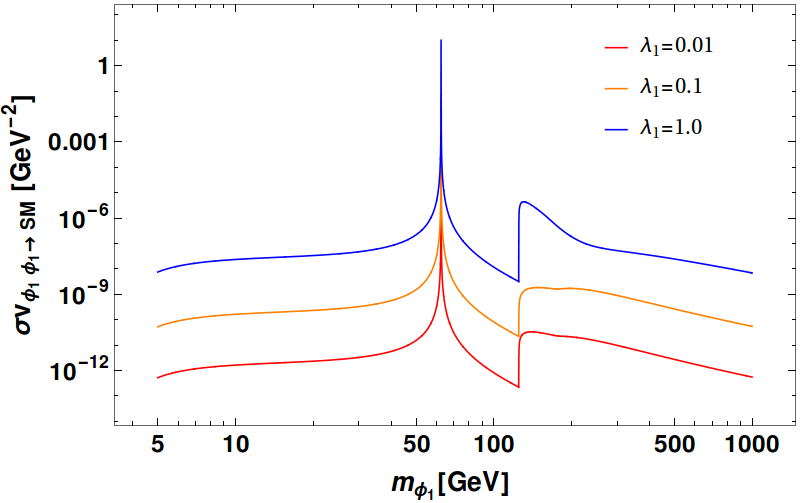}
 $$
 \caption{Top: $(\sigma v)_{\phi_1\phi_1 \to SM}$ is plotted as a function of $m_{\phi_1}$; Bottom Left: Annihilations to $WW,~ZZ,~hh$ are prohibited; Bottom right: Annihilations to $WW,~ZZ$ are prohibited. Three different values of $\lambda_1=\{0.01 ~(\rm Red),0.1~(\rm Orange),1.0~ (\rm Blue) \}$ are chosen from bottom to up respectively.}
 \label{fig:sigmavSM}
 \end{figure}
\begin{figure}[htb!]
$$
 \includegraphics[height=4.5cm]{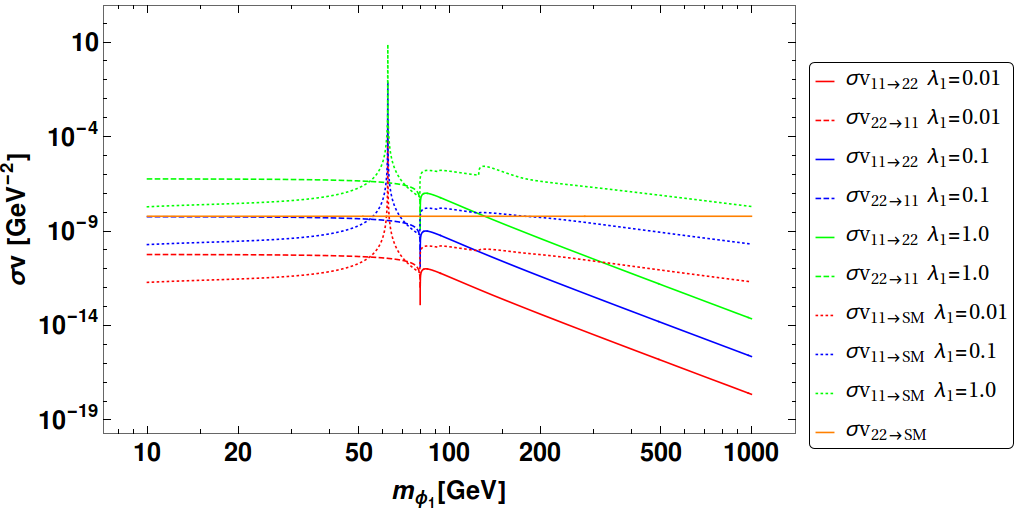}
 \includegraphics[height=4.5cm]{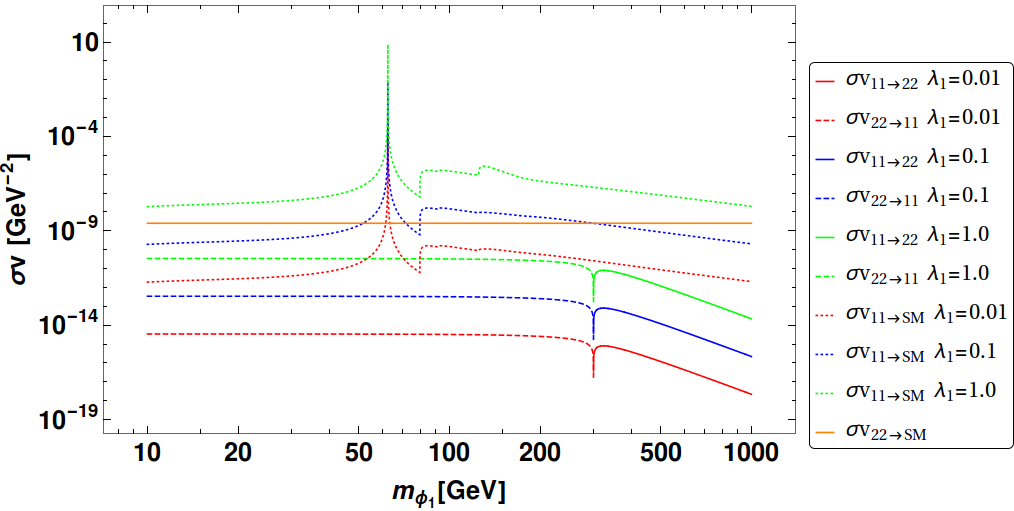}
 $$
 \caption{DM annihilations to SM is compared to DM-DM interactions as a function of $\mphia$ in absence of $\lambda_3$. $\mphib=80~ (\rm{left}),~300~(\rm{right})$ GeV are fixed. $\lambda_2=0.1$ is chosen for illustration while three different choices of $\lambda_1=\{0.01~(\rm Red),0.1~(\rm Blue),1.0~(\rm Green)\}$ (bottom to up respectively) are plotted.}
 \label{fig:compare-sigv1}
 \end{figure}

Our goal of this part of the analysis is to find the available parameter space of the two component DM framework in $\ZZp$ model from relic density constraint as well as indicate the departure in freeze-out due to interactions between two DM components. 
Multicomponent DM freeze-out is dependent on annihilations to SM and on the conversions between DM components. Annihilation cross-sections to SM for scalar singlet DM with Higgs portal coupling is well studied and understood, as pointed out in Eq.~\ref{eq:annihilationSM} and shown in Fig.~\ref{fig:sigmavSM}. We demonstrate the annihilation $\phi_1\phi_1 \to SM$ as a function of $\mphia$ in the top panel of Fig.~\ref{fig:sigmavSM} for three different values of $\la=\{0.01,0.1,1.0\}$ in red, orange and blue respectively. The Higgs resonance is clearly seen at $\mphia=m_h/2$, while there is a significant bump observed at $\mphia=80$ GeV, when $WW$ channel opens up. For large value of $\lambda_1=1.0$ (blue curve) in top panel of Fig.~\ref{fig:sigmavSM}, a small bump is seen at $\mphia \sim m_h=125$ GeV, indicating large annihilation contributions opening to $hh$ channel which has a larger sensitivity to $\lambda_1$ of the order of $\sim \mO(\la^4)$ (see Eq.~\ref{eq:annihilationSM}). To illustrate the contributions to different SM final states, in the bottom left panel we stop $WW,ZZ,hh$ annihilations to show the bump in annihilation cross-section at $\mphia \sim m_t=173$ GeV. In the bottom right panel, we similarly stop $WW,ZZ$ channel to show the contribution to $hh$ final state. Annihilation cross-section for $\phi_2\phi_2 \to SM$ is similar as a function of $\mphib$ and $\lambda_2$.
 \begin{figure}[htb!]
$$
 \includegraphics[height=5.0cm]{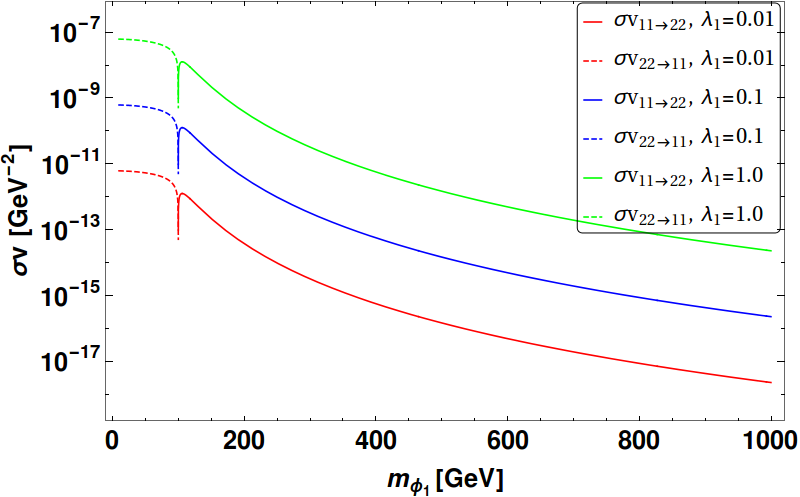}
 \includegraphics[height=5.0cm]{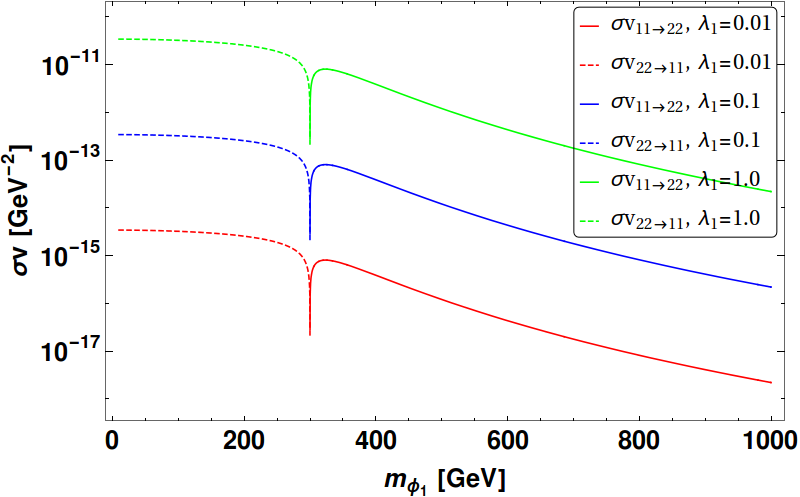}
 $$
 \caption{$(\sigma v)$ for DM conversions: $\phi_2\phi_2 \to \phi_1 \phi_1$ or $\phi_1\phi_1 \to \phi_2 \phi_2$ as a function of $\mphia$ for constant $\mphib=100$ GeV (left), 300 GeV (right). $\lambda_2=0.1$ and contact interaction $\lambda_3=0$ are kept fixed for illustration while three different choices of $\lambda_1=\{0.01~(\rm Red),0.1~(\rm Blue),1.0~(\rm Green)\}$ (bottom to up respectively) are plotted.}
 \label{fig:sigmavDM}
 \end{figure}
Interaction between DM components $\phi_1\phi_1 \to \phi_2 \phi_2$ or $\phi_2\phi_2 \to \phi_1 \phi_1$ occurs depending on $\mphia>\mphib$ or $\mphib>\mphia$. In Fig.~\ref{fig:sigmavDM}, the DM conversion is shown as a function of $\mphia$ keeping $\mphib$ fixed at 100 GeV (left) and 300 GeV (right). When $\mphia<\mphib$, $\phi_2\phi_2 \to \phi_1 \phi_1$ is present and it dies when $\mphia \sim \mphib$. When $\mphia>\mphib$, $\phi_1\phi_1 \to \phi_2 \phi_2$ process start contributing; the cross-section decreases with larger DM mass followed from the expressions in Eq.~\ref{eq:annihilationDM1}. A change in DM freeze out is possible when DM-DM interactions are of the same order as to the annihilations to SM. A comparison is illustrated in Fig.~\ref{fig:compare-sigv1}, where both the processes, annihilations to SM and DM-DM interactions are plotted together as a function of $\mphia$ in the limit of zero contact interaction, $\lambda_3=0$. The other DM mass is kept fixed, $\mphib=80$ GeV (left) and 300 GeV (right) of Fig.~\ref{fig:compare-sigv1}. We choose $\lambda_2=0.1$ and three different choices of $\la=\{0.01,0.1,1.0\}$ (red, blue and green respectively) for illustration. Fig.~\ref{fig:compare-sigv1} shows that for small $\mphib= 80$ GeV (left), DM-DM interaction is of the same order as with the annihilations to SM with $\lambda_3=0$ particularly for low $\mphia$. When $\mphib$ is larger, for example, 300 GeV (right), DM conversion cross-section gets smaller than annihilations to SM as expected (see Fig.~\ref{fig:sigmavDM} for example).

\begin{figure}[htb!]
$$
 \includegraphics[height=5.0cm]{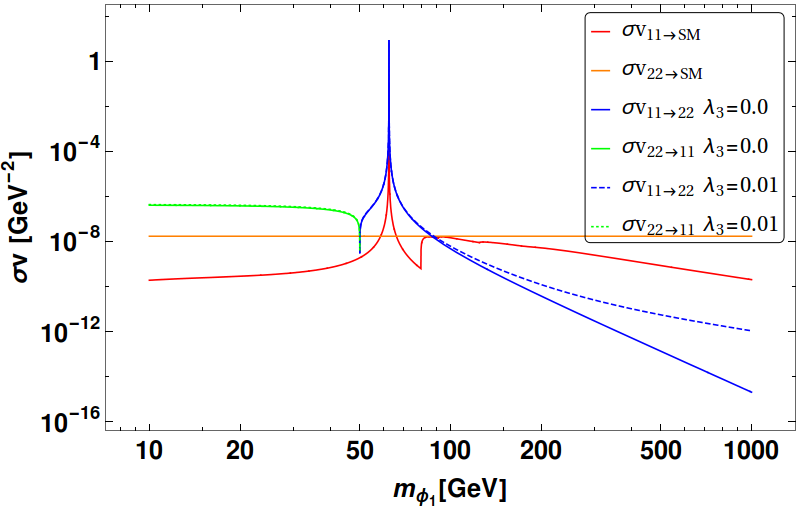}
  \includegraphics[height=5.0cm]{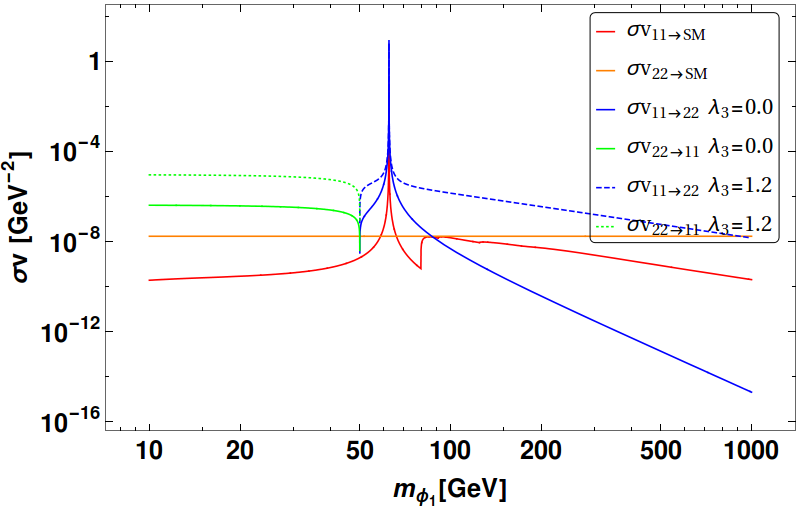}
 $$
 \caption{DM annihilation to SM is compared to DM interaction as a function of $\mphia$ in presence of non-zero $\lambda_3$. We choose $\lambda_3=0.01$ (left), $1.2$ (right); $\lambda_1= 0.1, \lambda_2=0.3,\mphib=50 ~{\rm GeV}$ are chosen for illustration.}
 \label{fig:compare-sigv2}
 \end{figure}

In Fig.~\ref{fig:compare-sigv2}, we show a comparison for all the relevant cross-sections in presence of non-zero $\lambda_3$. In the left panel, we choose small non-zero $\lambda_3=0.01$ and show that DM-DM interaction remains the same as $\lambda_3=0$ case, as both blue and green solid ($\lambda_3=0$) and dashed ($\lambda_3=0.01$) lines almost superpose on top of each other. We choose a small value of $\mphib=50$ GeV for illustrating this case. The cross-sections are varied as a function of $\mphia$. For $\mphia<\mphib$, $\phi_2 \phi_2 \to \phi_1 \phi_1$ occurs. After $\mphia>\mphib$, $\phi_1 \phi_1 \to \phi_2 \phi_2$ takes over and the Higgs resonance peak appears for $\mphia=\frac{m_h}{2}$ as can be seen in blue line. This of course coincides with the resonance peak for $\phi_1\phi_1 \to SM$ annihilations as shown in red. Only at a larger $\mphia> 100~{\rm GeV}$, small $\lambda_3$ starts showing up and the blue dashed curve separates out from the solid one in the left panel of Fig.~\ref{fig:compare-sigv2}. In the right panel of Fig.~\ref{fig:compare-sigv2}, the case for a moderately large DM contact interaction $\lambda_3=1.2$ has been depicted. This shows a sizeable difference in DM-DM interaction (see blue and green dashed and solid lines well separated) as expected. It is trivial to note that $\phi_2 \phi_2 \to SM$ remains constant with constant $\mphib,\lb$ shown by orange line in both Fig.~\ref{fig:compare-sigv1} and in Fig.~\ref{fig:compare-sigv2}. These together imply that a non-zero but moderate choice of $\lambda_3$ will alter the total annihilation cross-section of DM components and hence affect the freeze out and relic density of DM. 
\begin{figure}[htb!]
$$
 \includegraphics[height=4.9cm]{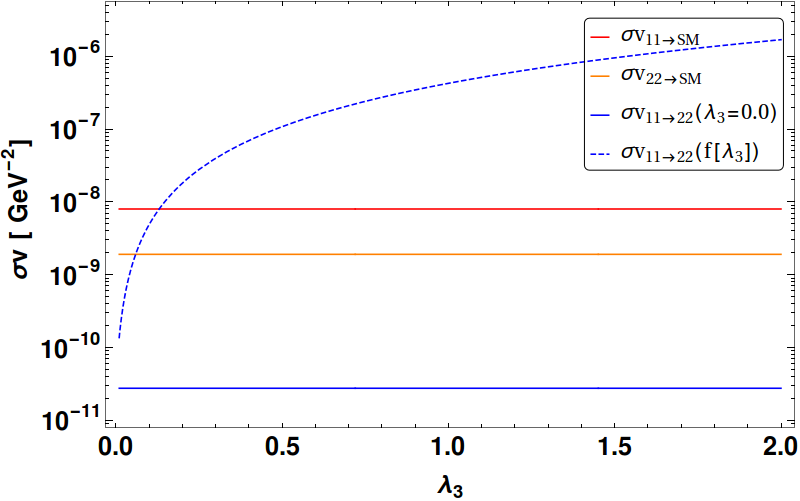}
  \includegraphics[height=4.9cm]{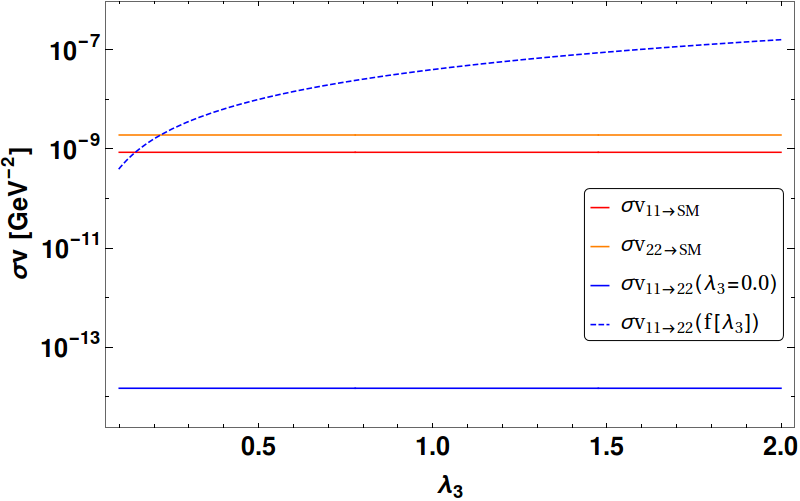}
 $$
 \caption{Variation of DM conversion as a function of $\lc$. Other parameters are chosen as follows: $\{\la,\lb,\mphia,\mphib\}=$  $\{0.1,0.1,150,50\}$ (left), 
 $\{0.1,0.1,500,50\}$ (right). All the masses are in GeV.}
\label{fig:lambda3}
 \end{figure}

\begin{figure}[htb!]
$$
 \includegraphics[height=4.8cm]{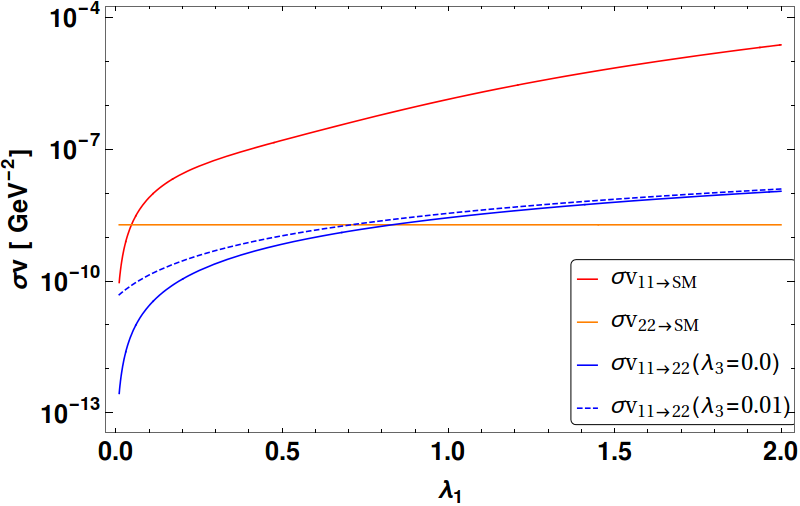}
  \includegraphics[height=4.8cm]{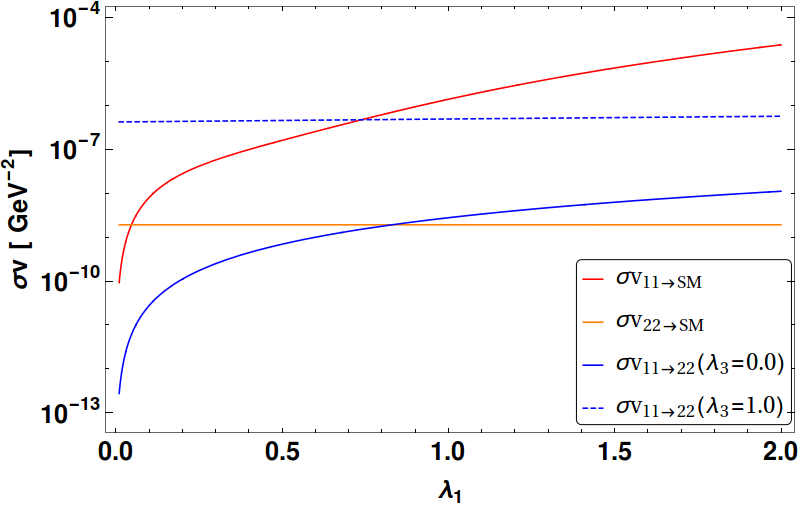}
 $$
 \caption{Variation of Annihilation cross-section and DM conversion as a function of $\la$. Other parameters are chosen as follows:
 $\{\lb,\lc,\mphia,\mphib\}=\{0.1,0.01,150,50\}$ (left), $\{0.1,1.0,150,50\}$ (right). All the masses are in GeV.}
 \label{fig:lambda1}
 \end{figure}

It is clear from Eq.~\ref{eq:annihilationSM} that annihilations $\phi_1\phi_1 \to SM$ is a quadratic function of $\lambda_1$. DM conversion $\phi_1\phi_1\to \phi_2\phi_2$ are although dominantly quadratic function of $\lambda_{1}$, $\lb$ and $\lambda_3$, there is an interference term proportional to $\lambda_1\lambda_2\lambda_3$ as can be seen from Eq.~\ref{eq:annihilationDM1}. However this interference term plays a crucial role when we choose any of those couplings negative. We do not analyse negative couplings in this report to be on the safe for vacuum stability of the potential (see Eq.~\ref{stab_conA}). Cross-sections as a function of $\lc$ and $\la$ are presented in Fig.~\ref{fig:lambda3} and Fig.~\ref{fig:lambda1} respectively. 

In left panel of Fig.~\ref{fig:lambda1}, for both $\lc=0$ and $\lc \neq 0$, $(\sigma v)_{\phi_1\phi_1\to \phi_2\phi_2}$ changes similarly with $\lambda_1$ as the chosen value of $\lc=0.01$ is very small and contact interaction term does not contribute much. However, on the right panel, we have instead chosen $\lambda_3=1.0$ (blue dashed line) and DM-DM interaction cross section do not change much with the change in $\lambda_1$ as the contact interaction term dominates over the Higgs mediated channel. This is also due to the fact that the terms proportional to $\la^2,~{\rm or}~\la$, have suppression from $1/(4\mphia^2-m_h^2)^2$, which the term proportional to $\lc^2$ do not contain (see Eq.~\ref{eq:annihilationDM1}). Hence a non-zero and large $\lc$ yields a dominant contribution to DM-DM conversion, making it a slow function of $\la$.

 \begin{figure}[htb!]
$$
 \includegraphics[height=5.05cm]{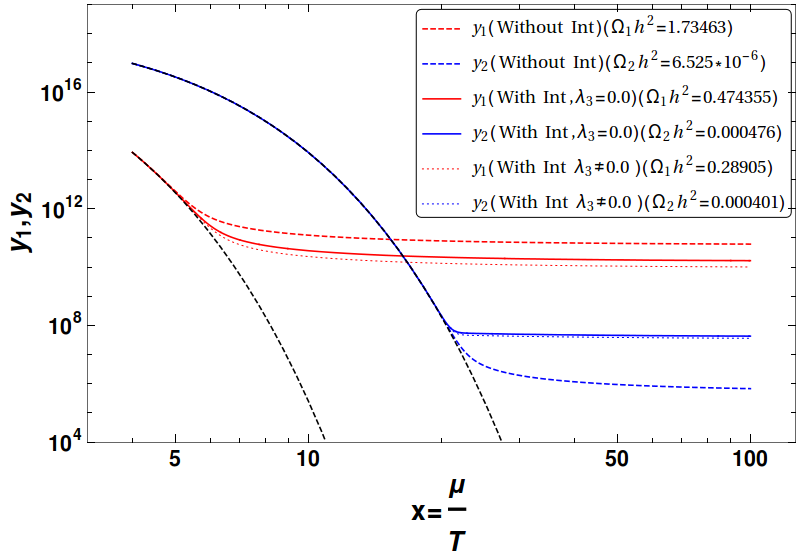}
\includegraphics[height=5.05cm]{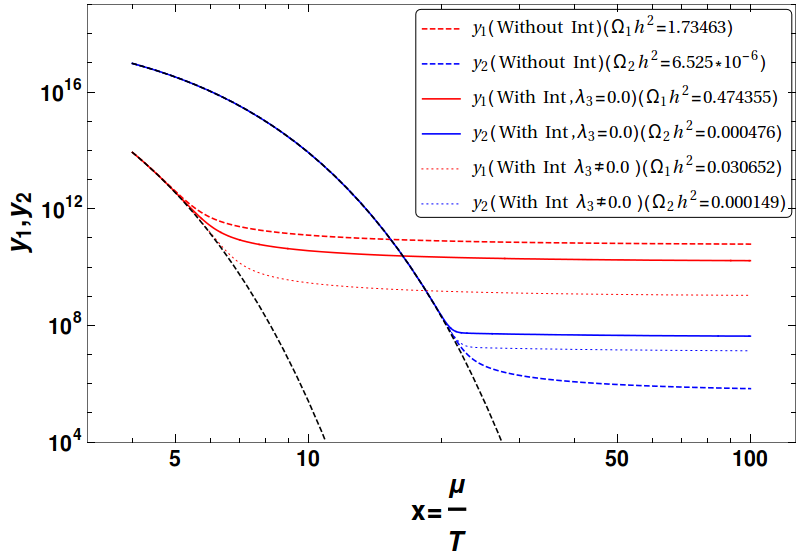}
$$
$$
 \includegraphics[height=5.05cm]{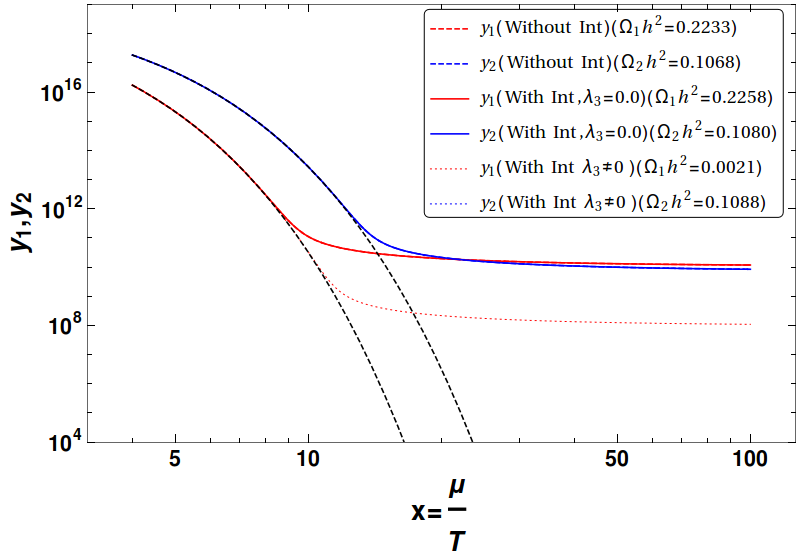}
\includegraphics[height=5.05cm]{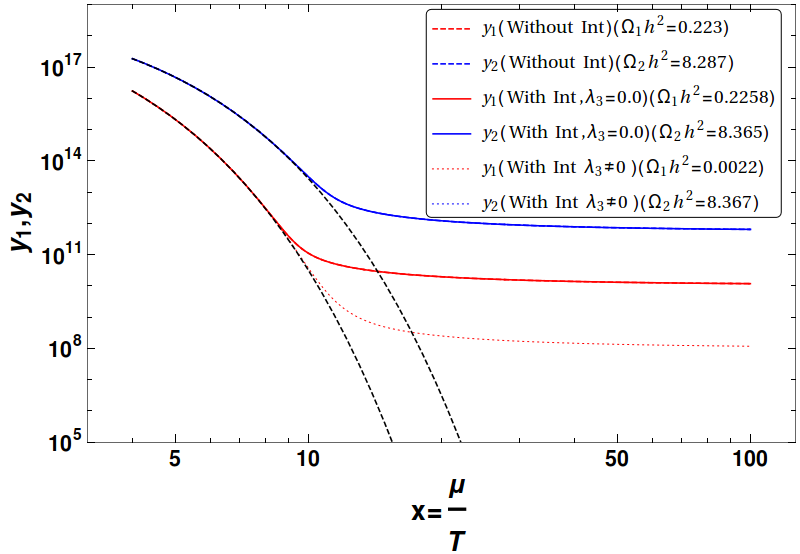}
$$
 \caption{Freeze out of $\phi_1,\phi_2$ (Red and Blue respectively) from equilibrium in $y-x$ plane for $\ZZp$ model. Dashed line indicates freeze out without any DM-DM interaction, solid line denotes $\phi_1-\phi_2$ interacting with $\lc=0$ and dotted line depicts interaction with non-zero $\lc$. The equilibrium distributions are indicated through black dashed lines. The parameters are chosen as follows: $\{\la,\lb,\lc,\mphia,\mphib\}=$  $\{0.01,3.2,0.01,150,60\}$ (top left), $\{0.01,3.2,0.1,150,60\}$ (top right), $\{0.1,0.1,2.0,500,350\}$ (bottom left), $\{0.1,0.01,2.0,500,350\}$ (bottom right). All the masses are in GeV.}
 \label{fig:freeze-out-ZZp}
\end{figure}
Let us now turn to freeze out of the DM components due to the combined effect of annihilation and DM conversions processes in $\ZZp$ model governed by Eq.~\ref{eq:BEQyx}. We show in Fig.~\ref{fig:freeze-out-ZZp} four different regions of parameter space where the decoupling of $\phi_1$ and $\phi_2$ has been indicated in red and blue lines from equilibrium distribution (black dashed line). We have chosen the case of $\mphia>\mphib$ for illustration. We classify three possibilities :
\begin{itemize}
\item No DM-DM interaction ($\sigma v_{\phi_1\phi_1 \to \phi_2\phi_2} =\sigma v_{\phi_2\phi_2 \to \phi_1\phi_1}= 0$, indicated by {\it dashed} lines) ,
\item DM-DM interaction with $\lc=0$ (Indicated by {\it solid} lines) ,
\item DM-DM interaction with $\lc \neq 0$ (Indicated by {\it dotted} lines) .
\end{itemize}
The plots in Fig.~\ref{fig:freeze-out-ZZp} are shown in $y-x$ plane (see Eq.~\ref{eq:BEQyx}). One of the crucial features that is observed for $\phi_1$ decoupling is as follows: {\it dashed} red lines appear on top and {\it dotted} red lines appear at the bottom with {\it solid} lines in the middle. This indicates the yield $y$ for $\phi_1$ DM and therefore the relic density is larger for the {\it dashed} line, smaller for the {\it solid} line and smallest for the {\it dotted} ones. One can correlate this feature as the {\it dotted} ones have largest annihilation contributions through DM-DM interactions with $\lc \neq 0$, the freeze out is delayed compared to solid and dashed lines. 
For the heavier component ($\phi_1$ here), we can appreciate the fact by simply assuming
\bea
\Omega_1h^2 \varpropto (\langle \sigma v\rangle_{\phi_1 \phi_1 \to {\rm SM}}+\langle \sigma v\rangle_{\phi_1 \phi_1 \to \phi_2 \phi_2})^{-1} ,
\eea
which we will explicitly demonstrate in the approximate analytic solutions in Sec \ref{ZZpA}.
On the other hand, for $\phi_2$, indicated in blue, {\it dashed, solid and dotted} lines have different freeze-out sequence. As $\phi_1 \phi_1 \to \phi_2 \phi_2$ is the only possibility with $\mphia>\mphib$, $\phi_2$ is produced from $\phi_1$. Hence, for $\phi_2$, the case without DM-DM interaction, shown in {\it dashed} line appears at the bottom with lowest yield, while {\it dotted} and {\it solid} lines yield larger DM density. The splitting between the {\it dashed, solid} and {\it dotted} lines depend on the strength of the couplings and also on the DM masses as can be observed for different parameter space regions in Fig.~\ref{fig:freeze-out-ZZp}. For example, we see that with small $\lambda_3 \sim 0.01$, the thick and the dotted lines are close enough (top left in Fig.~\ref{fig:freeze-out-ZZp}). Also, one can see that with small $\lambda_2 =0.01$ (bottom right), where the annihilations of $\phi_2 \phi_2 \to SM$ is very tiny, all the blue lines ({\it dashed, solid, dotted}) marge together as DM conversion play little role in decoupling of $\phi_2$, which is dominantly controlled by small interactions to SM yielding an early freeze out and large density. Also, with $\mphia>\mphib$, the annihilation cross-section for $\phi_1$ is smaller compared to $\phi_2$ due to the inverse mass square dependence, yielding an early freeze-out and hence a larger density for $\phi_1$. This is what is observed in two of the cases in upper panel of Fig.~\ref{fig:freeze-out-ZZp}. In bottom left panel, $\la,\lb$ are kept same with large $\lc$. This clearly shows that due to DM-DM conversion the $\phi_1$ freeze out is delayed with reduced yield $y_1$ (compare red dotted line with the red thick one). In same analogy, there is production of $\phi_2$ from DM-DM conversion, which should result in a visible upward shift of $\phi_2$ relic density. However, this is not visible. The reason for not being able to spot the shift is actually due to the log scaling of the figure along $y$ axis. The change of $y$ in the upper direction is much steeper and hence changes in the yield even upto an order of magnitude due to DM-DM conversion for large and non-zero $\lc$ is not clearly distinguishable for $\phi_2$. This is true for all the graphs in Fig.~\ref{fig:freeze-out-ZZp}.    

\begin{figure}[htb!]
$$
 \includegraphics[height=5.1cm]{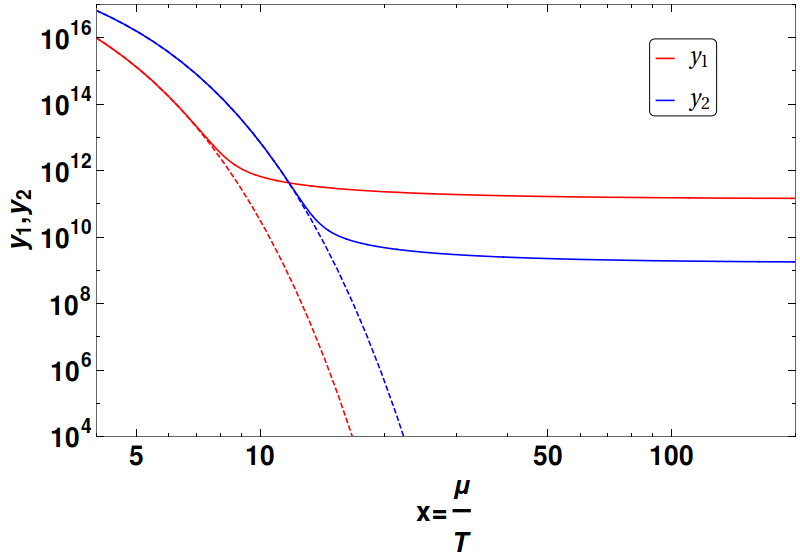}
 \includegraphics[height=5.1cm]{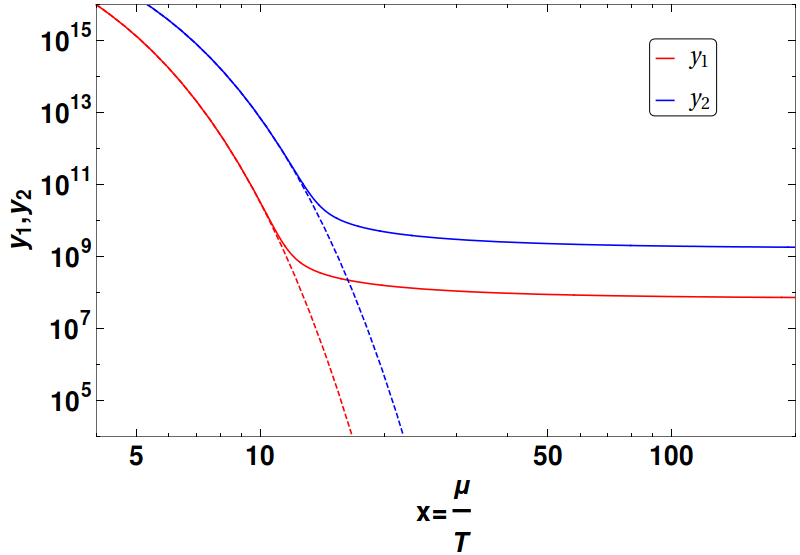}
$$
 \caption{ Freeze out of $\phi_1,\phi_2$ from equilibrium in $y-x$ plane for $\ZZp$ model. $\phi_1$ decoupling is shown in red thick line, $\phi_2$ in blue thick line. Equilibrium distributions are shown in dashed lines. Parameters chosen are as follows: $\{\la,\lb,\lc,\mphia,\mphib\}=$  $\{0.01,0.1,0.0,200,150\}$ (left), $\{0.01,0.1,1.0,200,150\}$ (right). All the masses are in GeVs.}
 \label{fig:Lfreeze-out-ZZp0}
\end{figure}

In order to highlight the importance of DM-DM interaction in the freeze-out of DM components we draw Fig.~\ref{fig:Lfreeze-out-ZZp0}, where on the left hand side, we choose $\lc=0$. Here, due to smaller $\la =0.01$, $\phi_1$ has an early freeze out and larger density. On the right hand side, we take large contact term $\lc=1.0$. With larger DM conversion $\phi_1 \phi_1 \to \phi_2 \phi_2$, the effective annihilation cross-section for $\phi_1$ is larger and yields a smaller DM density. Such examples play a crucial role in multipartite DM with DM-DM interactions.

\begin{figure}[htb!]
$$
 \includegraphics[height=5.1cm]{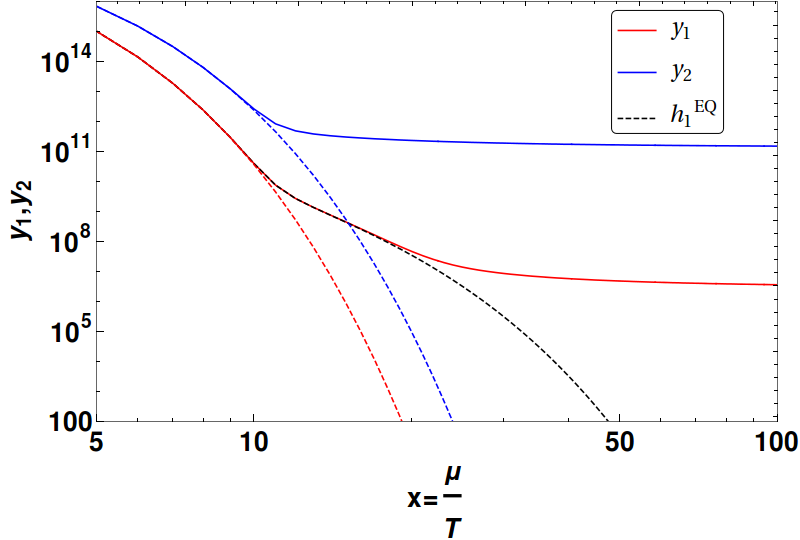}
 \includegraphics[height=5.1cm]{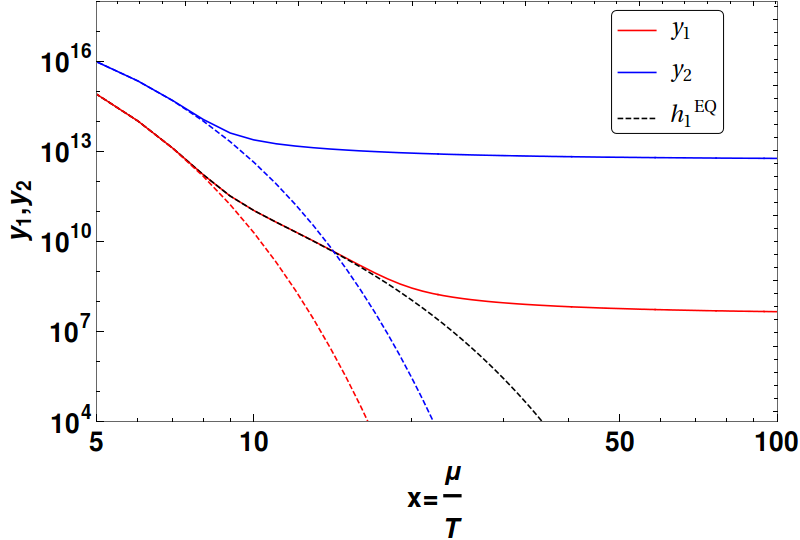}
$$
 \caption{ Freeze out of $\phi_1,\phi_2$ from equilibrium in $y-x$ plane for $\ZZp$ model. $\phi_1$ decoupling is shown in red thick line, $\phi_2$ in blue thick line. Equilibrium distributions are shown in dashed lines. Parameters chosen are as follows: $\{\la,\lb,\lc,\mphia,\mphib\}=$  $\{1.0,0.01,3.0,100,80\}$ (left), $\{0.1,0.001,1.0,120,90\}$ (right). All the masses are in GeVs.}
 \label{fig:Lfreeze-out-ZZp}
\end{figure}

However, in Fig.~\ref{fig:freeze-out-ZZp}, we have missed one important outcome in the freeze out of the two component DM scenario. Given $\mphia>\mphib$, if we can have an early freeze out of $\phi_2$ (by choosing smaller $\lb$), then $\phi_1$ can still sufficiently annihilate to $\phi_2$ remaining in thermal contact with SM, yielding a modified equilibrium before finally freezing out. This shows a bump in the freeze out of $\phi_1$ as has been shown in Fig.~\ref{fig:Lfreeze-out-ZZp}. The modified equilibrium can be identified from BEQ (with $\mphia > \mphib$) as $h_1^{EQ}$ and plotted in Fig.~\ref{fig:Lfreeze-out-ZZp} through black dashed lines to match the bump exactly as follows:
\bea
 \frac{d y_1}{dx}&=&-\frac{\langle \sigma v _{11\rightarrow SM}\rangle+\langle \sigma v _{11\rightarrow 22}\rangle}{x^2}[y_1^2-{h_1^{EQ}}^2];\nonumber \\
 {h_{1}^{EQ}}^2 &=& {y_{1}^{EQ}}^2[\frac{\langle \sigma v _{11\rightarrow SM}\rangle}{\langle \sigma v _{11\rightarrow SM}\rangle+\langle \sigma v _{11\rightarrow 22}\rangle}+
 \frac{\langle \sigma v _{11\rightarrow 22}\rangle}{\langle \sigma v _{11\rightarrow SM}\rangle+\langle \sigma v _{11\rightarrow 22}\rangle}(\frac{y_{2}}{y_{2}^{EQ}})^2] ~.
 \label{eq:BEQyxA} 
\eea
One can note here that in order for such a situation to arise, we have to have the lighter DM component freeze-out earlier, so that the heavier one can annihilate to the lighter component through DM-DM interaction and reaches a modified equilibrium. But for the lighter component to freeze out early, we need much smaller DM-SM coupling compared to the heavier one as has been shown in Fig.~\ref{fig:Lfreeze-out-ZZp}. This is a generic feature associated to interacting multicomponent DM frameworks as also have been pointed out earlier in scalar-fermion two component DM set up in \cite{scalar-fermion}. It is obvious that all the features of DM freeze out that have been discussed above with $\mphia>\mphib$, can also be extended for the case of $\mphib>\mphia$.
\begin{figure}[htb!]
$$
 \includegraphics[height=5.2cm]{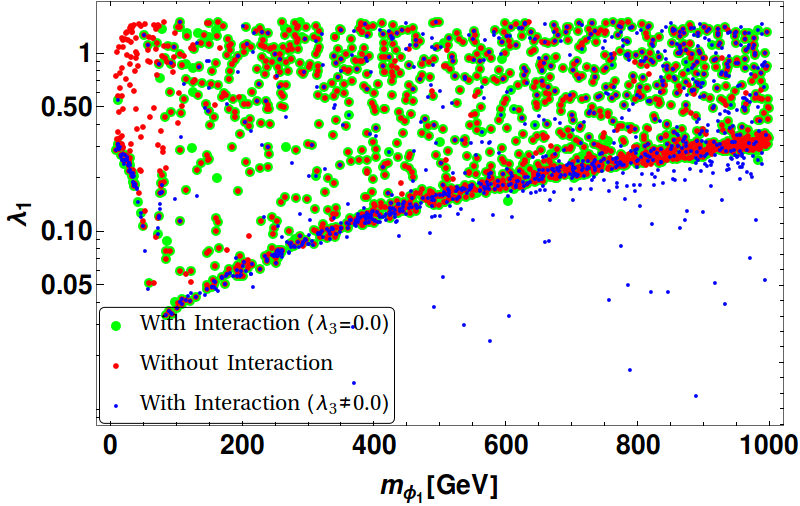}
 \includegraphics[height=5.2cm]{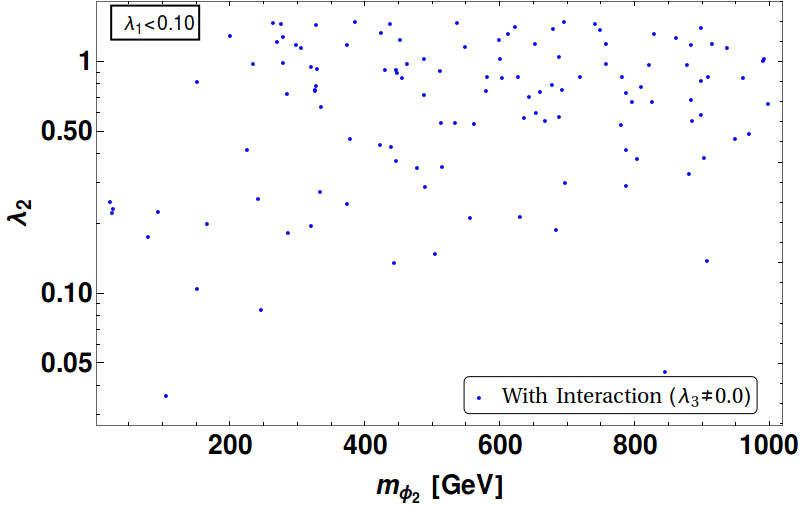}
 $$
 $$
 \includegraphics[height=5.2cm]{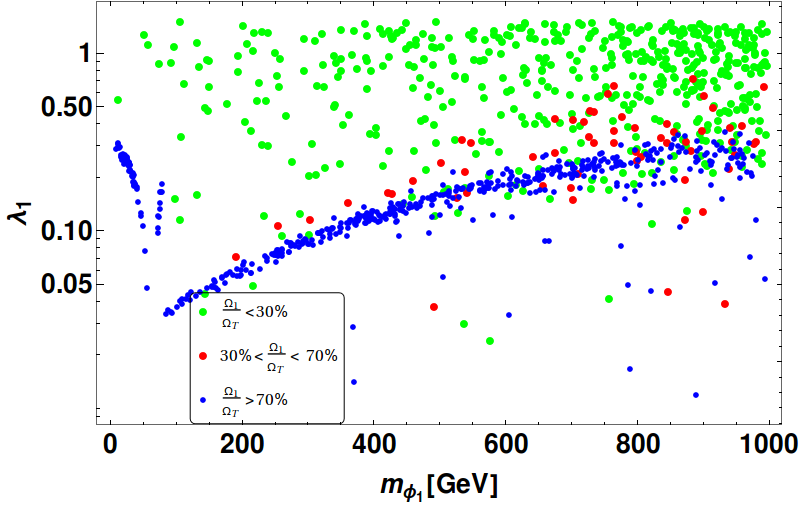}
  \includegraphics[height=5.2cm]{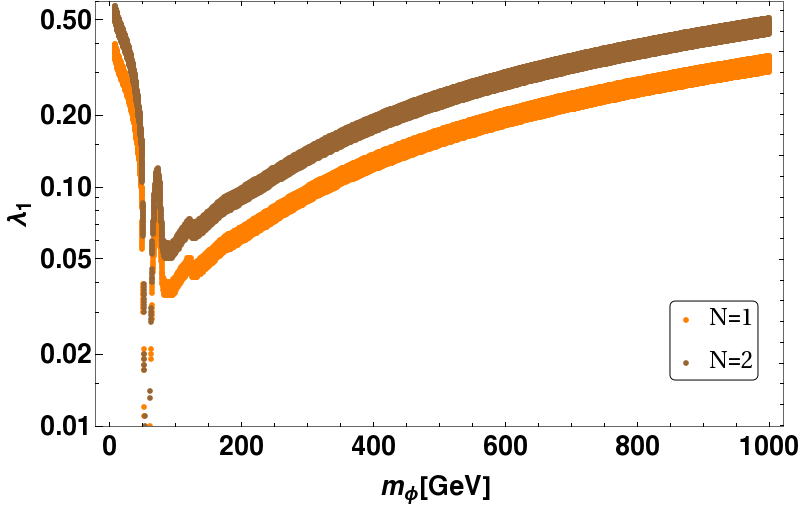}
 $$
 \caption{Top Left: Relic density allowed parameter space of $\ZZp$ model in $\mphia-\la$ plane for three different cases: without DM-DM interaction (Red points), DM-DM interaction with $\lc=0$ (Green points), DM-DM interaction with $\lc \neq 0$ (Blue points); Top right: The blue points of top left figure with $\la \le 0.1$ in $\mphib-\lb$ plane; Bottom left:  Allowed points with DM-DM interaction ($\lc \neq 0$) in $\mphia-\la$ plane showing different DM contributions, $\frac{\Omega_1}{\Omega_T}<30 \%$ (Green), $30\%<\frac{\Omega_1}{\Omega_T}<70 \%$ (Red), $\frac{\Omega_1}{\Omega_T}>70 \%$ (Blue); Bottom right: Relic density constraint on single component framework (Orange) and on two component case with $\mO(2)$ symmetry (Brown) are shown.} 
 \label{fig:m1l1}
\end{figure}

We will discuss now the outcome of the parameter space scan of this model that yields correct relic density. We scan essentially five dimensional parameter space of $\ZZp$ model as follows
\beq
\{10 ~{\rm GeV}<\mphia,\mphib< 1000~{\rm GeV},~0.005<\la, \lb<1.5,~ 0.1<\lc<3\} ~.
\eeq 
Note here that the specific values of the masses in the above limit do not convey anything special; excepting for covering a large range of DM masses allowed by relic density including the Higgs resonance region where annihilations of DM are mediated by Higgs portal coupling. We choose both possible mass hierarchies in uncorrelated way. We have confined ourselves to positive couplings only within a broad range to be compatible with vacuum stability. We solve the coupled BEQ for $\ZZp$ model in Mathematica (see Eq.~\ref{eq:BEQyx}) and then identify the allowed region of correct relic abundance satisfying WMAP~\cite{WMAP} constraint ~\footnote{The range we use corresponds to the 
WMAP results; the PLANCK constraints $0.112 \leq \Omega_{\rm DM} h^2 \leq 0.128$~\cite{PLANCK}, though more stringent, do not lead to 
significant changes in the allowed regions of parameter space.}
\bea
0.09\leq \Omega_{\rm DM} h^2 \leq 0.12 \, ~.
\label{eq:wmap.region}
\eea

The first scan of the allowed parameter space is presented in terms of $\mphia-\la$ plane in top left panel of Fig.~\ref{fig:m1l1}. Three different cases have been indicated in the scan: the case without any DM-DM interaction is indicated by red dots, the case with interaction, but assuming $\lambda_3=0$ is indicated by green dots and the case with $\lambda_3 \neq 0$ is indicated through blue dots. The image can be compared with a similar parameter space scan allowed by by relic density in the single component framework and the one with two components but protected by a $\mO(2)$ symmetry as shown in the bottom right panel of Fig.~\ref{fig:m1l1}. We see that a significantly larger parameter space is available to the two-component set up under $\ZZp$ compared to the single component or the two component case with $\mO(2)$ symmetry. We also see that the case without interaction (red dots) and the ones with $\lambda_3=0$ (green dots) yield identical allowed space in $\mphia-\la$ plane. Such a phenomena is observed because the interaction between DM components in absence of $\lambda_3$ occurs via $\la, \lb$ which also controls their respective annihilations to SM. Hence if one or the other or both of $\la, \lb$ is increased for large DM interaction, the annihilation to SM also gets significantly enhanced beyond the relic density limit. Hence, in absence of $\lambda_3$, the DM interaction is automatically controlled by the annihilation processes and that is why no new region of parameter space opens up within the relic density allowed region of parameter space. However, the situation alters in presence of $\lambda_3$, which indeed can expedite DM-DM interaction and alter relic density contribution of one without affecting the annihilations to SM for the other. Hence, we achieve points below $\la<0.1$, allowed by relic density (which otherwise is not available with no DM-DM interaction or with $\lc=0$) as shown by the blue points in the top left panel of Fig.~\ref{fig:m1l1}. The smaller values of $\la$ gets allowed for this case albeit small annihilations of $\phi_1$ to SM because $\phi_1\phi_1 \to \phi_2 \phi_2$ interaction supplements it in presence of non-zero $\lambda_3$, yielding $\phi_1$ relic density within the desirable range. In order to understand what happens to $\phi_2$ in such a situation with $\la<0.1$, we see the corresponding $\mphib-\lb$ scan of {\it these} points and observe that they lie in moderate to large $\lb$ region: $\sim\{0.1-1\}$ as shown in the top right panel of Fig.~\ref{fig:m1l1}. One can conclude hence, that the points below $\la<0.1$, although is dominated by the first component $\phi_1$, may have significant second component present. The DM composition of this two-component framework for correct relic density is shown in the bottom left panel of Fig.~\ref{fig:m1l1} in $\mphia-\la$ plane. We see that blue points where the first component dominates ($\frac{\Omega_1}{\Omega_T}>70 \%$) populate low $\la$ regions as expected with some minor presence of green ($\frac{\Omega_1}{\Omega_T}<30 \%$) and red points ($30\%<\frac{\Omega_1}{\Omega_T}<70 \%$). The allowed region with larger $\la$ is mostly dominated by $\phi_2$ green points, while equal contributions from both the DM components (red points) populate mostly the mid range of $\la$ in the bottom left panel of Fig.~\ref{fig:m1l1}. In the bottom right panel of Fig.~\ref{fig:m1l1}, we show the two component case with $\mO(2)$ symmetry (brown points), which requires larger coupling $\la$ for a given $m_\phi$ compared to the single component case (orange points). This happens because the individual DM density in $\mO(2)$ case is reduced to half of total DM density $\Omega_{1}=\Omega_2=\frac{\Omega_T}{2}$ (see Eq.~\ref{eq:relicON}); consequently the annihilation has to be twice as large than the single component case. We also figure out the resonance drop at $\mphia \sim \frac{m_h}{2}$ in all of the scans and resemblance of single component situation through larger population of points in $\la-\mphia$ scan. 

\begin{figure}[htb!]
$$
 \includegraphics[height=5.2cm]{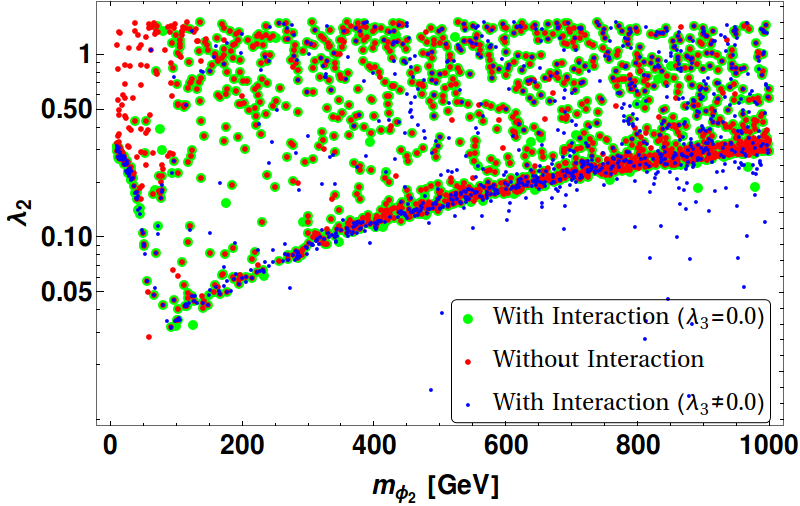}~ 
 \includegraphics[height=5.2cm]{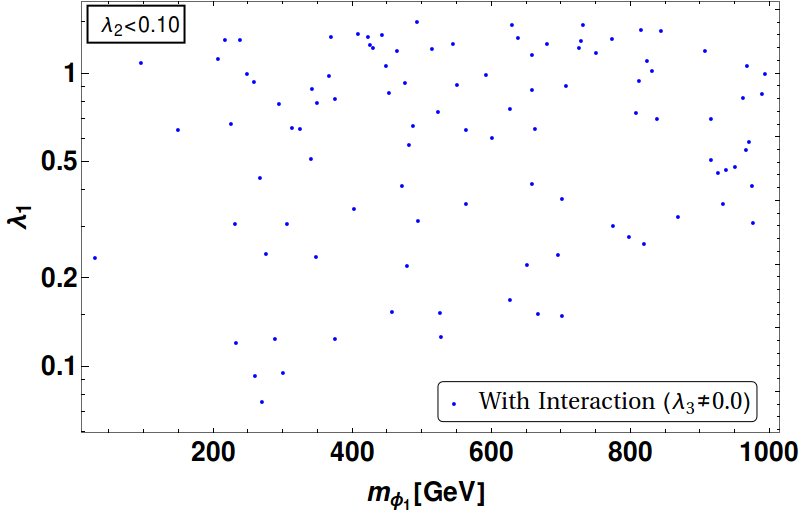}
 $$ 
 \caption{Left: Relic density allowed parameter space of $\mZ_2 \times \mZ^{'}_2$ in $\mphib-\lb$ plane for three different cases: without DM-DM interaction (Red points), DM-DM interaction with $\lc=0$ (Green points), DM-DM interaction with $\lc \neq 0$ (Blue points); Right: The blue points of left figure with $\lb <0.1$ in $\mphia-\la$ plane.}
 \label{fig:m2l2}
\end{figure}

A very similar situation arises when we recast the scan in $\mphib-\lb$ plane as shown in Fig.~\ref{fig:m2l2}. We once again point out that a clear distinction of points with non-zero $\lambda_3$ appearing below $\lb<0.1$ for $\mphib \sim \{300,~1000\}$ GeV, where the DM content is primarily dominated by the second ($\phi_2$) component.  
\begin{figure}[htb!]
$$
 \includegraphics[height=5.2cm]{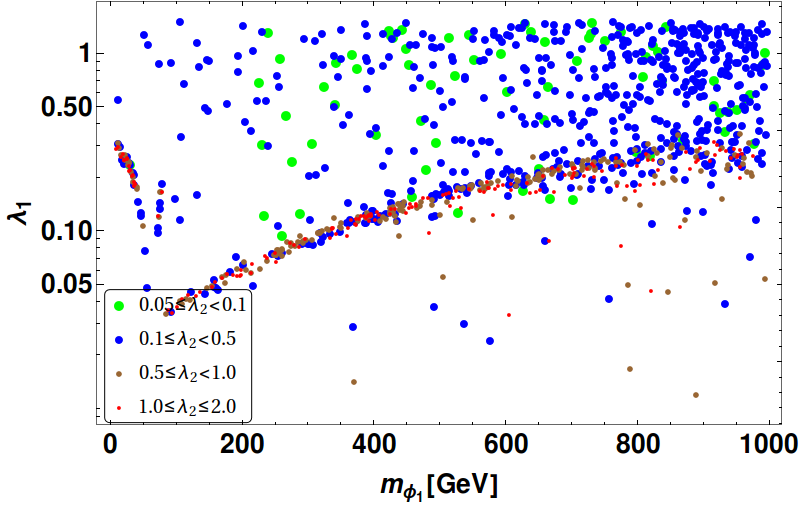}~ \includegraphics[height=5.2cm]{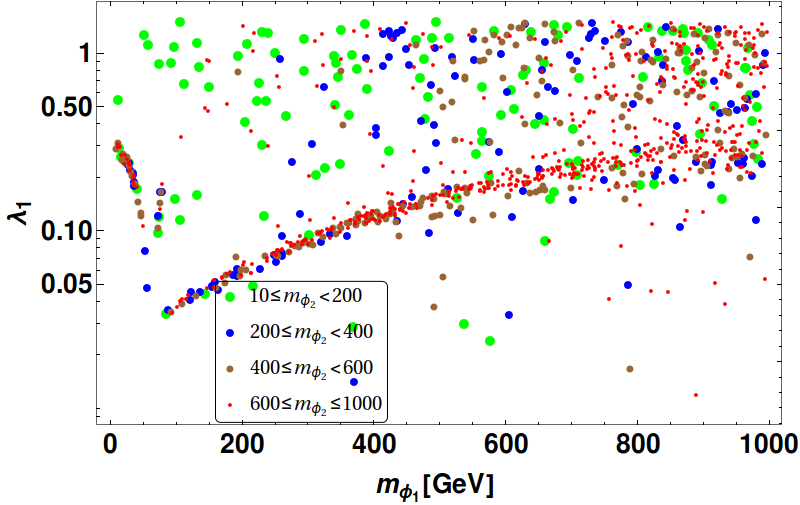}
 $$
 \caption{Relic density allowed parameter space of $\ZZp$ model in $\mphia-\la$ plane. The case with DM interaction ($\lambda_3\neq 0$) is depicted. Left: Different ranges of $\lb$: $\{0.05-0.1\}$ (Green), $\{0.1-0.5\}$ (Blue), $\{0.5-1\}$ (Grey), $\{1.0-2.0\}$ (Red) are shown. Right: Different ranges of $\mphib$: $\{10-200\}$ (Green), $\{200-400\}$ (Blue), $\{400-600\}$ (Grey), $\{600-1000\}$ (Red) are shown. All the masses are in GeV.}
 \label{fig:m3l3}
\end{figure}

In Fig.~\ref{fig:m3l3}, the $\mphia-\la$ parameter space allowed by relic density constraint is shown in terms of different ranges of $\lb$ (left) and $\mphib$ (right) for $\lc \neq 0$ case. In the left panel different ranges of $\lb$ are chosen as follows: $\{0.05-0.1\}$ (Green), $\{0.1-0.5\}$ (Blue), $\{0.5-1\}$ (Grey), $\{1.0-2.0\}$ (Red). Here we note that while the small $\lb(<0.5)$ (Blue and Green dots) span the most of the $\mphia-\la$ parameter space, larger values of $\lb(>0.5)$ tend to populate similar to the single component case. This can be understood as follows: due to very large annihilation cross-section with large $\lb$, $\phi_2$ barely contributes to relic density and it approximates to a single component case with $\phi_1$. In the right hand side of the Fig.~\ref{fig:m3l3} different mass ranges of the second component ($\mphib$) is indicated as $\{10-200\}$ (Green), $\{200-400\}$ (Blue), $\{400-600\}$ (Grey), $\{600-1000\}$ (Red), all in GeV. However, this doesn't show a necessary distinction amongst different $\phi_2$ mass ranges with all ranges couplings ($\lb$) included in the scan indicating $\mphib$ is insensitive to $\mphia-\la$ scan unless we choose a specific range of $\lb$. 

\begin{figure}[htb!]
$$
 \includegraphics[height=5.1cm]{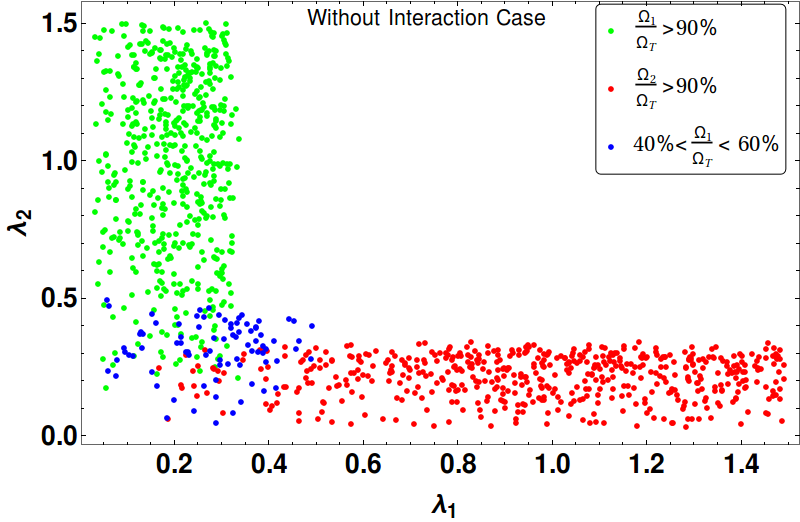}
 \includegraphics[height=5.1cm]{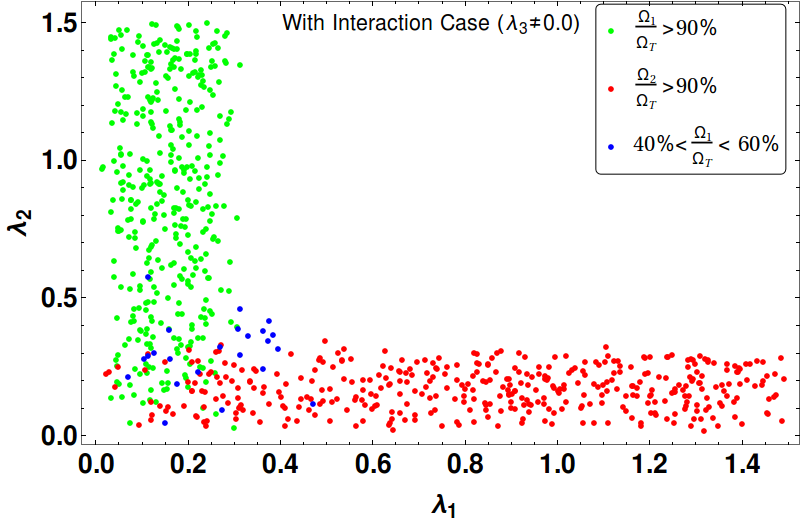}
 $$
$$ 
 \includegraphics[height=6.0cm]{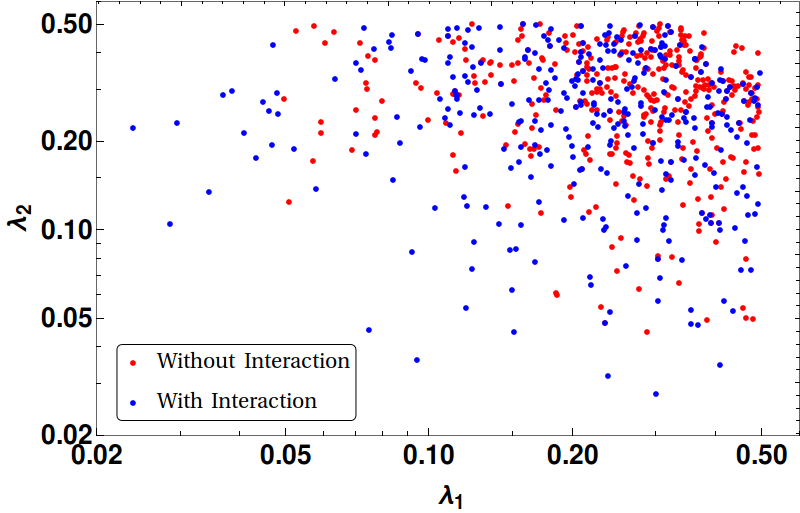}
 $$ 
 \caption{Relic density allowed parameter space of $\ZZp$ model in $\la-\lb$ plane. Top Left: The case without DM-DM interaction, Top Right: The case with interaction ($\lambda_3 \neq 0$) are shown. Different colors indicate the composition of the relic density; Blue: $40\%<\frac{\Omega_1}{\Omega_T}<60 \%$,  Green: $\frac{\Omega_1}{\Omega_T}> 90 \%$, Red: $\frac{\Omega_2}{\Omega_T}>90 \%$. Bottom: $\la-\lb$ scan for $\la,\lb<0.5$: Red: Without DM-DM interaction, Blue: With interaction ($\lc \neq 0$). }
 \label{fig:l1l2}
\end{figure}

Another possible representation of relic density allowed parameter space of the two component DM in $\ZZp$ framework is $\la-\lb$ plane as has been shown in Fig.~\ref{fig:l1l2}. Essentially this plane yields an L-shaped figure with the allowed region extending to as large possible value of the coupling as one wishes (upto perturbative limit). This is attributed to the presence of second DM component that the individual coupling of one DM with SM can be as large as possible while keeping the other in the correct ball park. In  Fig.~\ref{fig:l1l2}, we show three different regions: the ones indicated through red dots are those where $\phi_2$ constitutes $90 \%$ or more of the DM and understandably lives on the axis with larger $\la$; the ones in green dots show the region where $\phi_1$ constitutes $90 \%$ or more of the DM and extends along $\lb$; the ones with blue dots show the region where both of the components contribute in a similar way to relic density of the universe. When $\la$ is large, corresponding annihilation of the $\phi_1$ component is large enough to have a tiny contribution to relic density and vice versa. On the other had, when $40\%<\frac{\Omega_1}{\Omega_T}<60 \%$, there is a limit to which $\la-\lb$ can be large or small as seen in the blue dots accumulating in a small region at the corner of allowed $\la-\lb$ plane. The thickness of each leg is essentially determined by the width of the DM relic density ($\Omega h^2$: 0.09-0.12). It is also worthy of mentioning that large values of both $\la ~\&~ \lb$ are strongly disfavoured by direct search bounds as we will discuss later. The two cases of having no DM-DM interaction (top left panel of Fig.~\ref{fig:l1l2}) or  with interaction (top right panel of Fig.~\ref{fig:l1l2}) makes no visible difference unless one looks carefully into the smaller coupling regions. This is what has been displayed in the bottom panel of Fig.~\ref{fig:l1l2}, with $\la,\lb<0.5$. The cases of having DM-DM interaction is shown in blue while the red points do not have any interactions between the DM components. It is clear that the blue points tend to acquire smaller values than the red ones, showing that with DM-DM interaction (with $\lc \neq 0$), the two component DM models essentially gets allowed to smaller coupling strength marking a significant difference with the models without DM-DM interactions.
 
\section{Approximate Analytic Solution for Coupled BEQ in $\ZZp$ model}
\label{ZZpA}

In this section, we will evaluate approximate analytical solution for the coupled BEQ in $\ZZp$ framework and try to validate with the numerical solution obtained in previous section. Approximate analytical solutions help using them without actually solving the coupled differential equations numerically for such a model in general.
In order to obtain analytical solution we will assume $\mphia > \mphib$. The case for $\mphib > \mphia$ will be similar of course.  
BEQs for $\mphia > \mphib$  reads:
\begin{eqnarray}
 \frac{d y_1}{dx}&=&-\frac{1}{x^2}[\sigma_1(y_1^2-{y_1^{EQ}}^2)+\sigma_{12}(y_1^2-\frac{{y_1^{EQ}}^2}{{y_2^{EQ}}^2} y_2^2)] ~, \nonumber \\
 \frac{d y_2}{dx}&=&-\frac{1}{x^2}[\sigma_2(y_2^2-{y_2^{EQ}}^2)-\sigma_{12}(y_1^2-\frac{{y_1^{EQ}}^2}{{y_2^{EQ}}^2} y_2^2)]  ~;
 \label{BEQ:1gt2}
\end{eqnarray}
 where we rewrite the cross-sections for notational simplicity as: $\langle \sigma v _{11\rightarrow SM}\rangle \equiv \sigma_1 $, $\langle \sigma v _{22\rightarrow SM}\rangle \equiv \sigma_2 $ and  $\langle \sigma v _{11\rightarrow 22}\rangle \equiv \sigma_{12} $. Let us consider the difference between the actual yield from the equilibrium for $\phi_1$ as $\Delta_1=y_1-y_1^{EQ}$ and for $\phi_2$ as $\Delta_2=y_2-y_2^{EQ}$ which helps parametrising the freeze out of the DM components. In terms of $\Delta_{1,2}$ \cite{Kolb:1990vq}, the BEQs become :
 \bea
  \frac{d\Delta_1}{dx}+\frac{dy_1^{EQ}}{dx}&=&-\frac{1}{x^2}[\sigma_1(\Delta_1^2+2\Delta_1y_1^{EQ})+\sigma_{12}[(\Delta_1^2+2\Delta_1y_1^{EQ})-
  (\frac{y_1^{EQ}}{y_2^{EQ}})^2(\Delta_2^2+2\Delta_2y_2^{EQ})]] ~,\nonumber \\
   \frac{d\Delta_2}{dx}+\frac{dy_2^{EQ}}{dx}&=&-\frac{1}{x^2}[\sigma_2(\Delta_2^2+2\Delta_2y_2^{EQ})-\sigma_{12}[(\Delta_1^2+2\Delta_1y_1^{EQ})-(\frac{y_1^{EQ}}{y_2^{EQ}})^2 (\Delta_2^2+2\Delta_2y_2^{EQ})]] ~.\nonumber \\
   \label{eq:couple-delta}
 \eea
 Solving above set of coupled differential equations (Eq.~\ref{eq:couple-delta}) analytically is difficult. Hence, we would like to recast the equations to the nearest possible approximation,
 where solving two of these equations separately is viable. The term that can be neglected here is $\sigma_{12}(\frac{y_1^{EQ}}{y_2^{EQ}})^2$. $\frac{y_1^{EQ}}{y_2^{EQ}} \sim (\frac{\mphia}{\mphib})^{\frac{3}{2}}e^{-\frac{\mphia}{\mphib}x}< 1$ in the limit of $\mphia>\mphib$. However, when $\sigma_{12} \gg \sigma_{1,2}$ this approximation will fail. Generically, a moderate $\lambda_3$ will justify our approximation. With such approximations, the BEQ turns into
 \begin{eqnarray}
  \frac{d\Delta_1}{dx}+\frac{dy_1^{EQ}}{dx}=-\frac{1}{x^2}[\sigma_1(\Delta_1^2+2\Delta_1y_1^{EQ})+\sigma_{12}(\Delta_1^2+2\Delta_1y_1^{EQ})] ~,\nonumber \\
 \frac{d\Delta_2}{dx}+\frac{dy_2^{EQ}}{dx}=-\frac{1}{x^2}[\sigma_2(\Delta_2^2+2\Delta_2y_2^{EQ})-\sigma_{12}(\Delta_1^2+2\Delta_1y_1^{EQ})] ~.
 \end{eqnarray}
 At smaller values of $x$, i.e. before freeze-out of the DM components $x<x^i_f~(i=1,2)$, the number density follows the equilibrium distribution 
 very closely : $\frac{d\Delta_i}{dx}\rightarrow 0$. BEQs in such a situation
 \bea
 \frac{dy_1^{EQ}}{dx}&=&-\frac{1}{x^2}[(\sigma_1+\sigma_{12})(\Delta_1^2+2\Delta_1y_1^{EQ})] ~,\nonumber \\
  \frac{dy_2^{EQ}}{dx}&=&-\frac{1}{x^2}[\sigma_2(\Delta_2^2+2\Delta_2y_2^{EQ})-\sigma_{12}(\Delta_1^2+2\Delta_1y_1^{EQ})] ~.
  \label{eq:y1-y2-freeze}
\eea
Now, when we approach freeze-out, at $x\sim x_f^i$, one can parametrise the difference in the yield to scale the equilibrium distribution by some factor $c_i$ as
$\Delta_i[x_f^i]=c_iy_i^{EQ}[x_f^i]$ \cite{Kolb:1990vq}. Using this, one can easily find the freeze-out of the first component as  \cite{Kolb:1990vq}
 \begin{eqnarray}
  e^{({\frac{\mphia}{\mu})}{x_f^1}}=0.03828 M_{Pl}\frac{g}{\sqrt{g_*}}\frac{m_{\phi_1}(\sigma_1+\sigma_{12})(2+c_1)c_1}{\sqrt{\frac{\mphia}{\mu}}\sqrt{x_f^1}} ~.
 \end{eqnarray}
 Considering  $x_f^1 > \frac{3}{2}$, one finds approximately the freeze out of $\phi_1$
 \begin{eqnarray}
  x_f^1 &\approx& (\frac{\mu}{\mphia})~ \ln\left[0.03828 M_{Pl} \frac{g}{\sqrt{g^*}}\mphia \frac{(\sigma_1+\sigma_{12})(c_1+2)c_1}{\sqrt{\frac{\mphia}{\mu}}}\right] \nonumber \\
  &-&\frac{1}{2}(\frac{\mu}{\mphia})~ \ln\left[(\frac{\mu}{\mphia}) \ln[0.03828 M_{Pl} \frac{g}{\sqrt{g^*}}\mphia \frac{(\sigma_1+\sigma_{12})(c_1+2)c_1}{\sqrt{\frac{\mphia}{\mu}}}]\right] ~.
  \label{eq:x1f}
 \end{eqnarray}
 Similarly, 
  \begin{eqnarray}
  x_f^2 &\approx& (\frac{\mu}{\mphib}) \ln\left[0.03828 M_{Pl} \frac{g}{\sqrt{g^*}}\mphib \frac{\sigma_2(c_2+2)c_2}{\sqrt{\frac{\mphib}{\mu}}}\right] \nonumber \\
  &-&\frac{1}{2}(\frac{\mu}{\mphib}) \ln\left[(\frac{\mu}{\mphib}) \ln[0.03828 M_{Pl} \frac{g}{\sqrt{g^*}}\mphib \frac{\sigma_2(c_2+2)c_2}{\sqrt{\frac{\mphib}{\mu}}}]\right] ~.
  \label{eq:x2f}
 \end{eqnarray}
 We note here that the reduced freeze-out $x_f^i$ evaluated from the coupled BEQ has to be rescaled to the individual ones as follows:
 \beq
  x_{if}=\frac{m_{\phi_i}}{T_{if}}=\frac{m_{\phi_i}}{\mu}x_f^i ~,
  \label{eq:rescale-freeze}
  \eeq
 where $T_{if}$ is the freeze-out temperature of $i^{th} ~(i=1,2)$ DM component. In the limit of $\mphia>>\mphib:~ \frac{\mphib}{\mu}\to1$, the freeze-out of the second component reduces to  
 \bea
  x_{2f}\equiv x_f^2 &\approx& \ln \left[0.03828 M_{Pl} \frac{g}{\sqrt{g^*}}\mphib \sigma_2 (c_2+2)c_2\right] \nonumber \\
  &-&\frac{1}{2} \ln \ln \left[0.03828 M_{Pl} \frac{g}{\sqrt{g^*}}\mphib \sigma_2 (c_2+2)c_2 \right] ~,
   \label{eq:x2f-single}
 \eea 
 which exactly mimics the single component case \cite{Kolb:1990vq}.
 
  \begin{figure}[htb!]
$$
  \includegraphics[height=5.0cm]{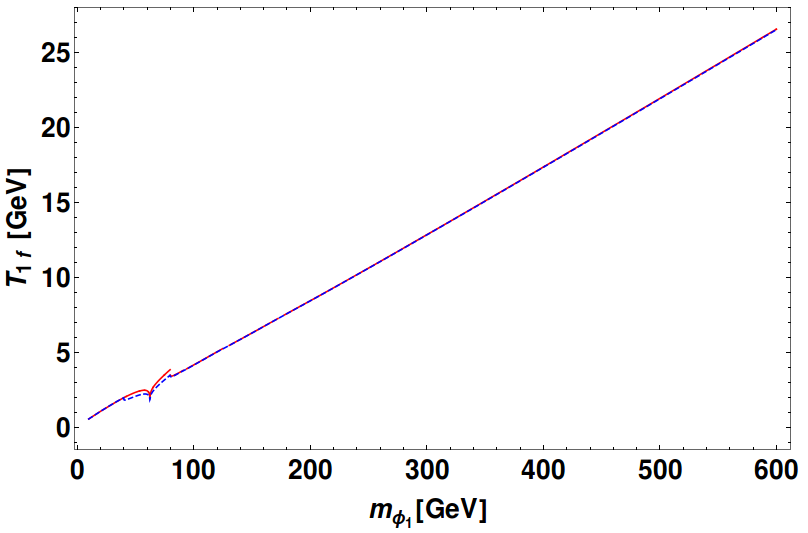}
  \includegraphics[height=5.0cm]{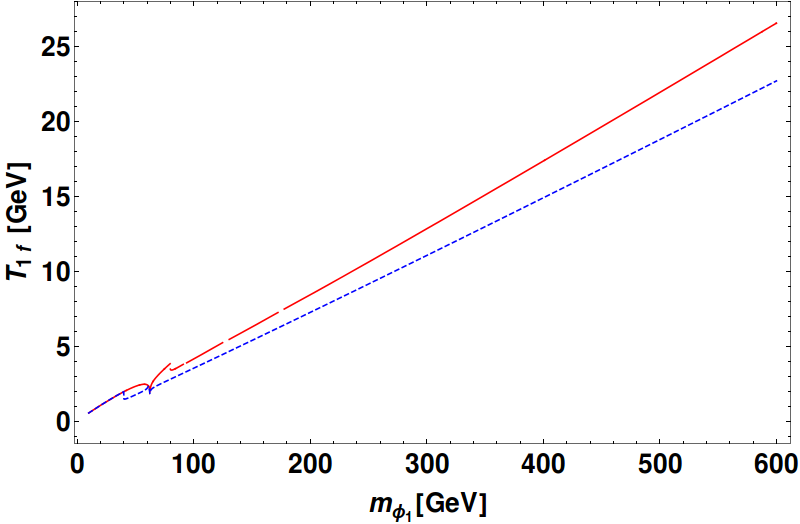}
 $$
 $$
 \includegraphics[height=5.0cm]{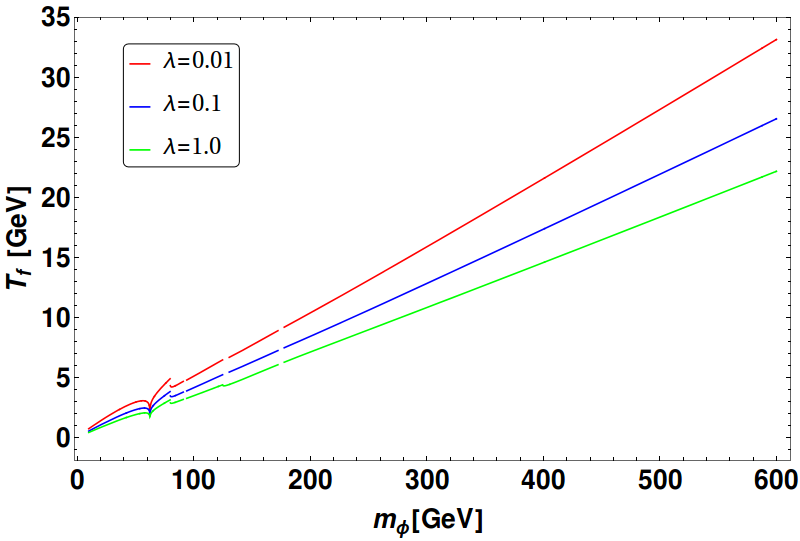}
 \includegraphics[height=5.0cm]{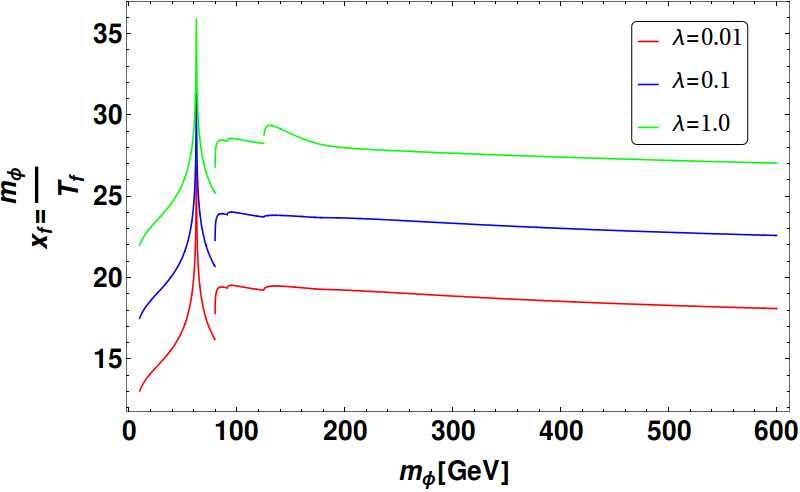}
$$
 \caption{ Top left: Freeze-out temperature of $\phi_1$ ($T_{1f}$) as obtained from analytical solution is plotted as a function of DM mass $\mphia$ for two cases: In absence of the second component $\phi_2$ (red) and in presence of $\phi_2$ and with DM-DM interaction (blue dotted) for $\{\la,\lb,\lc,\mphib\}=\{0.1,0.1,0.0,40\}$; Top right: Same as top left with $\{\la,\lb,\lc,\mphib\}=\{0.1,0.1,1.0,40\}$; Bottom left: Freeze-out temperature ($T_f$) of single component DM is shown for different choices of coupling $\lambda_1=\{0.01,0.1,1.0\}$  (red, blue and green respectively); Bottom right: $x_f$ versus $\m_\phi$ in the single component case with same choices of $\lambda$ as in bottom left. All masses are in GeVs.}
 \label{fig:freezT}
\end{figure}

Before proceeding further, let us study the dependence of freeze-out temperature of $\phi_1$ ($T_{1f}$) on DM mass $\mphia$ as is shown in Fig.~\ref{fig:freezT} as obtained from approximate analytical solutions of BEQ as in Eq.~\ref{eq:x1f},\ref{eq:x2f} and Eq.~\ref{eq:rescale-freeze}. Two cases have been compared here: single component case, in absence of $\phi_2$ (red) and two-component case with DM-DM interaction (blue dotted) for the following choices of parameters $\{\la,\lb,\lc,\mphib\}=\{0.1,0.1,0.0,40\}$ (top left), and $\{0.1,0.1,1.0,40\}$ (top right). It is clear from the figure that presence of $\lambda_3=1.0$ in the top right panel, shows a different freeze-out curve for the first component compared to the case when there is no other DM. This is simply because the presence of DM interaction enhances the effective annihilation cross-section and leads to early decoupling of the $\phi_1$ DM. Freeze out for the single component case with different choice of coupling $\lambda_1=\{0.01,0.1,1.0\}$ is shown in red blue and green respectively at the bottom left panel of Fig.~\ref{fig:freezT} ($T_f$ in the left and $x_f$ on the right bottom panel). Again, the larger is the coupling, the larger is annihilation cross-section and the earlier the DM freezes-out. Also, it is easy to appreciate that with larger DM mass the freeze out also gets delayed with smaller annihilation cross-sections. The resonance drop is also clearly visible in all the plots. Note that in Fig.~\ref{fig:freezT}, there are some discontinuities in the plots arising from singularities in plotting Eq.~\ref{eq:x2f-single} in Mathematica and contains no physics at all.

 After finding the freeze-out conditions of the DMs, the next goal is to find the respective relic densities. For that, we need to consider the epoch at $x\gg x_f^i$. For $x\gg x_f^i$: $y_i^{EQ}\rightarrow 0 $  and evidently $\Delta_i\rightarrow y_i$. Then BEQs read 

 \begin{eqnarray}
  \frac{d\Delta_1}{dx}&=&-\frac{1}{x^2}[(\sigma_1+\sigma_{12})\Delta_1^2] ~, \nonumber \\
  \frac{d\Delta_2}{dx}&=&-\frac{1}{x^2}[\sigma_2\Delta_2^2-\sigma_{12}\Delta_1^2] ~.
 \end{eqnarray}
Integrating the equations, we get 
\bea
\int_{\Delta_1(x_{1f})}^{\Delta_1(\infty)}-\frac{d\Delta_1}{\Delta_1^2}&=&(\sigma_1+\sigma_{12})\int_{x_{1f}}^{\infty}\frac{dx}{x^2} ~, \nonumber\\ 
\Delta_1(\infty)&=&\frac{1}{\frac{1}{\Delta_1(x_{1f})}+(\sigma_1+\sigma_{12})\frac{1}{x_{1f}}} ~.
\eea
We can evaluate the derivative of $y_{1}^{EQ}$ with respect to $x$ using the equilibrium distribution :
\bea
 \frac{dy_1^{EQ}}{dx}|_{x \to x_f ^1}= \left(\frac{3}{2x_f^1}-\frac{\mphia}{\mu}\right) y_1^{EQ} ~.
 \label{eq:dyeq-dx}
\eea

\begin{figure}[htb!]
$$
 \includegraphics[height=4.8cm]{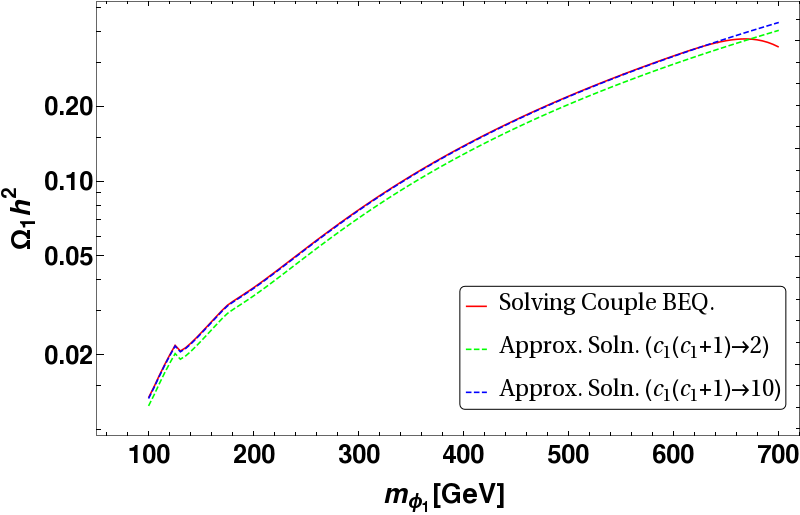}
 \includegraphics[height=4.8cm]{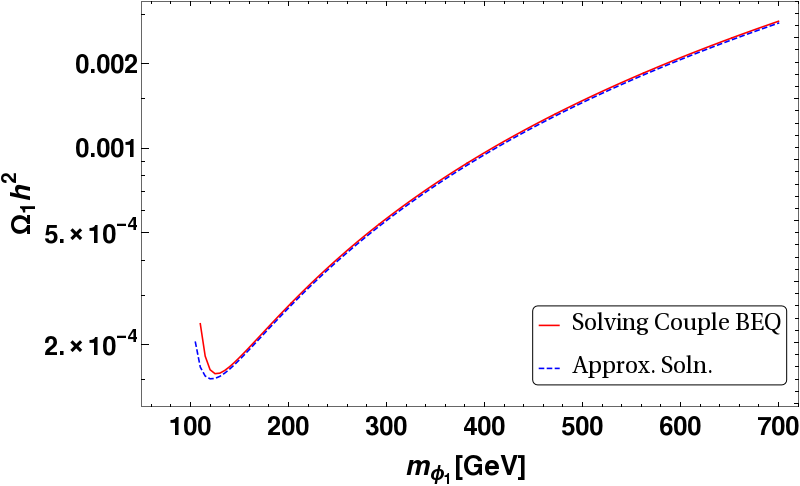}
 $$
 \caption{Relic density of the first DM component $\phi_1$ ($\Omega_1 h^2$) obtained from numerical solution (solid red line) and approximate analytic solution with different values of the factor of proportionality constant $c_1(c_1+1)\rightarrow$ {2 (green dashed line), 10 (blue dashed line)} are compared. The parameters chosen here are $\{\la,\lb,\lc,\mphib\}:\{0.1,0.1,0.0,100\}$(left) and $\{0.1,0.1,2.0,100\}$(right). Masses are in GeVs.} 
 \label{fig:compare1}
\end{figure}

Now, $\Delta_1(x_{1f})$ can be found easily by using Eq.~\ref{eq:y1-y2-freeze}, Eq.~\ref{eq:rescale-freeze} and Eq.~\ref{eq:dyeq-dx} as follows:
  \begin{eqnarray}
  \Delta_1(x_{1f})=\frac{\frac{\mu}{\mphia}(x_{1f}^2-\frac{3}{2}x_{1f})}{2(\sigma_1+\sigma_{12})} ~.
 \end{eqnarray}
 So that  we finally obtain the yield at a large $x \to \infty$ as
 \begin{eqnarray}
  y_1(\infty)\equiv\Delta_1(\infty)=\frac{1}{(\sigma_1+\sigma_{12})[(\frac{\mphia}{\mu})\left(\frac{2}{{x_{1f}}^2-\frac{3}{2}x_{1f}}\right)+\frac{1}{x_{1f}}]} ~.
 \end{eqnarray}
 The approximate solution thus obtained for $\phi_1$ is then compared to the numerical solution obtained using Mathematica and is shown in Fig.~\ref{fig:compare1}. For approximate solution we choose two different values of the proportionality constant $c_1(c_1+1)\rightarrow$
{2 (green dashed line), 10 (blue dashed line)} and numerical solution is shown in red and what we see is a good agreement which shows the robustness of the solution obtained here.
 However, solving the equation for second DM component $\phi_2$ is more tricky. In this case first DM component freeze-out before; so that $x_{1f}<x_{2f}$. Hence, here we can consider $\Delta_1(x)\rightarrow \Delta_1(x_{1f})$ as the region $x$ where we seek the freeze out of the second component lies beyond the freeze-out of the first component. BEQ for $\phi_2$ in such a case is 
 \begin{eqnarray}
  \frac{d\Delta_2}{dx}&=&-\frac{1}{x^2}[\sigma_2\Delta_2^2-\sigma_{12}\Delta_1^2(x_{1f})] ~.
 \end{eqnarray}
Integrating above equation we get 
\begin{eqnarray}
\int_{\Delta_2(x_{2f})}^{\Delta_2(\infty)}\frac{d\Delta_2}{\Delta_2^2-a^2}&=&-\sigma_2 \int_{x_{2f}}^{\infty}\frac{dx}{x^2} ~,\nonumber\\ 
y_2(\infty)\rightarrow\Delta_2(\infty)&=&a\left[\frac{1+\frac{\Delta_2(x_{2f})-a}{\Delta_2(x_{2f})+a}e^{-\frac{2a\sigma_2}{x_{2f}}}}
{1-\frac{\Delta_2(x_{2f})-a}{\Delta_2(x_{2f})+a}e^{-\frac{2a\sigma_2}{x_{2f}}}}\right] ~;
\label{eq:aprx2}
\end{eqnarray}
where $a=\sqrt{\frac{\sigma_{12}}{\sigma_2}}\Delta_1(x_{1f})$ and $\Delta_2(x_{2f})=\frac{(\frac{\mu}{\mphib})(x_{2f}^2-\frac{3}{2}x_{2f})}{2\sigma_2}$. 
%
Now, for $\mphib \ll \mphia$ (i.e $\frac{\mu}{\mphib}\to 1$) and $\sigma_2 \gg \sigma_{12}$: the terms $a=\sqrt{\frac{\sigma_{12}}{\sigma_2}}\Delta_1(x_{1f})\ll 1$. For this limit Eq.~\ref{eq:aprx2} becomes 

\begin{eqnarray}
y_2(\infty)\rightarrow\Delta_2(\infty)&=&a\left[\frac{(\Delta_2(x_{2f})+a)+(\Delta_2(x_{2f})-a)(1-\frac{2a\sigma_2}{x_{2f}})}{(\Delta_2(x_{2f})+a)-(\Delta_2(x_{2f})-a)(1-\frac{2a\sigma_2}{x_{2f}})}\right] \nonumber \\
&\approx& \frac{1}{\frac{1}{\Delta_2(x_{2f})}+\frac{\sigma_2}{x_{2f}}} 
\approx  \frac{1}{\sigma_2\big(\frac{2}{{x_{2f}}^2}+\frac{1}{{x_{2f}}}\big)} ~.
\label{eq:aprx2p}
\end{eqnarray}
Which exactly mimics the single component solution justifiably. It is good to remind the readers that the relic density of DM components in terms of the modified yield obtained using approximation solution can be written as: 

\begin{eqnarray}
 \Omega_i h^2&=&\frac{854.45\times10^{-13}}{\sqrt g_*} \frac{m_{\phi_i}}{\mu} y_i(\infty) ~.
\end{eqnarray}
\begin{figure}[htb!]
 $$
 \includegraphics[height=4.8cm]{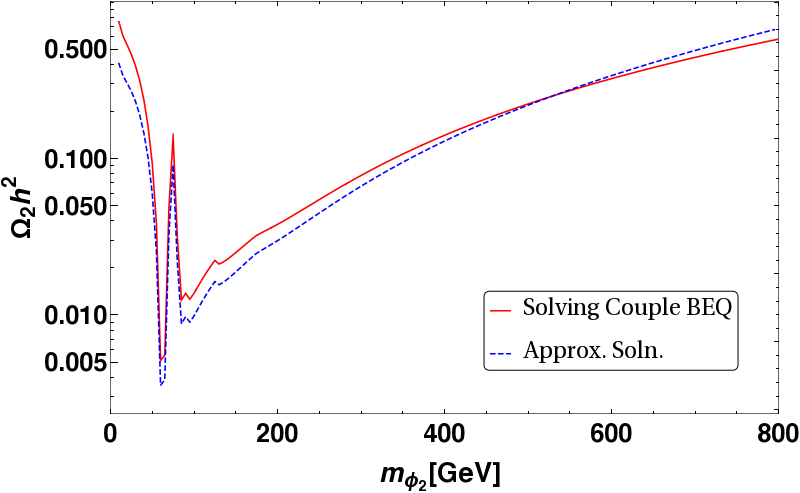}
 \includegraphics[height=4.8cm]{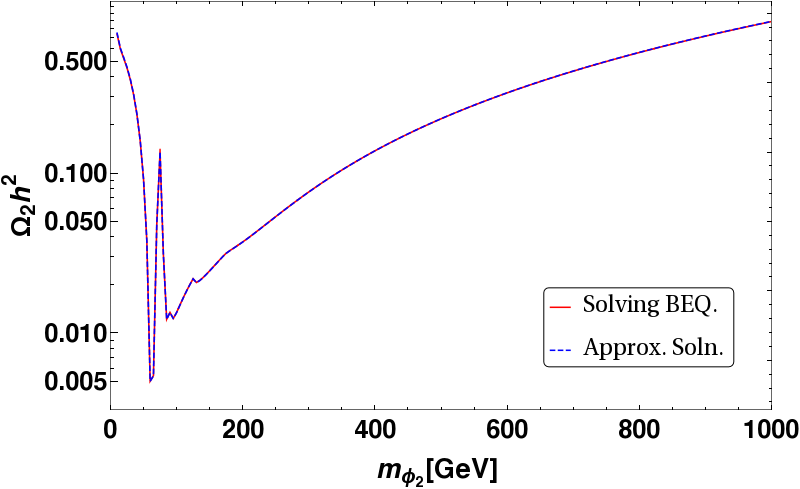}
$$
 \caption{Left: Relic density of $\phi_2$ ($\Omega_2 h^2$) obtained from numerical solution (solid red line) and approximate analytic solution (blue dashed line following Eq. \ref{eq:aprx2}) are compared in two component scenario. The parameters chosen here are: $\{\la,\lb,\lc,\mphia\}:\{0.1,0.1,0.0,800\}$. Right: Same in single component case is depicted following Eq. \ref{eq:aprx2p}. We chose $\lb=0.1$ for illustration. Masses are in GeVs.} 
 \label{fig:compare2}
\end{figure}
The approximate analytical solution for the second component ($\Omega_2 h^2$) is compared with the numerical solution obtained using Mathematica and is shown in Fig.~\ref{fig:compare2}. In the left panel we show the comparison for two component case, while on the right panel, we show the limit in which it behaves like a single component DM.


\section{Direct Detection in $\ZZp$ model}
\label{DD}

\begin{figure}[htb!]
$$
 \includegraphics[height=4.8cm]{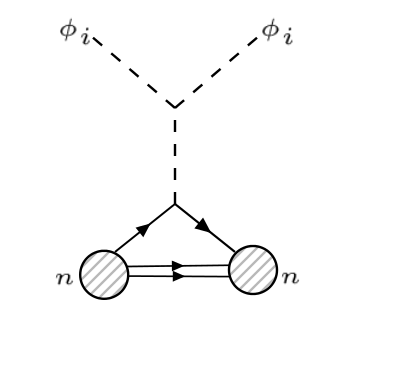}
  $$
 \caption{Feynman diagram for direct detection of scalar singlet DM.}
 \label{fig:direct}
\end{figure}

Scattering of scalar singlet DM with the detector nuclei originates from the Higgs-portal interaction. The quark level interaction in our model occurs via t channel diagram through Higgs exchange as shown in Fig.~\ref{fig:direct}. Two interaction vertices are $\frac{1}{2}\lambda_i v h \phi_i \phi_i$ and $\frac{m_q}{v}h\bar{q}q$, where $i=1,2$. Effective Lagrangian for direct search process can be written as \cite{Ghorbani,direct}

\begin{eqnarray}
 \mathcal{L}_{eff}&=&(\frac{1}{2}\lambda_i v)(\frac{m_q}{v})\frac{1}{m_h^2}\phi_i\phi_i\bar{q}{q}\nonumber\\
 &=&\alpha_q\phi_i\phi_i\bar{q}{q} ~.
\end{eqnarray}
 The matrix element of DM-quark elastic scattering in the limit of zero momentum transfer is ($Q\rightarrow 0$)
 \begin{eqnarray}
 \mathcal{M}_{\phi_i q}=\sum_{q} \alpha_q \langle (\phi_i)_f|\phi_i\phi_i|(\phi_i)_i\rangle\langle q_f|\bar{q}q|q_i\rangle ~.
 \end{eqnarray}
One can convert the quark level interaction $\Rightarrow$ neucleon level interaction as 
\begin{eqnarray}
 \sum_{q} \alpha_q \langle (\phi_i)_f|\phi_i\phi_i|(\phi_i)_i\rangle\langle q_f|\bar{q}q|q_i\rangle \Rightarrow 
 \alpha_n \langle (\phi_i)_f|\phi_i\phi_i|(\phi_i)_i\rangle\langle n_f|\bar{n}n|n_i\rangle ~,
 \end{eqnarray}
 where 
 \begin{eqnarray}
  \alpha_n &=& m_n\sum_{u,d,s} f_{T_q}^{(n)} \frac{\alpha_q}{m_q} + \frac{2}{27} f_{T_g}^{(n)} \sum_{q=c,t,b}\frac{\alpha_q}{m_q} \nonumber \\
  &=& m_n\sum_{u,d,s} f_{T_q}^{(n)} \frac{\alpha_q}{m_q} + \frac{2}{27}(1-\sum_{u,d,s} f_{T_q}^{(n)})\sum_{q=c,t,b}\frac{\alpha_q}{m_q} \nonumber \\
  &=& \frac{m_n \lambda_i}{m_h^2}[(f_{T_u}^{(n)}+f_{T_d}^{(n)}+f_{T_s}^{(n)})+\frac{2}{9}(f_{T_u}^{(n)}+f_{T_d}^{(n)}+f_{T_s}^{(n)})] ~;
  \end{eqnarray}
  
 where nucleon $n$ stands for both proton and neutron. For proton : $f_{T_u}^p=0.0153$ , $f_{T_d}^p=0.0191$ , $f_{T_s}^p=0.0447$ \cite{dd-form}. Hence, spin independent cross section of DM-Neucleon scattering reads\cite{s4}:
  \begin{eqnarray}
   \sigma_{n_i}^{SI}=\frac{\alpha_n^2 \mu_{n}^2}{4\pi m_{\phi_i}^2} ~,
    \end{eqnarray}
 where $\mu_n=\frac{m_n m_{\phi_i}}{m_n+m_{\phi_i}}$ ~.
 
For Two component DM case, the DM-nucleon cross-section has to be even further modified by the fraction of the particular component present in the universe as \cite{multi-singlet4}:
\begin{eqnarray}
 \sigma_{eff}^{SI}(n_i)= (\frac{\Omega_i}{\Omega_T})\sigma_{n_i}^{SI}=\frac{\Omega_i}{\Omega_T}\frac{\alpha_n^2 \mu_{n}^2}{4\pi m_{\phi_i}^2} ~~~~~~~~(i=1,2) ~.
 \label{eq:dd1}
 \end{eqnarray} 
    
\begin{figure}[htb!]
$$
 \includegraphics[height=5.0cm]{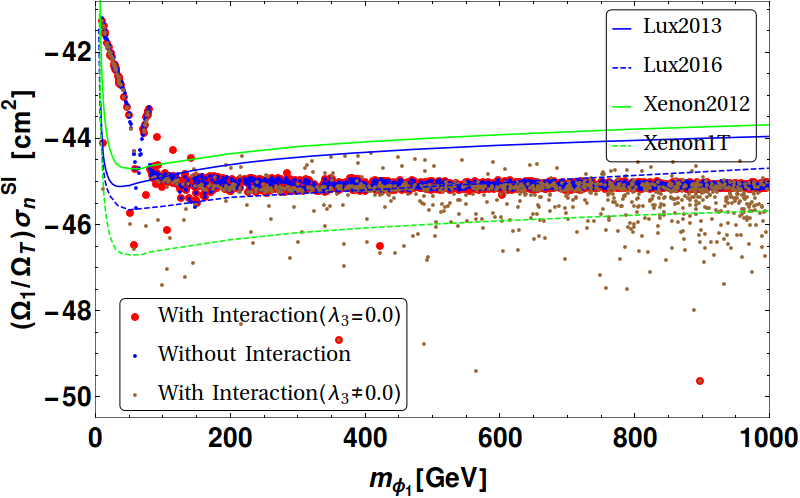} 
 \includegraphics[height=5.0cm]{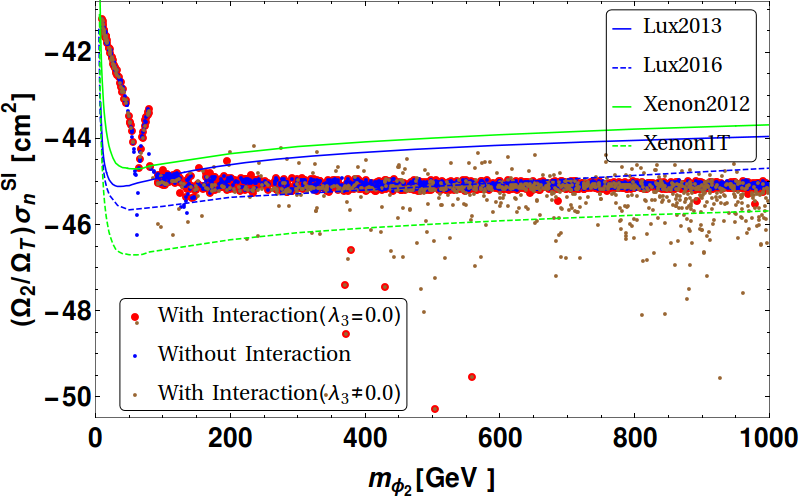}
 $$
 $$
 \includegraphics[height=6.0cm]{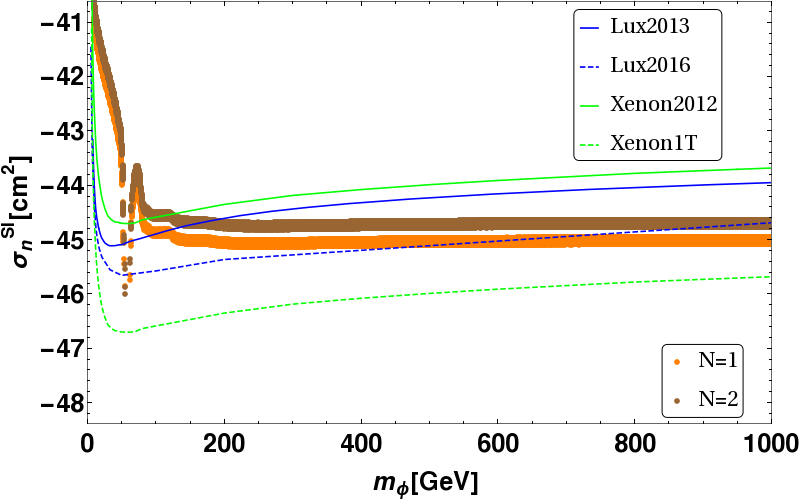}
 $$
 \caption{Spin-independent DM-nucleon effective cross-section (in Log scale) for relic density allowed points as a function of DM mass in $\ZZp$ model. Top Left: The case for $\phi_1$ is shown in $\mphia-\frac{\Omega_1}{\Omega_T}\sigma_n^{SI}$ plane; Top Right: The case for $\phi_2$ is shown in $\mphib-\frac{\Omega_2}{\Omega_T}\sigma_n^{SI}$ plane. Three different situations are indicated as follows: no DM-DM interactions by blue dots, DM interaction with $\lambda_3=0$ through red dots and DM interaction with $\lambda_3 \neq 0$ by grey dots. Bottom: Single component DM framework and two component case with $\mO(2)$ are shown for comparison. Limits from XENON 2012, LUX and LUX 2016 data have been indicated while the predictions of XENON1T is also provided.}
 \label{fig:dd1}
\end{figure}

\begin{figure}[htb!]
$$
 \includegraphics[height=5.0cm]{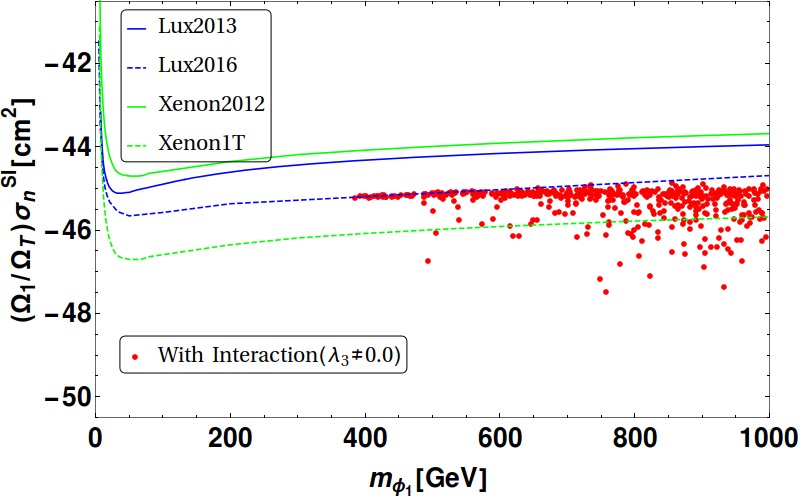}~
  \includegraphics[height=5.0cm]{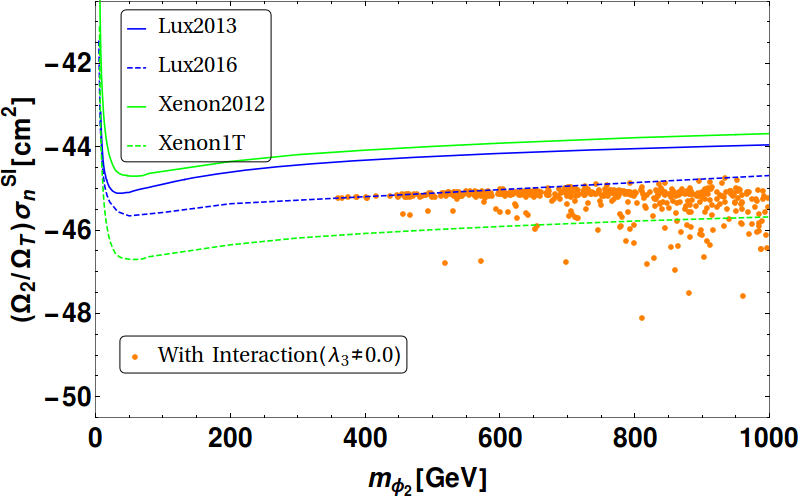}
 $$
 $$
 \includegraphics[height=6.0cm]{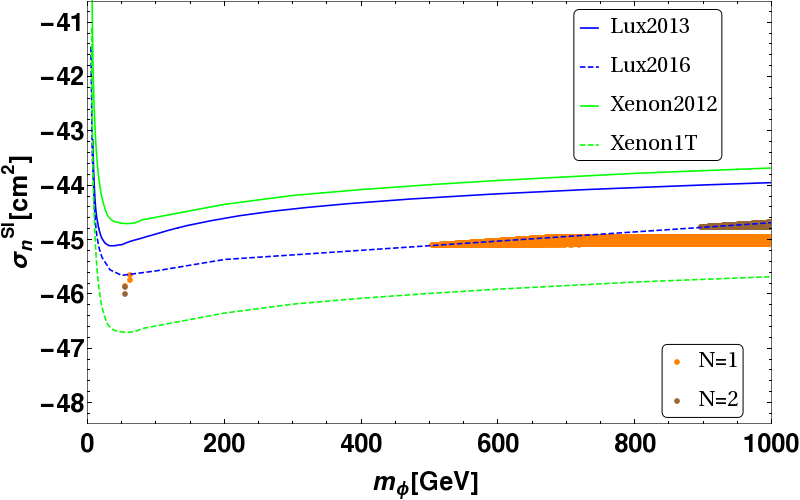}
 $$
 \caption{Points in $\ZZp$ model that satisfy relic density and direct search constraint for both $\phi_1$ and $\phi_2$ by LUX 2016 data in presence of DM-DM interactions with $\lambda_3 \neq 0$. Top left: Points in $\mphia-\frac{\Omega_1}{\Omega_T}\sigma_n^{SI}$ plane; Top right: Same points in $\mphib-\frac{\Omega_2}{\Omega_T}\sigma_n^{SI}$ plane. Bottom: Single component and two component DM in $\mO(2)$ cases are illustrated.}
 \label{fig:ddl31}
\end{figure}

It is of great importance to see how much of relic density allowed DM parameter space of $\ZZp$ model is allowed by the spin independent direct search constraints by XENON2012 \cite{Xenon1,Xenon2}, and updated LUX  data \cite{LUX,LUX2}. This is what is presented  in Fig.~\ref{fig:dd1}. The scattered points show DM-nucleon spin-independent cross-section ($\sigma_n^{SI}$) multiplied by effective scaling $\frac{\Omega_i}{\Omega_T}$ (as in Eq.~\ref{eq:dd1}) as a function of DM mass for relic density allowed points. The case for $\phi_1$ is shown in left while for $\phi_2$ is in the right side of Fig.~\ref{fig:dd1}. The cases for no DM-DM interaction (Blue), DM-DM interaction with $\lambda_3=0$ (Red) and DM-DM interaction with $\lambda_3 \neq 0$ (Grey) are separately shown. One can clearly see that the blue dots are placed below the LUX line but a significant part of them is discarded by the updated LUX data indicating that the model without DM-DM interaction is discarded in the low mass region excepting for the Higgs resonance $\mphia \sim m_h/2$. Although there are blue points allowed above $\sim 600$ GeVs, we will show that they are not simultaneously present for both $\phi_1$ and $\phi_2$. On the contrary, there are several points in red and grey which yields much smaller DM-nucleon cross-section that are placed below the updated LUX data and can be even delayed upto XENON1T. Hence, the multipartite model may live longer in presence of DM-DM interaction. This feature of multipartite DM models seems very interesting and hasn't been pointed out in analysis recent past. The reason for obtaining small DM-nucleon cross-section in presence of DM-DM interactions can be understood as follows: DM conversion helps heavier component to acquire small relic density (through larger annihilations due to DM conversion) without enhancing the coupling to SM ($\la$ or $\lb$); thus yielding DM-nucleon cross-section smaller than what one can achieve without having DM-DM interactions. What about the lighter component then ? It is easy to appreciate that it has to possess a larger density to account for correct abundance and hence requires smaller annihilation cross-section. In presence of DM-DM interaction, lighter component is also produced by the heavier ones requiring the annihilations of the lighter component to SM even smaller; hence, smaller becomes DM-SM coupling and direct search cross-section. Effective DM-nucleon cross-section can also get smaller by reducing the factor $\frac{\Omega_{1,2}}{\Omega_T}$. However, if one ratio is small, the other one must be large or close to one, which may not justify to keep both components on board. In Fig.~\ref{fig:dd1}, there is no correlation between the points in left and right. Here, we have plotted all the points allowed by relic density in both figures not bothering about one particular point in one picture correspond to what in the other. This is what we do systematically in the following scans. It is also very important to note the status of the single component framework of the scalar singlet DM and the two-component framework of $\mO(2)$ model as shown in the bottom panel of Fig.~\ref{fig:dd1}. We see that for the single component case, the DM is ruled out to approximately 500 GeV while for the two component DM in $\mO(2)$ case, $m_{DM} \ge 890$ GeV, excepting for the resonance region.

A correlated scan with $\lambda_3 \neq 0$ is shown in Fig.~\ref{fig:ddl31}. We plot relic density allowed points that satisfy LUX 2016 bounds for both $\phi_1$ and $\phi_2$. It turns out that now there are points even below XENON1T. This indicates that discovery of at least one component can get delayed even beyond XENON1T. The reason for such a phenomena to occur is easy to appreciate. In presence of $\lambda_3$, the DM abundance is within limit without having a large $\la,\lb$ even without compromising for one component to make up the whole relic density, thus having the effective DM-nucleon cross-section reduced for both the cases. The outcome also means, that the model yields a possibility of detecting one component in near future, while the other might be delayed beyond XENON1T. It is also important to note here that there are no points where both $\phi_1$ and $\phi_2$ direct search cross-section goes beyond XENON1T indicating that if nothing is seen till the sensitivity of XENON1T, the two component framework with $\ZZp$ is most likely discarded. In the bottom panel of Fig.~\ref{fig:ddl31}, we again compare the single component and two component $\mO(2)$ cases. 

\begin{figure}[htb!]
$$
 \includegraphics[height=5.0cm]{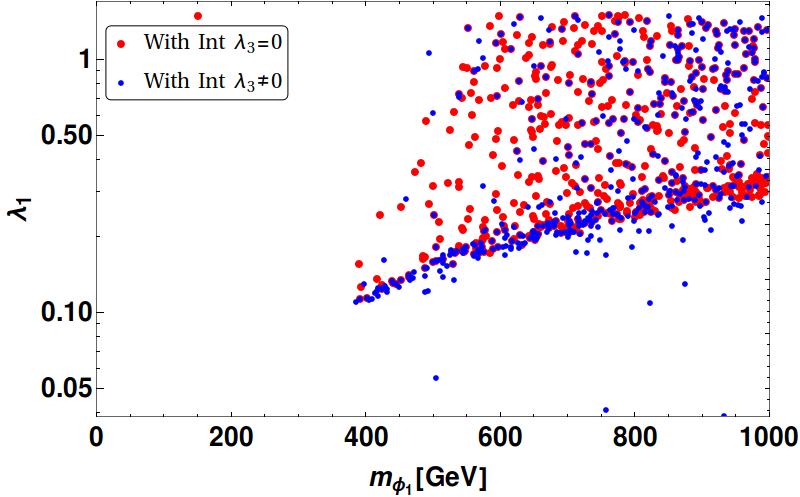}
  \includegraphics[height=5.0cm]{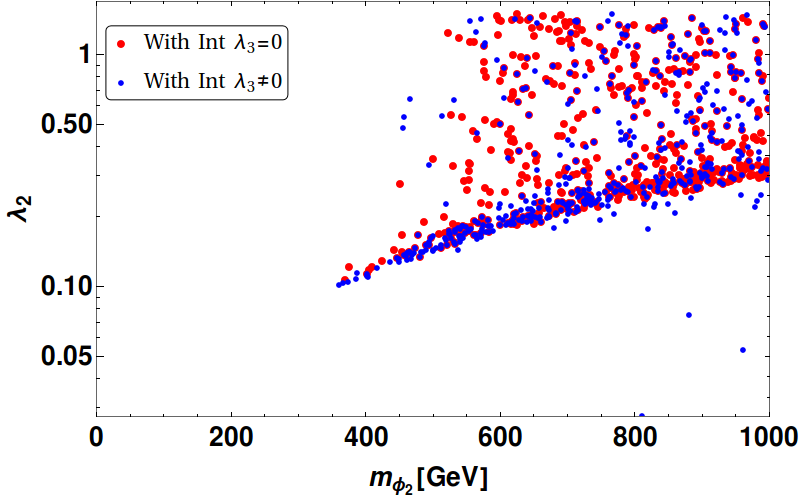}
  $$
  $$
  \includegraphics[height=5.0cm]{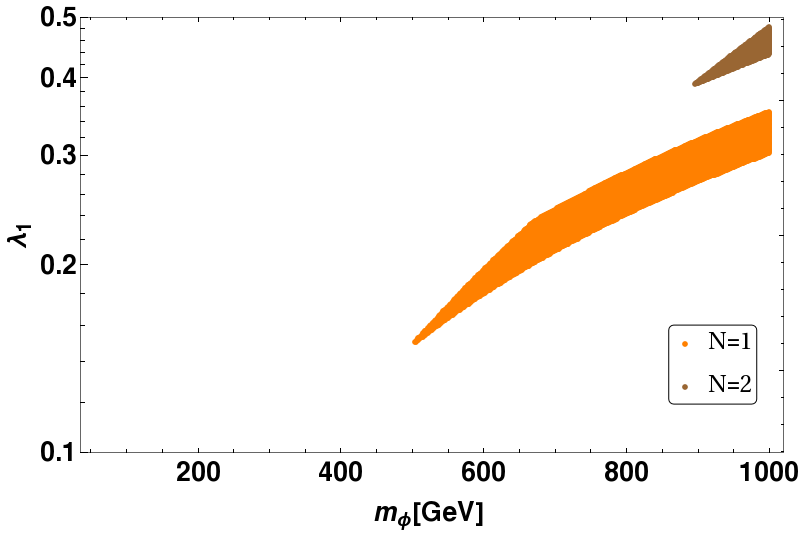}
 $$
 \caption{Points in $\ZZp$ model that satisfy relic density and direct search from updated LUX 2016 bound for both $\phi_1$ and  $\phi_2$ is projected in $\mphia-\la$ plane (top left) and in $\mphib-\lb$ (top right) plane. The case with $\lc=0$ is shown in red and $\lc \neq 0$ is shown in blue. The cases of single component and two component $\mO(2)$ framework are shown in the bottom panel in orange and brown respectively. We have omitted the resonance region in this figure for the sake of clarity.}
 \label{fig:ddlalb1}
\end{figure}

Relic density and direct search (LUX 2016) allowed points for both $\phi_1$ and $\phi_2$, for the case of $\lambda_3=0$ (red) and $\lambda_3 \neq 0$ (blue) are shown in $\mphia-\la$ plane (top left) and $\mphib-\lb$ plane (top right) of Fig.~\ref{fig:ddlalb1}. Again, those in single component and two component non-interacting cases in $\mO(2)$ are shown in the bottom panel for comparison. Required DM mass for $\ZZp$ model is around $\sim 350$ GeV where that of the single component case is $\sim 500$ GeV. There are broadly two reasons how one can evade single component limit on DM masses to allow even lower values from direct search constraints: (i) when both components contribute equally, direct search cross-sections for each component reduces to half of what it would be for single component scenario, (ii) in presence of large DM-DM interactions (preferably with large $\lambda_3$), some of the small coupling regions of single component DM framework which would have yielded larger relic density, yields right relic density and comes under direct search limit. However, the main flexibility in two component set up allows a larger region of DM-SM couplings ($\la,\lb$) spanning from a much lower to a higher value. Secondly, the two scenarios of DM-DM interactions with $\lc=0$ and $\lc \neq 0$ distinguishes themselves at small values of $\la,\lb$ as has already been discussed. The resonance region is anyway allowed in all versions of scalar singlet DM frameworks, which we have omitted in this figure for the sake of comparison.

\begin{table}[htb!]
\begin{tabular}{ |p{1.6cm}|p{4.0cm}|p{0.7cm}|p{0.7cm}|p{0.9cm}|p{0.9cm}|p{1.65cm}|p{1.65cm}| }
 \hline
  Benchmark Points  & {$\{\la,\lb,\lc,\mphia,\mphib\}$}& ${\Omega_1} h^2  $ & ${\Omega_2} h^2$ & $\frac{{\Omega_1}}{{\Omega_T}}(\%)$&$\frac{{\Omega_2}}{{\Omega_T}}
  (\%)$&${\sigma_1}_{eff}^{SI}$ ($10^{-46} cm^2$)  &  ${\sigma_2}_{eff}^{SI}$ ($10^{-46} cm^2$) \\
 \hline
 \hline
 BP1&$\{0.4,0.3,0.0,815,715\}$ & 0.04 & 0.06 & 40 & 60 & 7.7 & 8.4\\
 \hline
 BP2&$\{0.15,0.3,0.0,390,572\}$ & 0.06 & 0.04 & 60 & 40 & 7.8 & 7.3\\
 \hline
 BP3&$\{0.3,0.8,0.0,960,850\}$ & 0.10 & 0.01 & 91 & 9 & 7.0 & 6.3\\
 \hline
 BP4&$\{0.36,0.38,0.7,890,895\}$ & 0.06 & 0.06 & 50 & 50 & 6.2 & 7.4\\
 \hline
 BP5&$\{0.31,0.46,0.82,940,915\}$ & 0.05 & 0.06 & 45 & 55  & 4.3 & 1.0\\
 \hline
 BP6&$\{0.6,0.1,1.6,775,370\}$ & 0.003 & 0.117 & 2 & 98 & 2.8 & 7.2\\
 \hline
 \hline
\end{tabular}
\caption{Some benchmark points in $\ZZp$ model allowed by relic density and direct search. The input parameters, relic density of individual components and direct search cross-sections mentioned. All the masses are in GeV.}
  \label{tab:tableZZp}
\end{table}  

Finally some benchmark points have been mentioned for $\ZZp$ model in Tab.~\ref{tab:tableZZp} (BP1-BP6) where all the input parameters, individual relic densities and spin independent direct search effective cross-sections are mentioned. We choose examples of one component dominating over the other as well as the cases where they contribute equally to relic density. Here, BP2 is an example where single component limit could be evaded due to almost equal presence of both the components, reducing the direct search cross-section by a factor of half in each case. Another interesting example where single component limit was evaded is BP6. Here large $\lambda_3$ produces $\phi_1\phi_1\to\phi_2\phi_2$ significantly balancing the number density of $\phi_2$ through annihilation. In absence of DM-DM interactions, the combination $\{m_{\phi_2},\lambda_2\}=\{370 ~{\rm GeV},0.1\}$ would have ended yielding larger relic density ($\Omega_2>0.12$) and wouldn't be considered as a valid point due to over closure, although allowed by direct search.  

\begin{figure}[htb!]
$$
 \includegraphics[height=5.0cm]{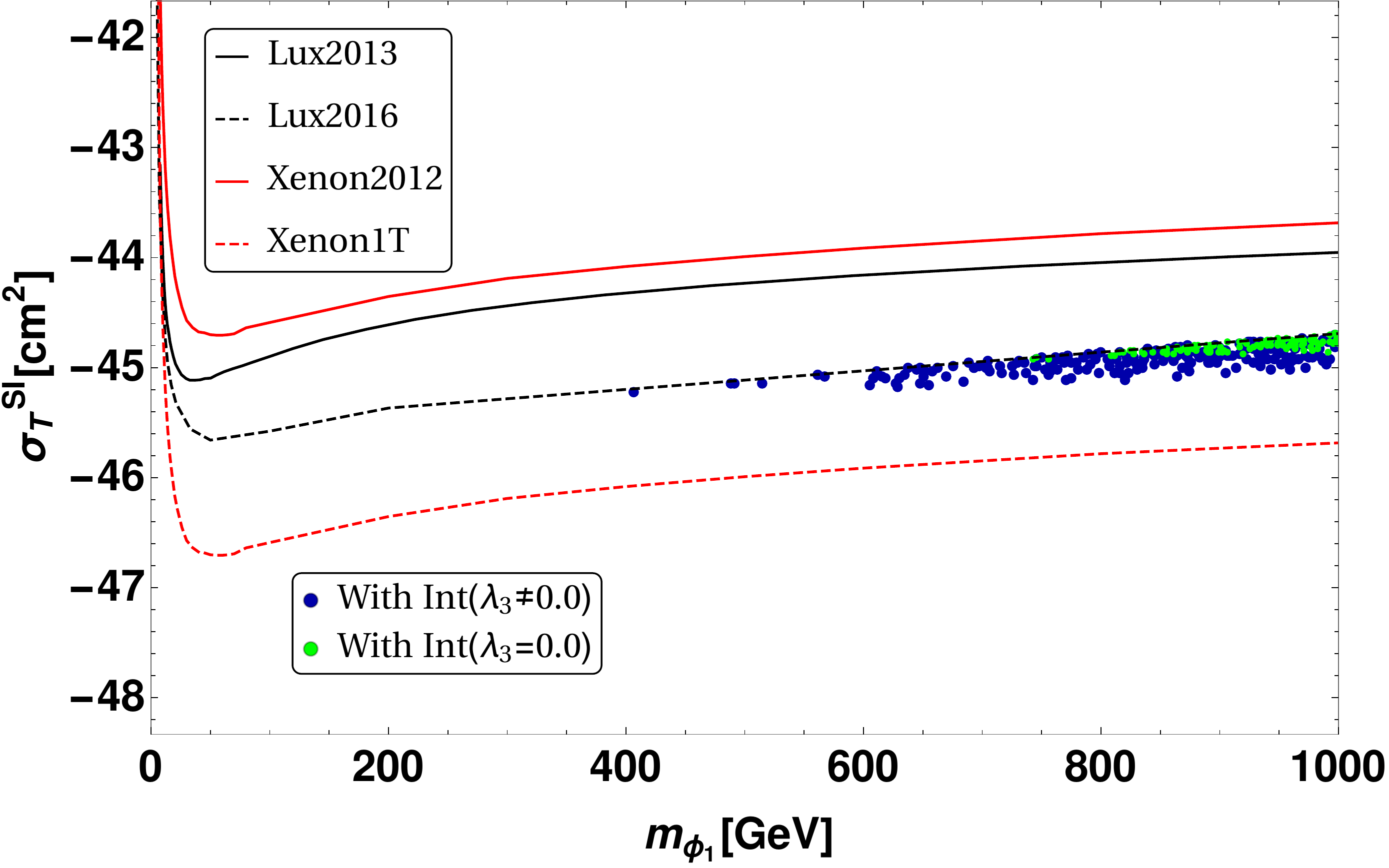}
  \includegraphics[height=5.0cm]{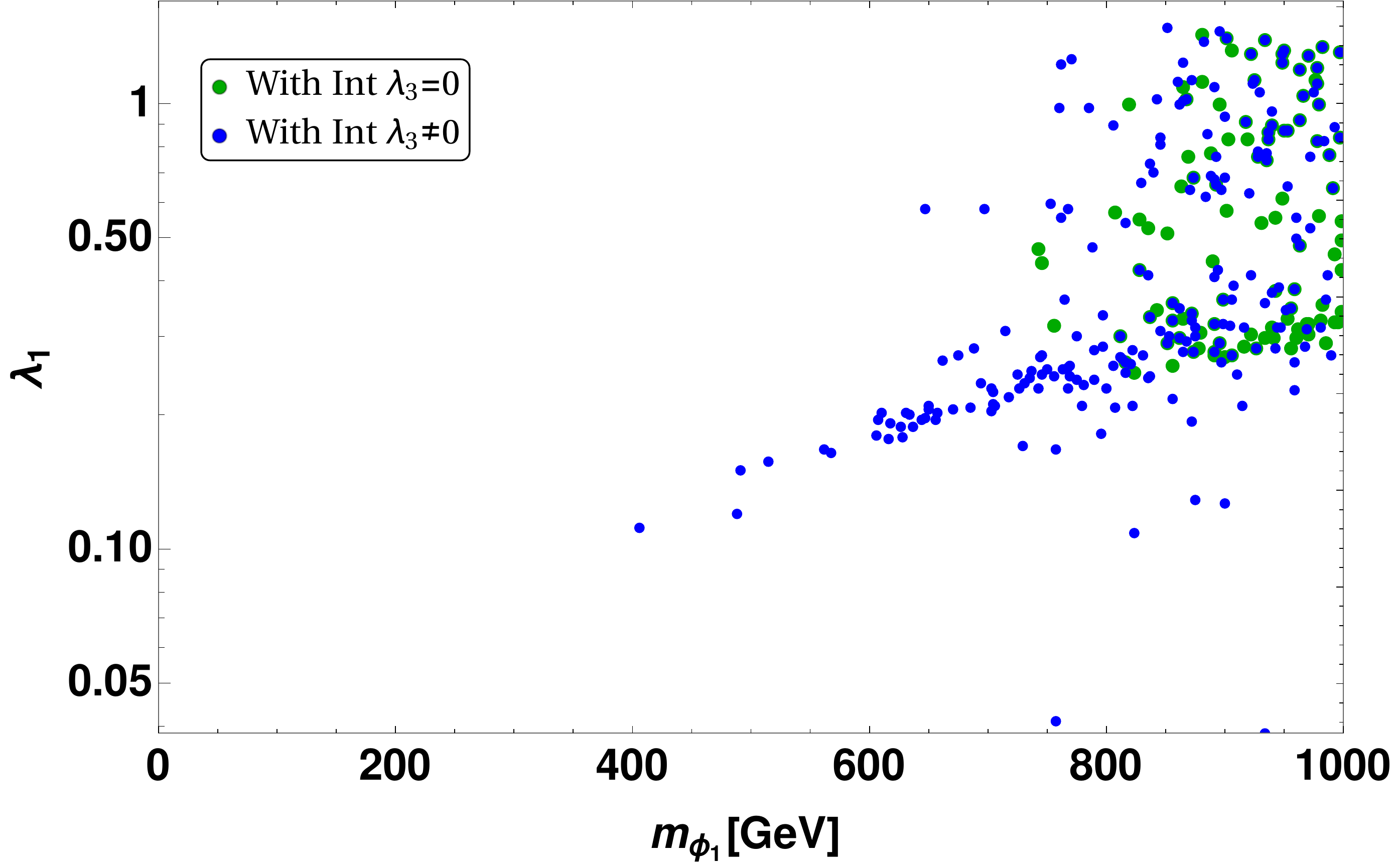}
  $$
 \caption{Direct search constraints on the two component model when $|\Delta m=\mphia-\mphib |\le300$ GeV behaves as degenerate DM scenario due to insensitivity of the detector. Left: Relic density allowed points in $\sigma_T^{SI}-\mphia$ plane satisfying LUX data; Right: Same points in $\mphia-\la$ plane. Blue points depict interacting scenario with $\lc\neq 0$, while the green points depict the case of $\lc=0$.}
 \label{fig:sig-tot}
\end{figure}

However, there is an important caveat to what we have discussed above in context of direct search of non-degenerate multicomponent DM model. Direct search of DM depends on the nuclear recoil of the detector. The rate of nuclear recoil is very less sensitive to DM mass above $\sim$400 GeV (as the allowed parameter space suggests here). For example, a DM of 400 GeV and 700 GeV yield similar rate so that they can hardly be distinguished. The effective recoil is then a sum of both the components interacting with the detector. In that circumstances, one needs to add the individual direct search effective cross-sections to evaluate the limit on the available parameter space of the model. Here, we assume that DMs with mass difference of $\le$ 300 GeV are indistinguishable. This is a simplified choice, the detailed analysis lies beyond the scope of the draft. Then, we choose the set of parameter space $\{\la,\lb,\lc, \mphia,\mphib\}$ which obeys relic density constraint for $|\Delta m=\mphia-\mphib |\le300$ GeV. We then add the individual direct search effective rates together to find the limit on $\phi_1$ as  $\sigma_T^{SI}= \frac{\Omega_1}{\Omega_T}\sigma_1^{DD}+\frac{\Omega_2}{\Omega_T}\left(\frac{m_{\phi_1}}{m_{\phi_2}}\right)\sigma_2^{DD}$. This follows from a simple derivation. The recoil rate in direct search \cite{recoil} is given by: 
\bea
\nonumber R&=&n_t\langle v \rangle n_{\phi}\sigma^{DD}\\
&=& \left(\frac{n_t \langle v \rangle \rho_c}{m_{\phi}}\right)\Omega_\phi \sigma^{DD} ~.
\label{eq:recoil}
\eea
where $n_t=\frac{N}{A}$ with $N$ is Avogadro number and $A$ is atomic mass of the target; rest of the variables are self explanatory. Then for the two component case: 
\bea
\nonumber R=\left(\frac{n_t \langle v \rangle \rho_c}{m_{\phi}}\right)\Omega_T \sigma_T^{SI}&=&\left(\frac{n_t \langle v \rangle \rho_c}{m_{\phi_1}}\right)\Omega_1 \sigma_1^{DD}+\left(\frac{n_t \langle v \rangle \rho_c}{m_{\phi_2}}\right)\Omega_2 \sigma_2^{DD} ~,\\
\nonumber &=& \left(\frac{n_t \langle v \rangle \rho_c}{m_{\phi_1}}\right)[\Omega_1\sigma_1^{DD}+\frac{m_{\phi_1}}{m_{\phi_2}}\Omega_2\sigma_2^{DD}]\\ 
\therefore && \sigma_T^{SI}= \frac{\Omega_1}{\Omega_T}\sigma_1^{DD}+\frac{\Omega_2}{\Omega_T}\left(\frac{m_{\phi_1}}{m_{\phi_2}}\right)\sigma_2^{DD} ~.
\label{eq:recoil-2comp}
\eea

 We plot $\sigma_T^{SI}$ it with respect to $\mphia$ in the left panel of Fig.~\ref{fig:sig-tot} to evaluate the direct search allowed region of parameter space. Blue points depict interacting scenario with $\lc \neq 0$, while the green points depict the case of $\lc=0$. On the right panel of Fig.~\ref{fig:sig-tot}, we show the same in $\mphia-\la$ plane. What we see that still with non-zero DM-DM interactions, one achieves a larger region of allowed parameter space than the single component set up assuming that a large mass gap of the DMs is insensitive to the detector, while the case for $\lc=0$ actually behaves like a degenerate two component scenario ($\mO_2$) for obvious reasons. The lower DM mass limit evaluated before is now shifted slightly to a higher value, so that points like BP2 may already be ruled out by LUX constraint.

\section{Higgs Invisible Decay Constraint in Two-Component set up}
\label{HiggsInv}

If DM masses are smaller than the Higgs mass as $m_{\phi_i}< m_h/2 $, then Higgs can decay to two DMs through the same vertex $h\rightarrow \phi_i\phi_i$ and will yield invisible decay width. There is a strong constraint on such a process, which in turn puts a constraint on DM-Higgs coupling for $m_{DM}<m_h/2$. 

\begin{figure}[htb!]
$$
 \includegraphics[height=7.5cm]{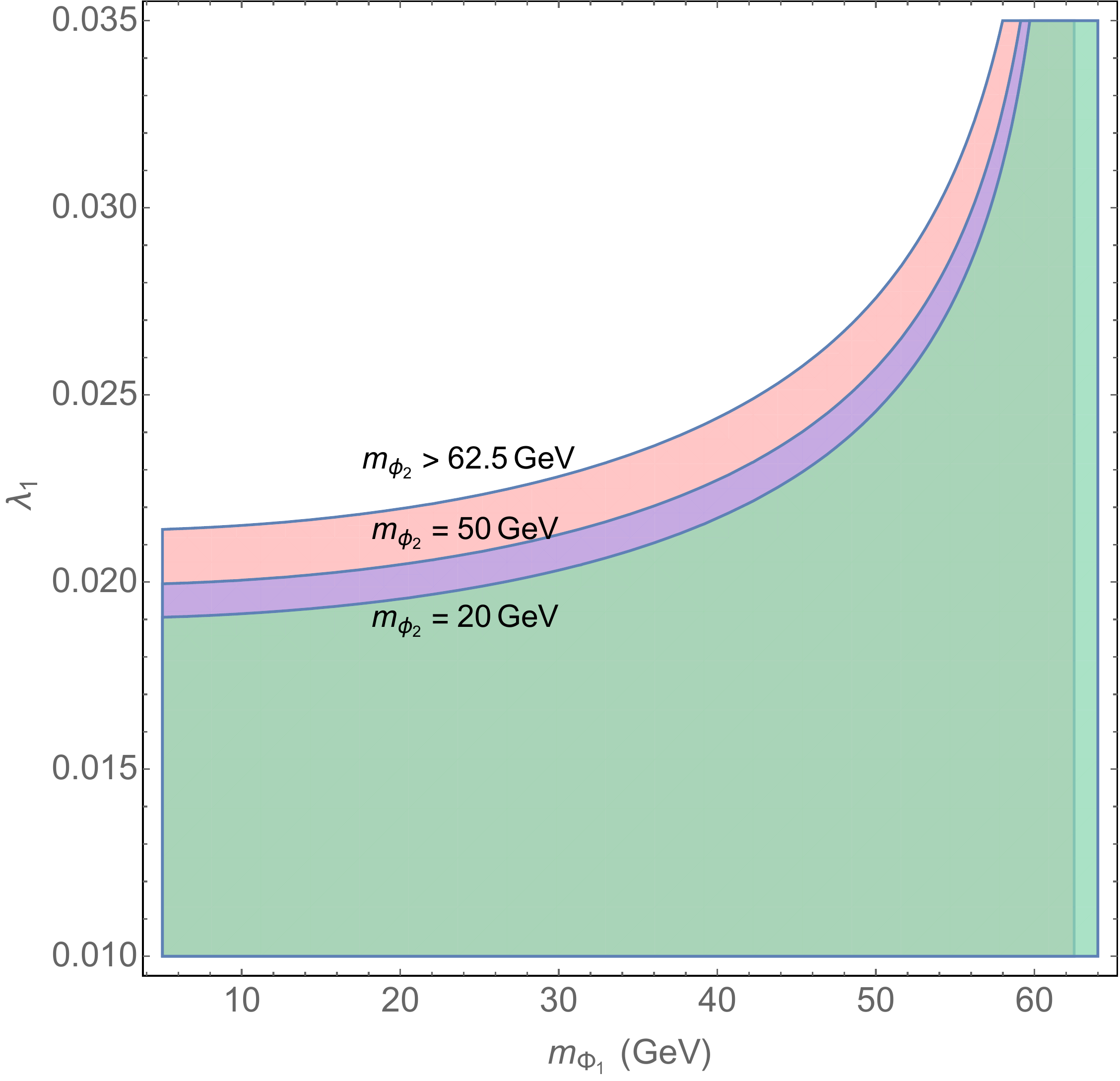}
 \includegraphics[height=7.5cm]{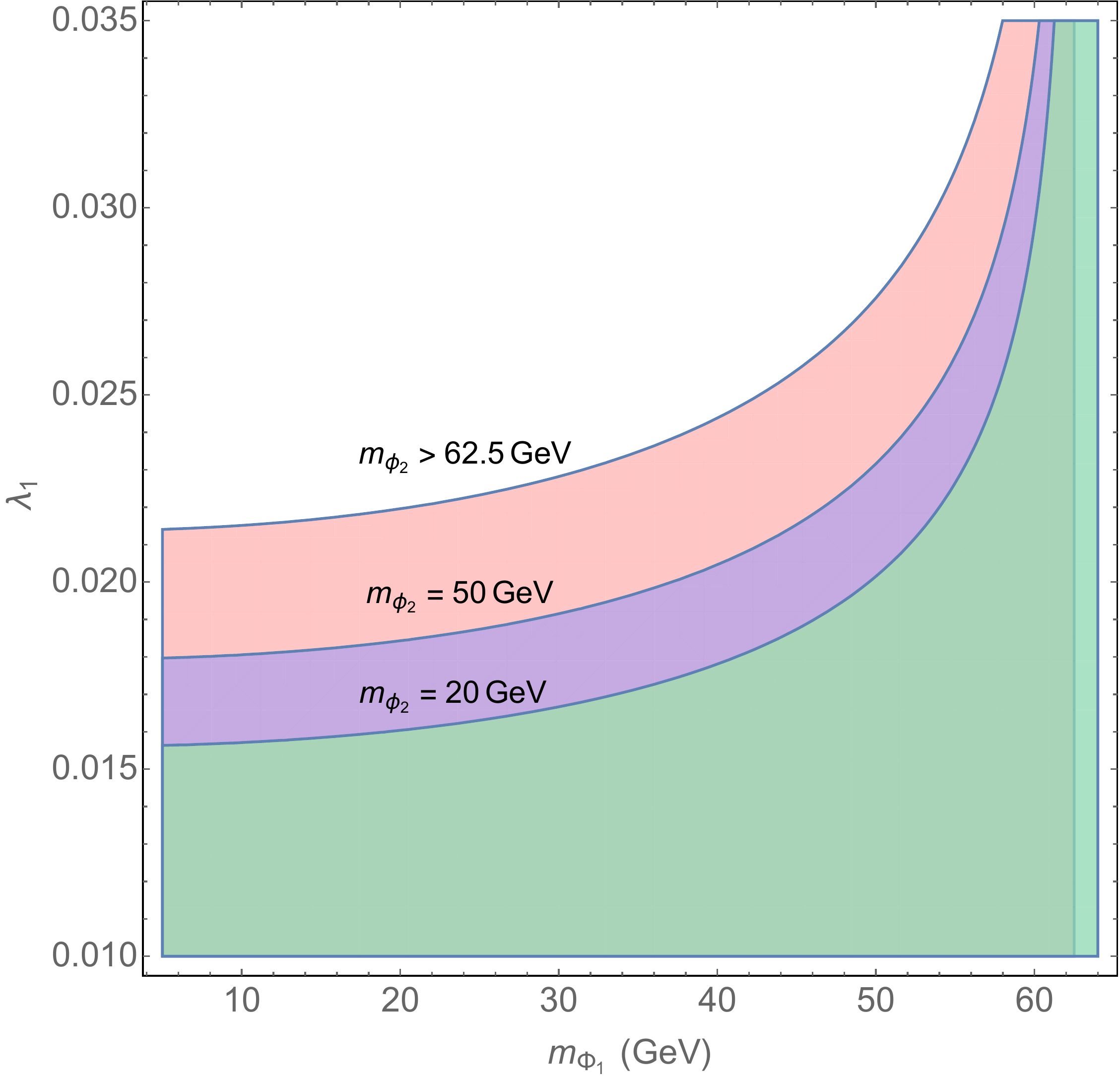}
 $$
 \caption{Constraints from invisible Higgs decay width on $\ZZp$ model in $\mphia-\la$ plane for $\lb=0.01$ (left) and $\lb=0.015$ (right) for $\mphib=20,~50~ \rm{ GeV ~and~ for~ \mphib}>62.5 $ GeV. For $\mphib=20$ GeV only the green region is allowed. For $\mphib=50$ GeV the lilac and green both are allowed and for $\mphib >62.5 $ GeV,  orange, lilac and green regions are allowed.}
 \label{fig:htoinv}
\end{figure}

The decay width of Higgs to one DM component ($\phi$) is given by \cite{s4,inv1}:
\beq
\Gamma_{h\to \phi \phi}= \frac{\lambda^2 v^2}{32 \pi m_h^2}(m_h^2-4m_{\phi}^2)^{1/2} ~.
\eeq
For the two component DM scenario in $\mZ_2 \times \mZ_2^{'}$ model, the total width will be 
\bea
\Gamma_{h \to inv}&=& \Gamma_{h\rightarrow \phi_1\phi_1 }+\Gamma_{h\rightarrow \phi_2\phi_2 } \nonumber \\
&=& \frac{\lambda_1^2 v^2}{32 \pi m_h^2}\sqrt{m_h^2-4m_{\phi_1}^2} 
+ \frac{\lambda_2^2 v^2}{32 \pi m_h^2}\sqrt{m_h^2-4m_{\phi_2}^2} ~.
\eea

From recent analysis at Large Hadron Collider (LHC), constraint on the invisible branching fraction of Higgs has become tighter $Br(h\rightarrow inv) \leq 0.35$ ~\cite{inv1,inv2}. Higgs decay width on the other hand is measured rather accurately at LHC and is given by $\Gamma_{h\to SM}=4.07$ MeV for Higgs mass $m_h=125~ {\rm GeV}$. For the two component set up this then indicates  
\begin{eqnarray}
Br(h\rightarrow \phi_1\phi_1)+Br(h\rightarrow \phi_2\phi_2)\leq 0.35 ~,\nonumber\\
\Rightarrow \Gamma_{h\rightarrow \phi_1\phi_1 }+\Gamma_{h\rightarrow \phi_2\phi_2 }\leq 2.2~ {\rm MeV} ~. \nonumber
\end{eqnarray}

\begin{figure}[htb!]
$$
 \includegraphics[height=7.5cm]{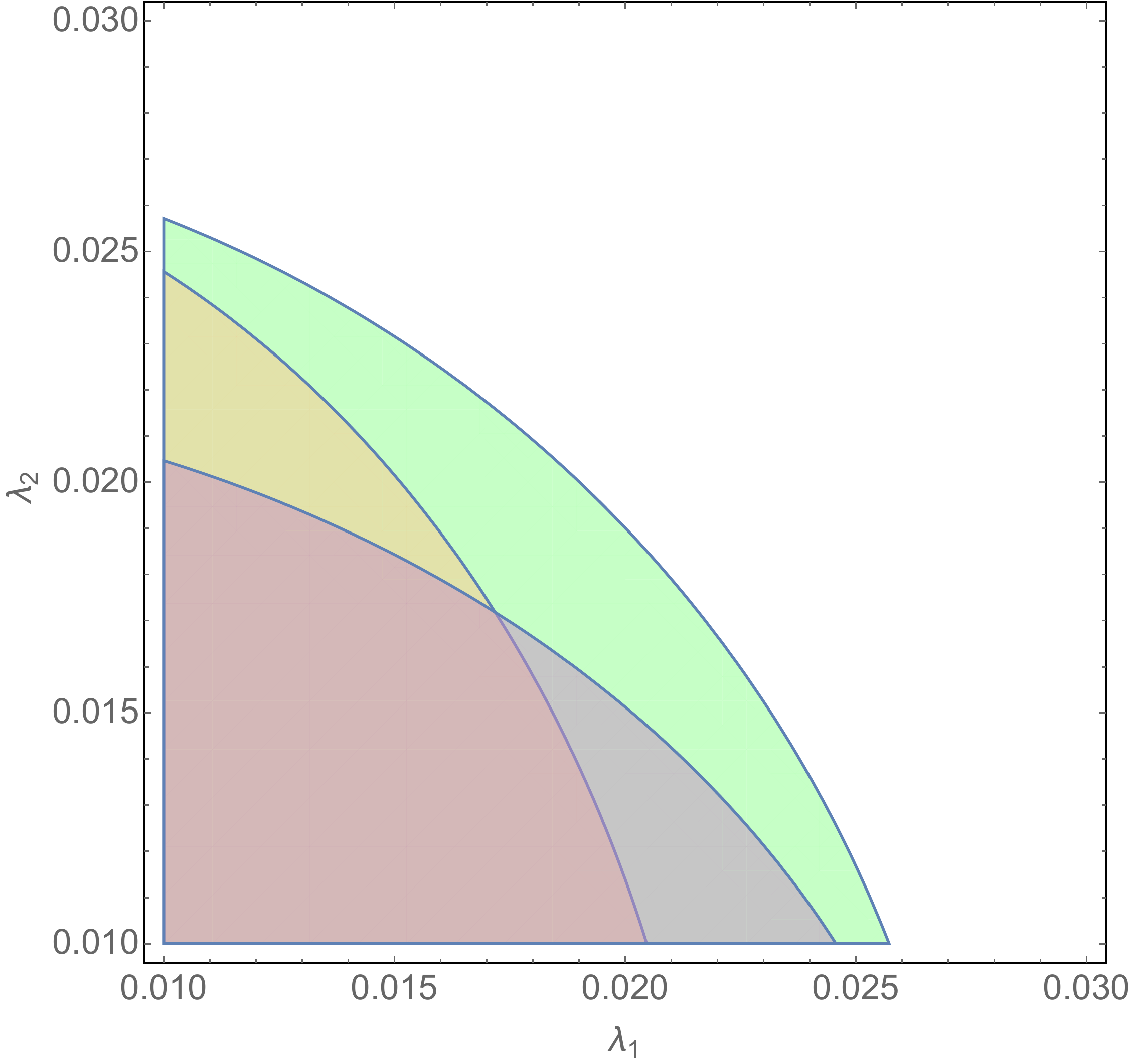}
 $$
 \caption{Constraints from invisible Higgs decay width on $\ZZp$ model in $\la-\lb$ plane for $\{\mphia,\mphib\}=\{50,50\}~(\rm{Green ~and ~below}), \{20,50\}~(\rm{Yellow ~and ~below}),\{50,20\}~(\rm{Grey~ and~ below})$. All masses are in GeVs.}
 \label{fig:htoinvl1l2}
\end{figure}

Clearly the constraints on individual DM-SM coupling is tighter in two component framework than the single component one. The constraint in $\ZZp$ model is presented in Fig.~\ref{fig:htoinv} in the plane of $\mphia-\la$ for specific choices of $\lb=0.01~ (\rm{left}), ~0.015~(\rm{right})$ and $\mphib=20,~50~ \rm{ GeV ~and~ for~ \mphib}>62.5 $ GeV. For $\mphib=20$ GeV, only the green region is allowed. For $\mphib=50$ GeV the lilac and green both are allowed indicating that the constraint is slightly loose on $\la$. For $\mphib >62.5 $ GeV, it effectively boils down to a single component scenario and all the regions below the curve is allowed. With larger $\lb$, the constraints on $\la$ is tighter as is clear from the RHS of Fig.~\ref{fig:htoinv}. Another possible representation of the invisible decay width constraint on $\ZZp$ model will be in $\la-\lb$ plane for different combinations of $\mphia, \mphib$ as shown in Fig.~\ref{fig:htoinvl1l2}. Here, we have chosen three different combinations of DM masses $\{\mphia,\mphib\}=\{50,50\}~(\rm{Green ~and ~below}), \{20,50\}~(\rm{Yellow ~and ~below}),\{50,20\}$ ({Grey and below}) showing that smaller the DM mass is, tighter is the corresponding coupling. 

To summarise, invisible Higgs decay constraint is tighter on multipartite DM models than the single component one as the invisible Higgs decay width to individual components get narrower if the DM component have masses $m_{\phi_i}< m_h/2 $. Such constraints are very important to take into account when we look into the Higgs resonance region and below ($m_{\phi_i} \le m_h/2$). What we conclude is that the invisible decay of Higgs rules out a viable scalar singlet DM for ($m_{\phi_i} < m_h/2$), as there is a clash between the constraints from invisible decay to that of DM relic density. Relic density requires the DM-SM couplings much larger ($\geq 0.1$) than what the invisible decay predicts (see for example, Fig.~\ref{fig:m1l1}). Direct search constraints also disfavour this region of parameter space for the singlet scalar DM (see for example, Fig. ~\ref{fig:ddl31}).

\section{Conclusions}

We have analysed the two-component scalar singlet DM scenarios coupled to SM with Higgs portal coupling, in particular with $\ZZp$  symmetry in the light of relic density and direct search constraints. We point out that DM-DM interaction plays a crucial role and yields a much larger region of parameter space from relic density and direct detection constraints. We also demonstrate the effect of DM-DM interaction in the freeze out of DM components and in particular how it alters the DM density from the non-interacting or single component scenario. A generic feature of interacting multicomponent DM framework turns out to be reducing the required DM-SM coupling compared to the non-interacting or single component cases, as has been demonstrated for a two component scenario. With more than two DM components, the changes in the available parameter space will be in regions where the DM components contribute equally. However, the lowest DM-SM coupling for one DM component that can be achieved is determined by two component framework itself. 

 On the other hand, direct search bound from updated LUX data constraints single component scalar singlet DM scenario upto DM mass $\sim$ 500 GeV and the two component case with $\mO(2)$ symmetry upto $\sim$ 890 GeV excepting for Higgs resonance region. It is through multipartite models such as $\ZZp$, scalar singlet DM can still survive in a large parameter space. The presence of non-zero DM-DM interaction coupling $\lambda_3$ can delay the direct detection of one component upto XENON1T while the other (lighter component) is expected to be unravelled soon. DM masses larger than $\sim$ 400 GeV seems to be allowed in $\ZZp$ case even if we assume that direct search is insensitive to a large mass gap (say, $\sim$ 300 GeV) between the DM components, so that the effective direct search cross-sections will be added for both the components. Other mechanisms of hiding DM from direct search bound includes co-annihilation, semi-annihilation etc., which we will elaborate elsewhere. Here, we also discuss approximate analytical solution for the $\ZZp$ case which closely agree to the numerical solutions in the moderate DM-DM interaction regions.  
  
 We propose a few benchmark points allowed by relic density and direct search for discovery potential at LHC. It is indeed important to study constraints coming from non-observation of missing energy signatures of such DM models at the LHC. The common notion is that the limits from collider are still weaker than the one from direct search, particularly for Higgs portal interactions. However one needs to carefully study the multicomponent framework in context of LHC and see if any new feature or correlation emerges. Such a study lies beyond the scope of this report, and will be taken in a systematic manner in future publication(s). Indirect search also plays a crucial role constraining the dark matter models, but again less constraining to multicomponent frameworks \cite{multi-singlet2}. 
  
It is important at the end to ask whether it is possible to unravel a multicomponent DM scenario in future experiments and how. We argue that the very existence of a scalar singlet DM in future detection may hint towards a multipartite framework to satisfy relic density and direct search constraints. Given the knowledge of one DM mass and coupling with SM, one can loosely predict the other DM mass and coupling to account for correct relic density and direct search limits. We plan to elaborate more in this direction from the point of view of signatures at the collider. \\

{\bf Acknowledgements:} SB would like to acknowledge the hospitality at IIT Kharagpur and illuminative discussions with Dr. Tirtha Shankar Ray. SB also acknowledges the DST-INSPIRE project with grant no PHY/P/SUB/01 at IIT Guwahati. PG  would like to acknowledge Dr. Sunando Patra at IIT Guwahati for technical help in using mathematica. We acknowledge the discussion with Dr. Partha Sarathi Mandal from department of Mathematics at IIT Guwahati for the approximate analytical solution obtained in this work. 


\end{document}

%% file: mylatex.tex
\usepackage{bbm}                                                  

\newcommand{\nc}{\newcommand}
\nc{\beq}{\begin{equation}}  \nc{\eeq}{\end{equation}}
\nc{\bea}{\begin{eqnarray}}  \nc{\eea}{\end{eqnarray}}
\nc{\baa}{\begin{array}}     \nc{\eaa}{\end{array}}
\nc{\bit}{\begin{itemize}}   \nc{\eit}{\end{itemize}}
\nc{\ben}{\begin{enumerate}} \nc{\een}{\end{enumerate}}
\nc{\bce}{\begin{center}}    \nc{\ece}{\end{center}}
\nc{\bpm}{\begin{pmatrix}}   \nc{\epm}{\end{pmatrix}}
\nc{\bvt}{\begin{verbatim}}  \nc{\evt}{\end{verbatim}}
\def\to{\rightarrow}

\def\boldoverdot{\,{\raise6pt\hbox{\bf.}\!\!\!\!\>}}

\def\lcal{{\cal L}}

\def\ocal{{\cal O}}

\def\ee{{\bf e}}

\def\ZZ{{\bf Z}}
\usepackage{bm}
\def\diag{\hbox{\diag}}

\def\gev{\hbox{GeV}}

\def\m{\hbox{m}}

\def\doubleundertext#1{
{\undertext{\vphantom{y}#1}}\par\nobreak\vskip-\the\baselineskip\vskip4pt%
\undertext{\hbox to 2in{}}}
\def\inbox#1{\vbox{\hrule\hbox{\vrule\kern5pt
     \vbox{\kern5pt#1\kern5pt}\kern5pt\vrule}\hrule}}
\def\sqr#1#2{{\vcenter{\hrule height.#2pt
      \hbox{\vrule width.#2pt height#1pt \kern#1pt
         \vrule width.#2pt}
      \hrule height.#2pt}}}

\def\today{\ifcase\month\or
  January\or February\or March\or April\or May\or June\or
  July\or August\or September\or October\or November\or December\fi
  \space\number\day, \number\year}
\def\pmb#1{\setbox0=\hbox{#1}%
  \kern-.025em\copy0\kern-\wd0
  \kern.05em\copy0\kern-\wd0
  \kern-.025em\raise.0433em\box0 }

\def\pmbb#1{\setbox0=\hbox{#1}%
  \kern-.02em\copy0\kern-\wd0
  \kern.04em\copy0\kern-\wd0
  \kern-.02em\raise.03464em\box0 }
\def\sumprime_#1{\setbox0=\hbox{$\scriptstyle{#1}$}
  \setbox2=\hbox{$\displaystyle{\sum}$}
  \setbox4=\hbox{${}'\mathsurround=0pt$}
  \dimen0=.5\wd0 \advance\dimen0 by-.5\wd2
  \ifdim\dimen0>0pt
  \ifdim\dimen0>\wd4 \kern\wd4 \else\kern\dimen0\fi\fi
\mathop{{\sum}'}_{\kern-\wd4 #1}}